\documentclass[aps,prd,10pt,notitlepage,nofootinbib]{revtex4-1}

\usepackage[utf8]{inputenc}
\usepackage{amsmath,amssymb,amsfonts}
\usepackage{newtxtext,newtxmath}

\usepackage{bbold}
\usepackage{bm}
\usepackage{graphicx}
\usepackage[usenames,dvipsnames]{xcolor}
\usepackage{color}
\usepackage[colorlinks=true,linkcolor=blue,urlcolor=blue,citecolor=blue]{hyperref}

\usepackage{slashed}
\usepackage[english]{babel}
\usepackage{dcolumn}
\usepackage{pifont}
\usepackage{dsfont,mathrsfs}
\usepackage{cancel}
\usepackage{bigints}
\usepackage{accents}
\usepackage{soul}
\usepackage{multirow}

\usepackage{amsmath, amssymb}
\usepackage{bbold}
\usepackage{makecell}
\usepackage{simpler-wick}
\usepackage{setspace}
\usepackage{scalerel}
\usepackage[normalem]{ulem}
\usepackage{hyperref} 
\usepackage{amsmath}

\usepackage{tikz}
\usetikzlibrary{svg.path}
\definecolor{orcidlogocol}{HTML}{A6CE39}
\tikzset{
	orcidlogo/.pic={
		\fill[orcidlogocol] svg{M256,128c0,70.7-57.3,128-128,128C57.3,256,0,198.7,0,128C0,57.3,57.3,0,128,0C198.7,0,256,57.3,256,128z};
		\fill[white] svg{M86.3,186.2H70.9V79.1h15.4v48.4V186.2z}
		svg{M108.9,79.1h41.6c39.6,0,57,28.3,57,53.6c0,27.5-21.5,53.6-56.8,53.6h-41.8V79.1z M124.3,172.4h24.5c34.9,0,42.9-26.5,42.9-39.7c0-21.5-13.7-39.7-43.7-39.7h-23.7V172.4z}
		svg{M88.7,56.8c0,5.5-4.5,10.1-10.1,10.1c-5.6,0-10.1-4.6-10.1-10.1c0-5.6,4.5-10.1,10.1-10.1C84.2,46.7,88.7,51.3,88.7,56.8z};
	}
}
\newcommand{\orcidlink}[1]{\href{https://orcid.org/#1}{\mbox{\scalerel*{\begin{tikzpicture}[yscale=-1,transform shape]\pic{orcidlogo};\end{tikzpicture}}{|}}}}

\newcommand{\nn}{\nonumber}

\newcommand{\FB}[1]{\left(#1\right)}

\newcommand{\SB}[1]{\left\{#1\right\}}
\newcommand{\TB}[1]{\left[#1\right]}
\newcommand{\AB}[1]{\left<#1\right>}

\newcommand{\munu}{{\mu\nu}}

\allowdisplaybreaks

\begin{document}
	\title{Speed of Sound in Magnetized Nuclear Matter}
	
	\author{Rajkumar Mondal\orcidlink{0000-0002-1446-6560}$^{a,c}$}
	\email{rajkumarmondal.phy@gmail.com}

	\author{Nilanjan Chaudhuri\orcidlink{0000-0002-7776-3503}$^{a,c}$}
	\email{nilanjan.vecc@gmail.com}
	\email{n.chaudhri@vecc.gov.in}
	

	\author{Pradip Roy$^{b,c}$}
	\email{pradipk.roy@saha.ac.in}
		
	\author{Sourav Sarkar\orcidlink{0000-0002-2952-3767}$^{a,c}$}
	\email{sourav@vecc.gov.in}

	\affiliation{$^a$Variable Energy Cyclotron Centre, 1/AF Bidhannagar, Kolkata - 700064, India}
	\affiliation{$^b$Saha Institute of Nuclear Physics, 1/AF Bidhannagar, Kolkata - 700064, India}
	\affiliation{$^c$Homi Bhabha National Institute, Training School Complex, Anushaktinagar, Mumbai - 400085, India} 
	

\begin{abstract}	
	 Employing the non-linear Walecka model we investigate the characteristics of nuclear matter under the influence of a background magnetic field at a finite temperature and baryon chemical potential. In the presence of the magnetic field the spinodal lines and the critical end point (CEP) undergo changes in the $T-\mu_B$ plane. The squared speed of sound exhibits anisotropic behavior, dividing into parallel and perpendicular components. Additionally, the presence of a magnetic field induces anisotropy in the isothermal compressibility. It is found that the parallel component is smaller than the perpendicular one for all values of temperature, chemical potential and magnetic field indicating that the equation of state is stiffer along the magnetic field direction.
\end{abstract}
\maketitle

\section{Introduction}
The speed of sound, represented as $C_x$, is an important quantity intrinsic to all thermodynamic systems. In the context of fluid dynamics, it denotes the speed at which a longitudinal compression wave propagates through the medium. Mathematically, it is calculated as the square root of the ratio between a change in pressure $(p)$ and a corresponding shift in energy density $(\epsilon)$, while holding $x$ as a constant parameter employed in the calculation. Hence, it establishes a direct correlation with the thermodynamic properties of the system, including its equation of state (EoS).

In the domain of dense nuclear matter the speed of sound holds particular significance for neutron star research. The variation of $C_x$ with density has a substantial impact on the mass-radius relationship, cooling rate,  the maximum possible mass of neutron star~\cite{Ozel:2016oaf} and tidal deformability. 
Analysis of current neutron star data indicates a substantial increase in $C^2_x$ at densities $(n_B)$ beyond the nuclear saturation density $(n_0)$~\cite{Bedaque:2014sqa,Tews:2018kmu,McLerran:2018hbz,Fujimoto:2019hxv}. 
Moreover, as indicated in Ref.~\cite{Jaikumar:2021jbw}, the speed of sound has crucial impact on the frequencies of gravitational waves generated by the $g$-mode oscillation of a neutron star.

Currently, the sole experimental method available for studying strongly interacting hot and/or dense matter in the laboratory is through relativistic heavy-ion collisions in which a new state of transient matter called quark-gluon plasma (QGP) is expected to be formed as a result of phase transition/crossover from hadronic matter at high temperature and/or density~\cite{STAR:2005gfr}. 
The first-principle lattice QCD (LQCD) calculations indicates that the transition from hadronic matter to QGP is a smooth crossover at high temperature and low baryon chemical potential~\cite{Borsanyi:2020fev,Borsanyi:2013hza,Aoki:2006we,HotQCD:2014kol}. 
A first-order phase transition is predicted by few models at large chemical potential with a critical endpoint connecting with a crossover transition~\cite{Fukushima:2008wg,Qin:2010nq,Fu:2019hdw}. 
The investigation of nuclear matter at high baryon number densities $n_B$, as conducted by programs like the Beam Energy Scan at the Relativistic Heavy Ion Collider (RHIC), plays a particularly vital role in the search for the QCD critical point. During the space-time evolution of the QGP the speed of sound plays a crucial role emerges in characterizing the EoS which is an essential input to the hydrodynamic equations. The sensitivity of the speed of sound on  temperature, density, chemical potential, etc. provides crucial insights: it exhibits a local minimum at a crossover transition, while it reaches zero at the critical point and along the corresponding spinodal lines. At a vanishing baryon chemical potential $(\mu_B=0)$,  LQCD demonstrates a minimum in the speed of sound $C_s$ at temperature $T_0=156.5\pm1.5$ MeV, signifying crossover transition between hadron gas and QGP~\cite{HotQCD:2018pds}. 
The speed of sound in QCD matter has been computed using various methods including LQCD ~\cite{Borsanyi:2020fev,Philipsen:2012nu,Aoki:2006we,HotQCD:2014kol}, the (Polyakov–)Nambu–Jona-Lasinio [(P)NJL] model ~\cite{He:2022kbc,Marty:2013ita}, the quark-meson coupling model \cite{Abhishek:2017pkp,Schaefer:2009ui}, the hadron resonance gas (HRG) model \cite{Venugopalan:1992hy,Bluhm:2013yga}, the field correlator method (FCM) \cite{Khaidukov:2018lor,Khaidukov:2019icg} and the quasiparticle model \cite{Mykhaylova:2020pfk}. 
It is conjectured that in non-central HICs, a very strong magnetic field is  generated due to the rapid movement of the electrically charged spectators during the initial phase of the collision. The estimated strength of this magnetic field is around $10^{15-18}$ Gauss~\cite{Kharzeev:2007jp,Skokov:2009qp,Tuchin:2013apa} and it experiences a rapid decay within a few fm/c. However, owing to the finite conductivity (approximately a few MeV) of the produced medium, the decay of the magnetic field is significantly delayed,  allowing a non-zero magnetic field to persist even during the subsequent hadronic phase, following the phase transition/crossover from the QGP~\cite{Kalikotay:2020snc,Gursoy:2014aka,Inghirami:2016iru}. Such type of matter can also be found in the outer layer of magnetars~\cite{Duncan:1992hi,Thompson:1993hn}. The presence of a large magnetic field could lead to significant modifications in the properties of the hadronic matter, leading to extensive research in this area, a few notable examples of which we mention here. The impact of magnetic fields on transport properties within the hadronic medium has been studied in Refs.~\cite{Kadam:2014xka,Das:2019pqd,Dash:2020vxk,Ghosh:2022xtv,Das:2019wjg,Kalikotay:2020snc}. Additionally, estimates of shear and bulk viscosity from magnetically modified hadronic matter have been explored in various approaches in  Refs.~\cite{Kadam:2014xka,Das:2019pqd,Dash:2020vxk,Ghosh:2022xtv}. Furthermore, the effects of magnetic fields on the electrical conductivity of a strongly interacting hadron gas have been investigated in Refs.~\cite{Das:2019wjg,Kalikotay:2020snc}. The present authors have also studied the dilepton production rate from magnetized hadronic medium in Refs.~\cite{Mondal:2023vzx,Mondal:2023ypq}.
Ref.~\cite{Mukherjee:2018ebw} describes the effect of a constant background magnetic field on nucleon mass in a strongly interacting medium in the weak field approximation within the Walecka model.
Recently in Ref.~\cite{He:2022yrk} the authors studied the speed of sound and liquid-gas phase transition in nuclear matter at finite temperature and density (chemical potential) using the nonlinear Walecka model. In this work, we will study the the nature of speed of sound and liquid-gas phase transition in nuclear matter in presence of background magnetic field in the frame work of nonlinear Walecka model. In addition we also investigate the isothermal compressibility in the presence of magnetic field.

The article is structured as follows: Section \ref{WM} provides a concise overview of the general formalism of the nonlinear Walecka model, while Section \ref{SpdSound} presents expressions for the sound speed in various thermodynamical situations. In Section \ref{Numerical}, we investigate the results, followed by a summary and conclusion in Section \ref{SC}. Additional details can be found in the appendix.
\section{Walecka Model}\label{WM} 

The Lagrangian density of the Walecka model, describing the nucleons-meson system, is given by:
\begin{eqnarray}
	\mathcal{L}=\mathcal{L_{\rm em}}+\mathcal{L_N}+\mathcal{L_{\rm mes}}+\mathcal{L_I}~~.
\end{eqnarray} 
Here, 
$\mathcal{L_{\rm em}}=-\frac{1}{2}F_{\munu}F^{\munu}$  represents the free field part where $F_{\munu}$ is the field tensor corresponding to the external magnetic field, $\mathcal{L_N}$ describes nucleons in a magnetic field, $\mathcal{L_{\rm mes}}$ denotes the free mesons and their self-interactions and $\mathcal{L_I}$ contains the interactions between nucleons mediated by the $\sigma$ and $\omega$ mesons:
\begin{eqnarray}
	\mathcal{L_N}=\bar{\psi}\FB{i\gamma^\mu D_\mu-m_N+\gamma^0\mu_B}\psi ~~,
\end{eqnarray}
\begin{eqnarray}
\mathcal{L_I}=g_\sigma\bar{\psi}\sigma\psi-g_\omega\bar{\psi}\gamma^\mu\omega_\mu\psi~~,
\end{eqnarray}
\begin{eqnarray}
	\mathcal{L_{\rm mes}}=\frac{1}{2}(\partial_\mu\sigma\partial^\mu\sigma-m_\sigma^2\sigma^2)-\frac{b}{3}m_N(g_\sigma\sigma)^3-\frac{c}{4}(g_\sigma\sigma)^4-\frac{1}{4}\omega_\munu\omega^\munu+\frac{1}{2}m_\omega^2\omega_\mu\omega^\mu
\end{eqnarray}
where $\omega_\munu=\partial_\mu\omega_\nu-\partial_\nu\omega_\mu$, $\psi$ represents the nucleon isospin doublet, the covariant derivative $D_\mu=\partial_\mu+ieA_\mu$ with $A_\mu=(0, yB, 0, 0)$ representing a homogeneous background magnetic field in the $z$ direction and $e$ is the electric charge of the proton. 

We apply the mean-field approximation to calculate the free energy, i.e, we neglect the fluctuations around the background mesonic fields assumed to be uniform in space-time. Therefore, the dynamical mass and the effective chemical potential of nucleon are
\begin{eqnarray}
	M=m_N-g_\sigma\bar{\sigma}~,~~~~~~\mu^\star=\mu_B-g_\omega{\bar{\omega}}_0
\end{eqnarray}
where $\bar{\sigma}=\AB{\bar\psi\psi}$ and ${\bar{\omega}}_0=\AB{\bar\psi\gamma^0\psi}$.

The model parameters - $g_\sigma,~g_\omega,~b,~c$ - are fitted in mean-field approximation to reproduce the properties of nuclear matter at saturation in absence of magnetic field such as the saturation density $n_0=0.153~\rm fm^{-3}$, compression modulus $K = 240~\rm MeV$, binding energy $E_{\rm bind}=-16.3~\rm MeV$ and the effective nucleon mass $M=0.8m_N$. This leads to the chemical potential $\mu_0=922.7~\rm MeV$ at saturation. The parameters are listed in table \ref{Table1}.~\cite{Haber:2014ula}
\begin{table*}[htb!]
	\caption{\label{Table1}Parameters in Walecka Model}
	\begin{tabular}{ccccccccc}
		\\\hline\hline \\
		$m_\sigma$(MeV)~~~~~~~~~~&$m_\omega$(MeV)~~~~~~~~~&$m_N$(MeV)~~~~~~~~~~~~~&$g_\sigma$~~~~~~~~~&$g_\omega$~~~~~~~&$b$~~~~~~~~~~~~~~&$c$\\
		\\\hline\\
		550~~~~~~~~~~&782~~~~~~~&939~~~~~~~~~~~~~~~~&8.1617~~~~~~~~~~&~~~8.5062~~~~~~~~~~&$1.0784\times 10^{-2}$~~~~~~~&$-6.2205\times 10^{-3}$\\
		\hline\hline
	\end{tabular}   
\end{table*}

In this model the free energy has the form
\begin{eqnarray}
	\Omega=\frac{B^2}{2}+U+\Omega_N~~.
\end{eqnarray}
$\frac{(eB)^2}{2}$ stems from the magnetic field pointing in the $z$ direction, i.e, $\boldsymbol{B}(0,0,B)$, $U$ is the tree-level potential given by
\begin{eqnarray}
	U=\frac{1}{2}m_\sigma^2\bar{\sigma}^2+\frac{b}{3}m_N(g_\sigma\bar{\sigma})^3+\frac{c}{4}(g_\sigma\bar{\sigma})^4-\frac{1}{2}m_\omega^2{\bar{\omega}}_0^2
\end{eqnarray}
 and $\Omega_N$ is the nucleonic contribution to the free energy. It depends on the dynamical nucleon mass $M(B, \mu, T)$ which is determined by minimizing the free energy. We can decompose $\Omega_N(M, B, \mu, T)$ as
 \begin{eqnarray}
	\Omega_N=\Omega_{\rm sea}+\Omega_{\rm TM}
 \end{eqnarray}
 where $\Omega_{\rm sea}$ contains pure vacuum as well as the magnetic field dependent vacuum contributions and $\Omega_{\rm TM}$ is the thermo-magnetic (TM) contribution in free energy. We can write the expressions as
 \begin{eqnarray}
 	\Omega_{\rm sea}&=&-2\int\frac{d^3k}{\FB{2\pi}^3}E-\frac{eB}{2\pi}\sum_{n=0}^{\infty}\alpha_n\int\frac{dk_z}{2\pi}E_n\\
 	&=&\Omega_{\rm vac}+\Omega_{\rm vac}^B
 \end{eqnarray}
 where $\alpha_n=2-\delta_{n0}$ arises due to the spin degeneracy of each Landau level ($n$),  $E=\sqrt{M^2+k^2}$, $E_n=\sqrt{M^2+k_z^2+2neB}$ for the spin $\frac{1}{2}$ fermion with electric charge $e$.
The $eB-$dependent vacuum contribution can be expressed as (see Appendix \ref{A1} for details)
\begin{eqnarray}
 \Omega_{\rm vac}^B&=&-\frac{(eB)^2}{2\pi^2}\SB{\zeta'\FB{-1,x}+\frac{1}{4}x^2+\frac{1}{2}x(1-x)~\text{ln}~x}
 \end{eqnarray}
 where $x=\frac{M^2}{2eB},$~ $\zeta'(-1,x)=\frac{d\zeta(z,x)}{dz}|_{z=-1}$, $\zeta(z,x)$ is Hurwitz zeta function and
 \begin{eqnarray}
	 \Omega_{\rm TM}=2\beta^{-1}\int\frac{d^3k}{\FB{2\pi}^3}\SB{\text{ln}(1-f^+)+\text{ln}(1-f^-)}+\beta^{-1}\frac{eB}{2\pi}\sum_{n=0}^{\infty}\alpha_n\int\frac{dp_z}{2\pi}\SB{\text{ln}(1-f_n^+)+\text{ln}(1-f_n^-)}~~.
 \end{eqnarray}
Here $f^\pm=f(E\mp\mu^\star)$, $f_n^\pm=f(E_n\mp\mu^\star)$, $\beta=\frac{1}{T}$, $f(x)=\frac{1}{e^{x/T}+1}$ is the Fermi distribution function and "+(-)"  corresponds to the fermion(anti-fermion) distribution function. 
		Now, minimizing the free energy, i.e, putting 
\begin{eqnarray}
 	\frac{\partial\Omega}{\partial\bar{\sigma}}=\frac{\partial\Omega}{\partial\bar{\omega}_0}=0~~,
\end{eqnarray}
we get the equations
\begin{eqnarray}
 	n_s&=&-\frac{M-m_N}{\FB{g_\sigma/m_\sigma}^2}+bm_N\FB{m_N-M}^2+c\FB{m_N-M}^3+\frac{eB}{2\pi^2}M\SB{x(1-\text{ln}~x)+\frac{1}{2}\text{ln}\frac{x}{2\pi}+\text{ln}~\Gamma(x)}\label{TansEq.ns}\label{ns1}~~,\\ 	
 	n_B&=&\frac{\mu-\mu^\star}{\FB{g_\omega/m_\omega}^2}\label{nB1}
\end{eqnarray}
where we have defined the scalar and baryon densities as
\begin{eqnarray}
	n_s&=&2\int\frac{d^3k}{(2\pi)^3}\frac{M}{E}\FB{f^{+}+f^{-}}+\frac{eB}{2\pi}\sum_{n=0}^{\infty}\alpha_n\int\frac{dk_z}{(2\pi)}\frac{M}{E_n}\FB{f_n^{+}+f_n^{-}}\label{ns2}~~,\\
	n_B&=&2\int\frac{d^3k}{(2\pi)^3}\FB{f^{+}-f^{-}}+\frac{eB}{2\pi}\sum_{n=0}^{\infty}\alpha_n\int\frac{dk_z}{(2\pi)}\FB{f_n^{+}-f_n^{-}}\label{nB2}~~.
\end{eqnarray}
Both $n_s$ and $n_B$ contain contributions from neutrons and protons, while the term within second brackets in Eq.\eqref{TansEq.ns} is the renormalized term coming from the derivative $\frac{\partial\Omega_{\rm sea}}{\partial M}$ which contributes only for protons. Furthermore, the expressions for entropy density and magnetization are as follows:
\begin{eqnarray}
	s&=&-\frac{\partial\Omega}{\partial T}
	=-2\int\frac{d^3k}{(2\pi)^3}\SB{\text{ln}\FB{1-f^+}+\text{ln}\FB{1-f^-}-\frac{E}{T}(f^++f^-)+\frac{\mu^\star}{T}(f^+-f^-)}\nn\\&&-\frac{eB}{2\pi}\sum_{n=0}^{\infty}\alpha_n\int\frac{dk_z}{2\pi} \left\lbrace  \text{ln}\FB{1-f_n^+}\right. \hspace{0cm}\left.+\text{ln}\FB{1-f_n^-}-\frac{E_n}{T}(f_n^++f_n^-)+\frac{\mu^\star}{T}(f_n^+-f_n^-)\right\rbrace~,
\end{eqnarray}
\begin{eqnarray}\label{Magnetization}
	\mathcal{M}&=&-\frac{\partial\Omega}{\partial B}=-B-\frac{eB}{2\pi^2}x\SB{x(1-\text{ln}x)+\text{ln}\Gamma(x)+\frac{1}{2}\text{ln}\frac{x}{2\pi}}\nn\\&&-\frac{T}{2\pi}\sum_{n=0}^{\infty}\alpha_n\int\frac{dp_z}{2\pi}\SB{\text{ln}(1-f_n^+)+\text{ln}(1-f_n^-)}-\frac{eB}{2\pi}\sum_{n=0}^{\infty}\alpha_n\int\frac{dp_z}{2\pi}\frac{n}{E_n}\SB{f_n^++f_n^-}~.
\end{eqnarray}

\section{Speed of Sound}\label{SpdSound}

The general definition of the speed of sound requires the specification of a constant quantity $x$ such as entropy density $s$, $s/n_B$, $T$, $\mu_B$ etc. during the propagation of the compression wave through a medium. The squared speed of sound $(C^2_x)$ is defined as
\begin{eqnarray}
	C^2_x=\FB{\frac{\partial p}{\partial\epsilon}}_x~.
\end{eqnarray}
Here, $p$ represents pressure and $\epsilon$ denotes energy density.

In relativistic HICs, the created ideal fluid evolves with constant $s/n_B$. This conclusion can be derived in hydrodynamics due to the conservation of energy and baryon number. Therefore, it is important to compute the squared sound speed $C^2_{s/n_B}$ along the isentropic curve.
%
%
%
The other definitions of speed of sound found in literature at constant baryon number density or entropy are usually used to describe the intermediate stages of hydrodynamic evolution.
%
%
%
%
%
Furthermore, it is also interesting to compute the squred speed of sound with constant temperature $T$ and chemical potential $\mu_B$. 
%
%

In this paper, we will investigate the speed of sound in nuclear matter subjected to a background magnetic field in the $T-\mu_B$ plane. The expressions for energy density and the longitudinal and transverse components of the pressure are as follows~\cite{Ferrer:2010wz,Ferrer:2022afu}:
%
\begin{eqnarray}
\epsilon&=&\Omega+Ts+\mu_Bn_B~~,\\
p^{\parallel}&=&-\Omega~~,~~~p^{\perp}=p^\parallel-B\mathcal{M}~~.
\end{eqnarray}
Correspondingly, the speed of sound becomes anisotropic due to the presence of magnetic field.
We will specify the speed of sound using different thermodynamic relations expressed in terms of temperature $T$ and baryon chemical potential $\mu_B$ as:
\begin{eqnarray}
{C_x^2}(T,\mu_B)&=&{C_x^2}^{(\parallel)}(T,\mu_B)=\FB{\frac{\partial p_\parallel}{\partial \epsilon}}_x\nn\\
&=&\frac{\FB{\frac{\partial{p^\parallel}}{\partial T}}_{\mu_B}\FB{\frac{\partial x}{\partial\mu_B}}_T-\FB{\frac{\partial{p^\parallel}}{\partial\mu_B}}_T\FB{\frac{\partial x}{\partial T}}_{\mu_B}}{\FB{\frac{\partial\epsilon}{\partial T}}_{\mu_B}\FB{\frac{\partial x}{\partial\mu_B}}_T-\FB{\frac{\partial\epsilon}{\partial\mu_B}}_T\FB{\frac{\partial x}{\partial T}}_{\mu_B}}\label{C2xParallel}~~,
\end{eqnarray}
\begin{eqnarray}
{C_x^2}^{(\perp)}(T,\mu_B)&=&\FB{\frac{\partial p_\perp}{\partial \epsilon}}_x={C_x^2}^{(\parallel)}-B\FB{\frac{\partial\mathcal{M}}{\partial\epsilon}}_x\nn\\
&=&{C_x^2}^{(\parallel)}-B\frac{\FB{\frac{\partial\mathcal{M}}{\partial T}}_{\mu_B}\FB{\frac{\partial x}{\partial\mu_B}}_T-\FB{\frac{\partial\mathcal{M}}{\partial\mu_B}}_T\FB{\frac{\partial x}{\partial T}}_{\mu_B}}{\FB{\frac{\partial\epsilon}{\partial T}}_{\mu_B}\FB{\frac{\partial x}{\partial\mu_B}}_T-\FB{\frac{\partial\epsilon}{\partial\mu_B}}_T\FB{\frac{\partial x}{\partial T}}_{\mu_B}}\label{C2xPerpendicular}
\end{eqnarray}
where ${C_x}={C_x}^{(\parallel)}$ and ${C_x}^{(\perp)}$ are the sound velocities along and perpendicular to the magnetic field direction respectively. Using the thermodynamic relation in Appendix~\ref{A2}, we can further write down the sound velocity along the magnetic field as:

\begin{eqnarray}
	{C_{s/n_B}^{2(\parallel)}}&=&\frac{n_Bs\FB{\frac{\partial s}{\partial \mu_B}}_T-s^2\FB{\frac{\partial n_B}{\partial \mu_B}}_T-n_B^2\FB{\frac{\partial s}{\partial T}}_{\mu_B}+sn_B\FB{\frac{\partial n_B}{\partial T}}_{\mu_B}}{\FB{sT+\mu_Bn_B}\SB{\FB{\frac{\partial s}{\partial\mu_B}}_T\FB{\frac{\partial n_B}{\partial T}}_{\mu_B}-\FB{\frac{\partial s}{\partial T}}_{\mu_B}\FB{\frac{\partial n_B}{\partial \mu_B}}_T}}\label{C2sbynB}~~,\\
	{C_{n_B}^{2(\parallel)}}&=&\frac{s\FB{\frac{\partial n_B}{\partial\mu_B}}_T-n_B\FB{\frac{\partial n_B}{\partial T}}_{\mu_B}}{T\SB{\FB{\frac{\partial s}{\partial T}}_{\mu_B}\FB{\frac{\partial n_B}{\partial \mu_B}}_T-\FB{\frac{\partial s}{\partial\mu_B}}_T\FB{\frac{\partial n_B}{\partial T}}_{\mu_B}}}~~,\\
	{C_s^{2(\parallel)}}&=&\frac{s\FB{\frac{\partial s}{\partial\mu_B}}_T-n_B\FB{\frac{\partial s}{\partial T}_{\mu_B}}_{\mu_B}}{\mu_B\SB{\FB{\frac{\partial s}{\partial\mu_B}}_T\FB{\frac{\partial n_B}{\partial T}}_{\mu_B}-\FB{\frac{\partial s}{\partial T}}_{\mu_B}\FB{\frac{\partial n_B}{\partial \mu_B}}_T}}~~,\\
	{C_T^{2(\parallel)}}&=&\frac{n_B}{T\FB{\frac{\partial s}{\partial\mu_B}}_T+\mu_B\FB{\frac{\partial n_B}{\partial \mu_B}}_T}~~,\\
	{C_{\mu_B}^{2(\parallel)}}&=&\frac{s}{T\FB{\frac{\partial s}{\partial T}}_{\mu_B}+\mu_B\FB{\frac{\partial n_B}{\partial T}}_{\mu_B}}\label{C2MuB}
\end{eqnarray}
and sound velocity perpendicular to the magnetic field as:
\begin{eqnarray}
{C_{s/n_B}^{2(\perp)}}&=&{C_{s/n_B}^{2(\parallel)}}-B\frac{n_B\SB{\FB{\frac{\partial\mathcal{M}}{\partial T}}_{\mu_B}\FB{\frac{\partial s}{\partial\mu_B}}_T-\FB{\frac{\partial\mathcal{M}}{\partial\mu_B}}_T\FB{\frac{\partial s}{\partial T}}_{\mu_B}}-s\SB{\FB{\frac{\partial\mathcal{M}}{\partial T}}_{\mu_B}\FB{\frac{\partial n_B}{\partial\mu_B}}_T-\FB{\frac{\partial\mathcal{M}}{\partial\mu_B}}_T\FB{\frac{\partial n_B}{\partial T}}_{\mu_B}}}{\FB{sT+\mu_Bn_B}\SB{\FB{\frac{\partial s}{\partial\mu_B}}_T\FB{\frac{\partial n_B}{\partial T}}_{\mu_B}-\FB{\frac{\partial s}{\partial T}}_{\mu_B}\FB{\frac{\partial n_B}{\partial \mu_B}}_T}}\label{C2sbynBP}~~,\\
C_{n_B}^{2(\perp)}&=&{C_{n_B}^{2(\parallel)}}-B\frac{\FB{\frac{\partial\mathcal{M}}{\partial T}}_{\mu_B}\FB{\frac{\partial n_B}{\partial\mu_B}}_T-\FB{\frac{\partial\mathcal{M}}{\partial \mu_B}}_{T}\FB{\frac{\partial n_B}{\partial T}}_{\mu_B}}{T\SB{\FB{\frac{\partial s}{\partial T}}_{\mu_B}\FB{\frac{\partial n_B}{\partial \mu_B}}_T-\FB{\frac{\partial s}{\partial\mu_B}}_T\FB{\frac{\partial n_B}{\partial T}}_{\mu_B}}}~~,\\
{C_s^{2(\perp)}}&=&{C_{s}^{2(\parallel)}}-B\frac{\FB{\frac{\partial\mathcal{M}}{\partial T}}_{\mu_B}\FB{\frac{\partial s}{\partial\mu_B}}_T-\FB{\frac{\partial\mathcal{M}}{\partial \mu_B}}_{T}\FB{\frac{\partial s}{\partial T}_{\mu_B}}_{\mu_B}}{\mu_B\SB{\FB{\frac{\partial s}{\partial\mu_B}}_T\FB{\frac{\partial n_B}{\partial T}}_{\mu_B}-\FB{\frac{\partial s}{\partial T}}_{\mu_B}\FB{\frac{\partial n_B}{\partial \mu_B}}_T}}~~,\\
{C_T^{2(\perp)}}&=&{C_T^{2(\parallel)}}-B\frac{\FB{\frac{\partial\mathcal{M}}{\partial \mu_B}}_{T}}{T\FB{\frac{\partial s}{\partial\mu_B}}_T+\mu_B\FB{\frac{\partial n_B}{\partial \mu_B}}_T}~~,\\
{C_{\mu_B}^{2(\perp)}}&=&{C_{\mu_B}^{2(\parallel)}}-B\frac{\FB{\frac{\partial\mathcal{M}}{\partial T}}_{\mu_B}}{T\FB{\frac{\partial s}{\partial T}}_{\mu_B}+\mu_B\FB{\frac{\partial n_B}{\partial T}}_{\mu_B}}\label{C2MuBPeprp}~~,
\end{eqnarray}
where the analytical expressions for the derivatives $\FB{\frac{\partial\mathcal{M}}{\partial T}}_{\mu_B}$, $\FB{\frac{\partial\mathcal{M}}{\partial\mu_B}}_T$, $\FB{\frac{\partial s}{\partial \mu_B}}_T$, $\FB{\frac{\partial s}{\partial T}}_{\mu_B}$, $\FB{\frac{\partial n_B}{\partial \mu_B}}_T$, $\FB{\frac{\partial n_B}{\partial T}}_{\mu_B}$ are provided in Appendix \ref{Appendix.ES}. 

In the case of zero magnetic field, the isothermal compressibility of the medium  is defined by the relation 
\begin{eqnarray}
	K_T=\frac{1}{n_B}\frac{\frac{\partial n_B}{\partial\mu_B}}{\frac{\partial p}{\partial\mu_B}}=\frac{1}{n_B^2}\FB{\frac{\partial n_B}{\partial\mu_B}}_T~.
\end{eqnarray}
Due to the anisotropy in pressure  $K_T$ splits into parallel and perpendicular components with respect to the magnetic field direction. The expressions for  $K_T^{(\parallel)}$ and $K_T^{(\perp)}$ are  given by
\begin{eqnarray}
K_T^{(\parallel)}&=&\frac{1}{n_B^2}\FB{\frac{\partial n_B}{\partial\mu_B}}_T~~,\label{KT_para}\\
K_T^{(\perp)}&=&\frac{1}{n_B\SB{n_B-B\FB{\frac{\partial\mathcal{M}}{\partial\mu_B}}_T}}\FB{\frac{\partial n_B}{\partial\mu_B}}_T.\label{KT_per}
\end{eqnarray}
\section{Numerical Results}\label{Numerical}
In this section  we will perform a numerical analysis of thermodynamic variables in magnetized nuclear matter. It is well known that the energy-momentum tensor shows anisotropies due to the breaking of spatial rotational symmetry when a finite value of magnetic field is present. Hence, the pressure becomes dependent on the direction of the background magnetic field~\cite{Ferrer:2022afu,Ferrer:2010wz}. Since we are interested in the thermodynamic properties of magnetized nuclear matter, we will choose two representative values $eB=0.02 \rm~GeV^2$ and $eB=0.05 \rm~GeV^2$ along with $eB=0$ which will provide us the opportunity to explore the interplay between the magnetic field and the thermal effects. In this section all the numerical calculations with finite values of background magnetic field will be evaluated by considering upto 1000 Landau levels ensuring the convergence of the results.
\subsection{Mass of Nucleon}
\begin{figure}[h] 
	\includegraphics[angle = -90, scale=0.23]{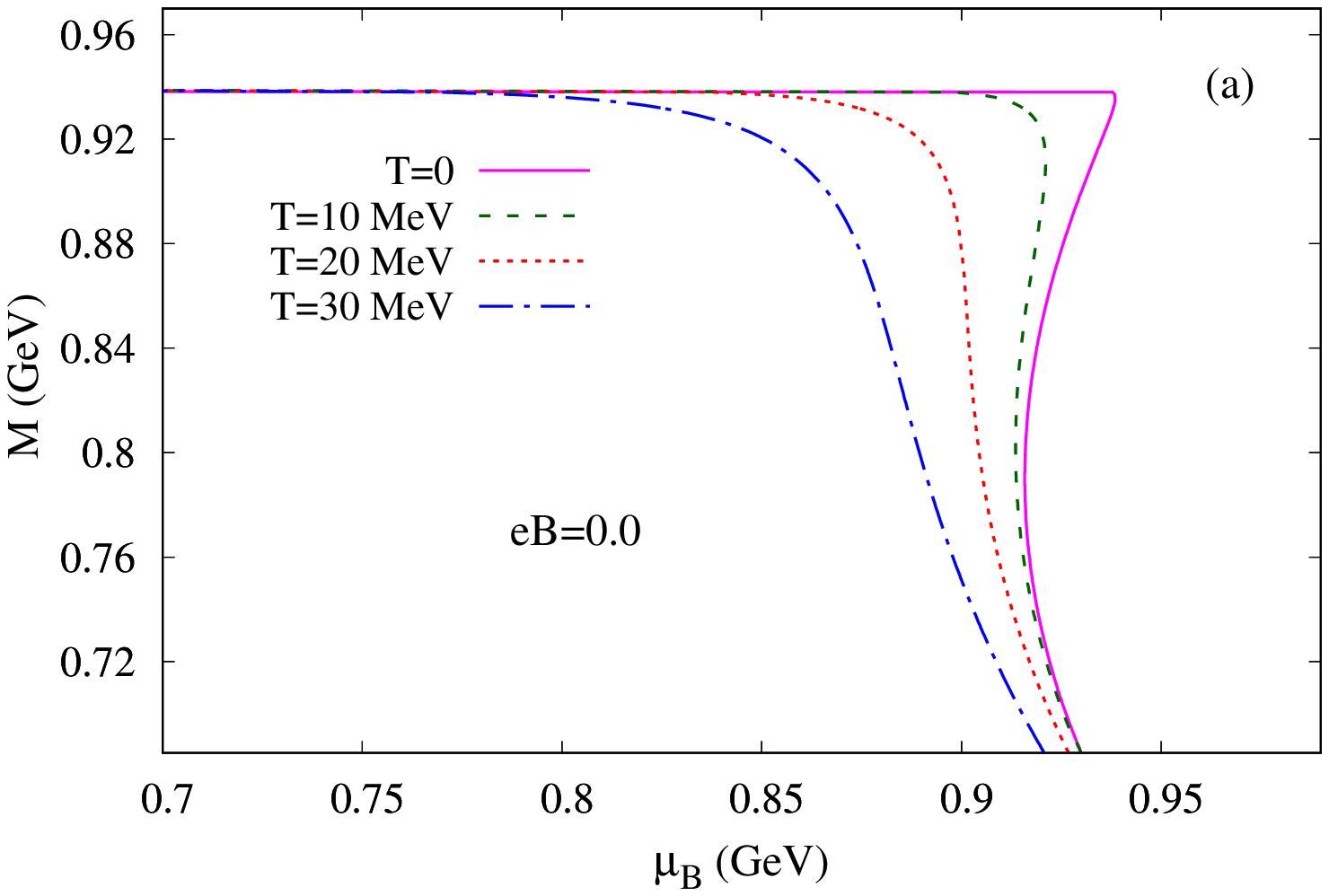}
	\includegraphics[angle = -90, scale=0.23]{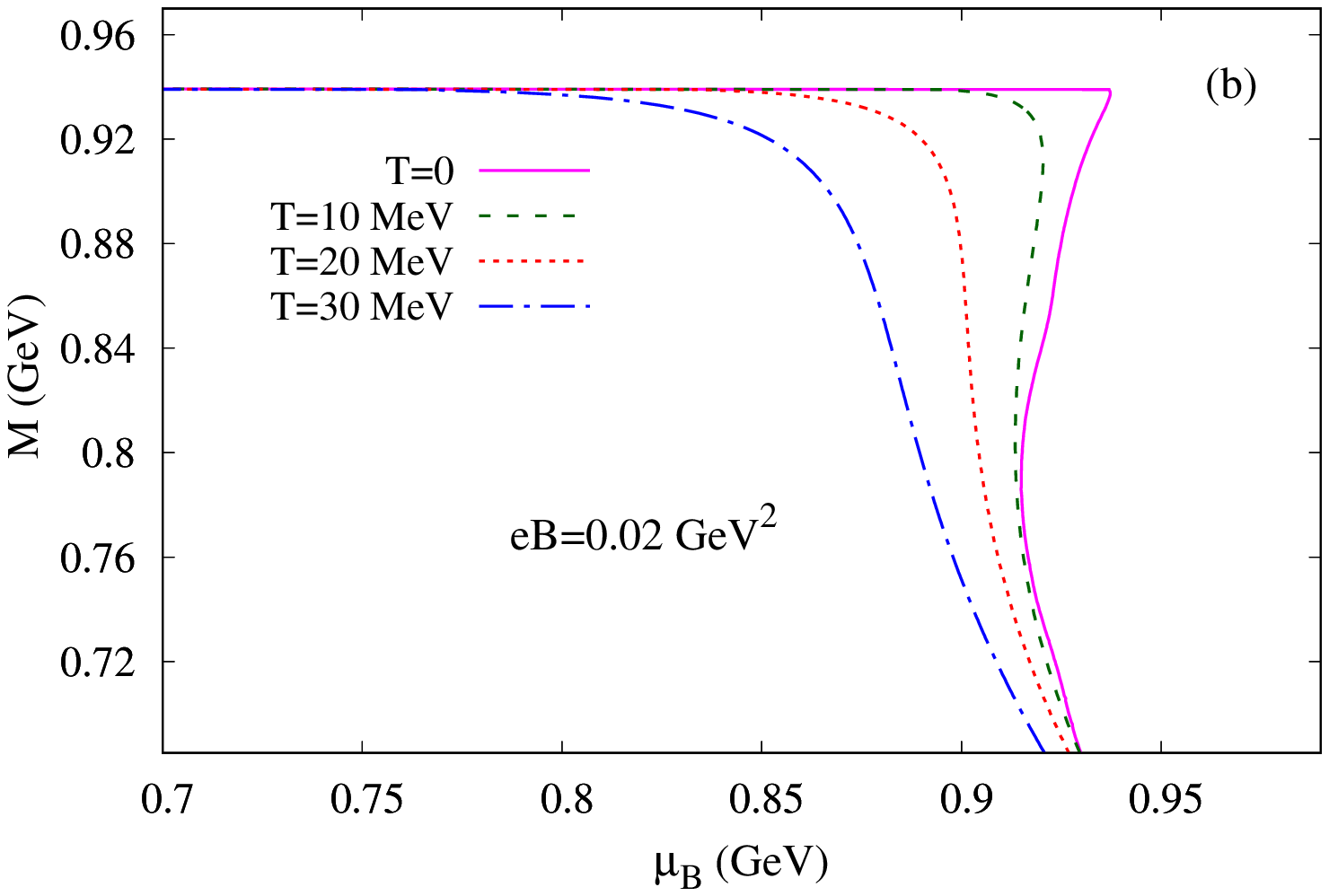}
	\includegraphics[angle = -90, scale=0.23]{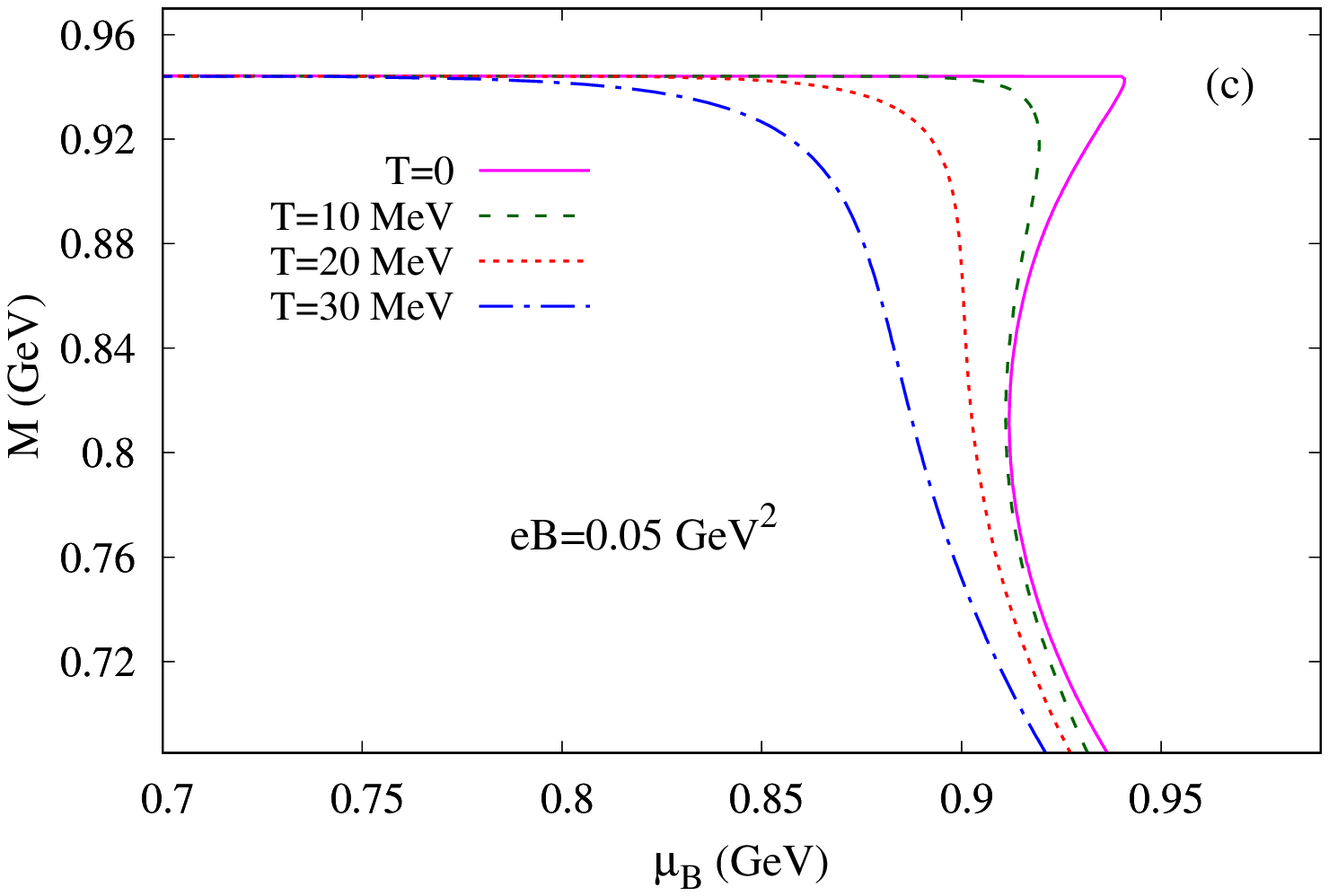}
	\caption{The effective nucleon mass $(M)$ as function of baryon chemical potential ($\mu_B$) for different values of temperatures $T=0,~10,~20,~30$ MeV at $(a)$ $eB=0$, $(b)$ $eB=0.02\rm~GeV^2$, $(c)$ $eB=0.05\rm~GeV^2$.}\label{Fig.MuM}
\end{figure}

We begin this subsection with an investigation of the effective mass of nucleons which is a function of chemical potential, temperature, and magnetic field strength. Hence, it is necessary to find self-consistent solutions for Eqs.\eqref{ns1} and~\eqref{nB1}. In Figs.~\ref{Fig.MuM}$(a)-(c)$, we show the variation of nucleon mass with chemical potential for several values of temperatures $T=0,~10,~20,~30~\rm MeV$ at different magnetic field strengths $eB=0,~0.02,~0.05~\rm GeV^2$ in the vicinity of liquid-gas phase transition.  At $T=0$, if the effective chemical potential $\mu^\star$ is smaller than the effective energy of the nucleon, then both the scalar density $n_s$ and baryon density $n_B$ will be zero. This can be seen from Eqs.~\eqref{ns2} and \eqref{nB2}.
Hence, at $T=0$, there is no contribution in effective nucleon mass from the medium.  Now, since according to Eq.~\eqref{nB1}, $\mu_B=\mu^\star$ at $T=0$ the same holds for $\mu_B$. Therefore, the solution for $M$ remains independent of $\mu_B$ leading to a horizontal line in the Figs.~\ref{Fig.MuM}$(a)-(c)$ (see magenta line).  The medium terms contribute when effective chemical potential $\mu^\star$ is larger than the effective energy of the nucleon.  In a certain range of $\mu_B$ three solutions are present and hence three distinct values for the nucleon mass exist. In this regime, a first-order phase transition presumably occurs.
As the temperature rises, the medium-term contributions tend to reduce the effective nucleon mass as depicted in the figures. Additionally, the existence of multiple solutions for the nucleon mass disappears at a particular value of $T$ and $\mu_B$ called critical end point (CEP) beyond which the phase transition goes towards the crossover regime. This result is consistent with the previous  findings in Ref.~\cite{Haber:2014ula}.
\begin{figure}[h] 
	\includegraphics[angle = -90, scale=0.23]{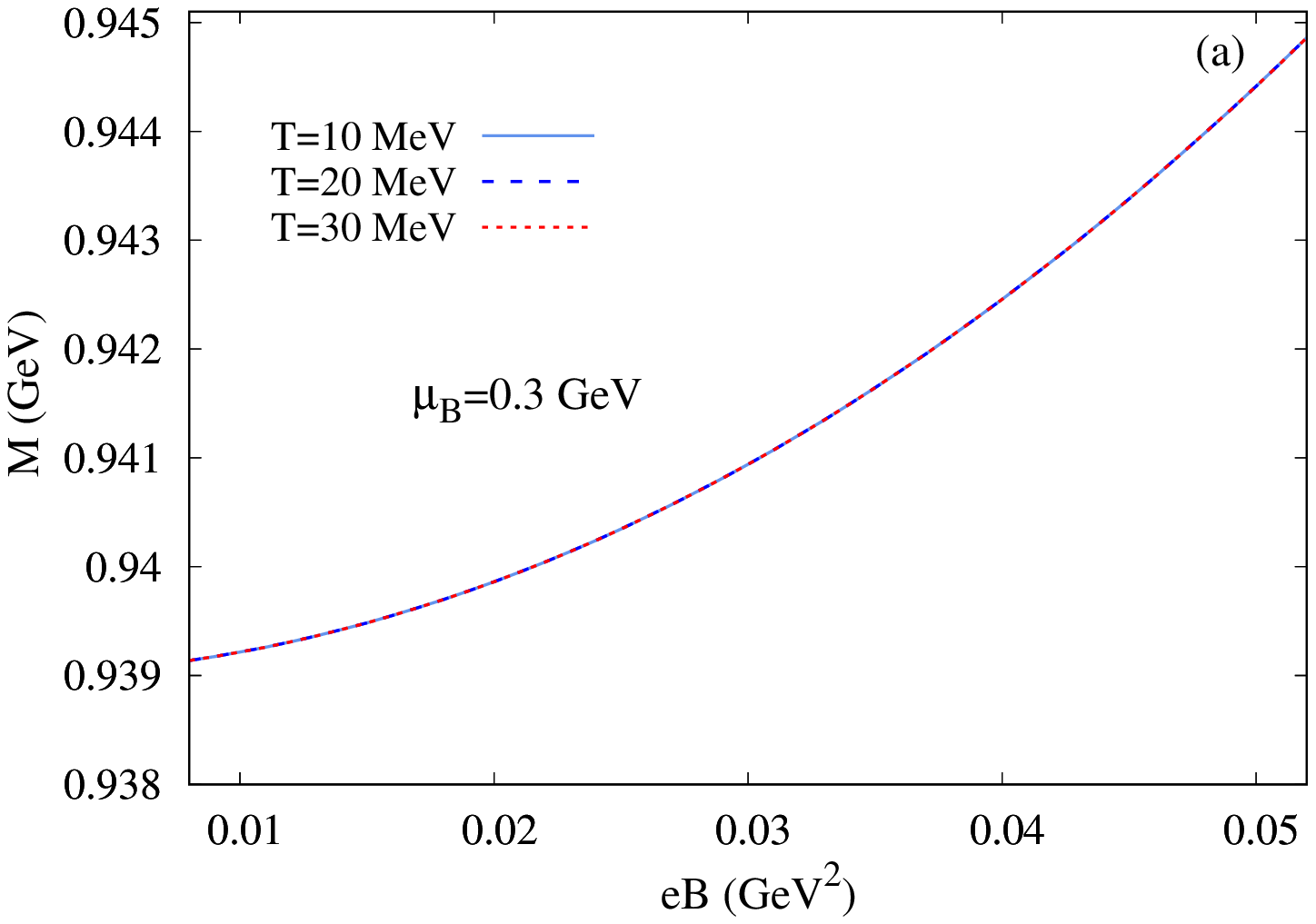}
	\includegraphics[angle = -90, scale=0.23]{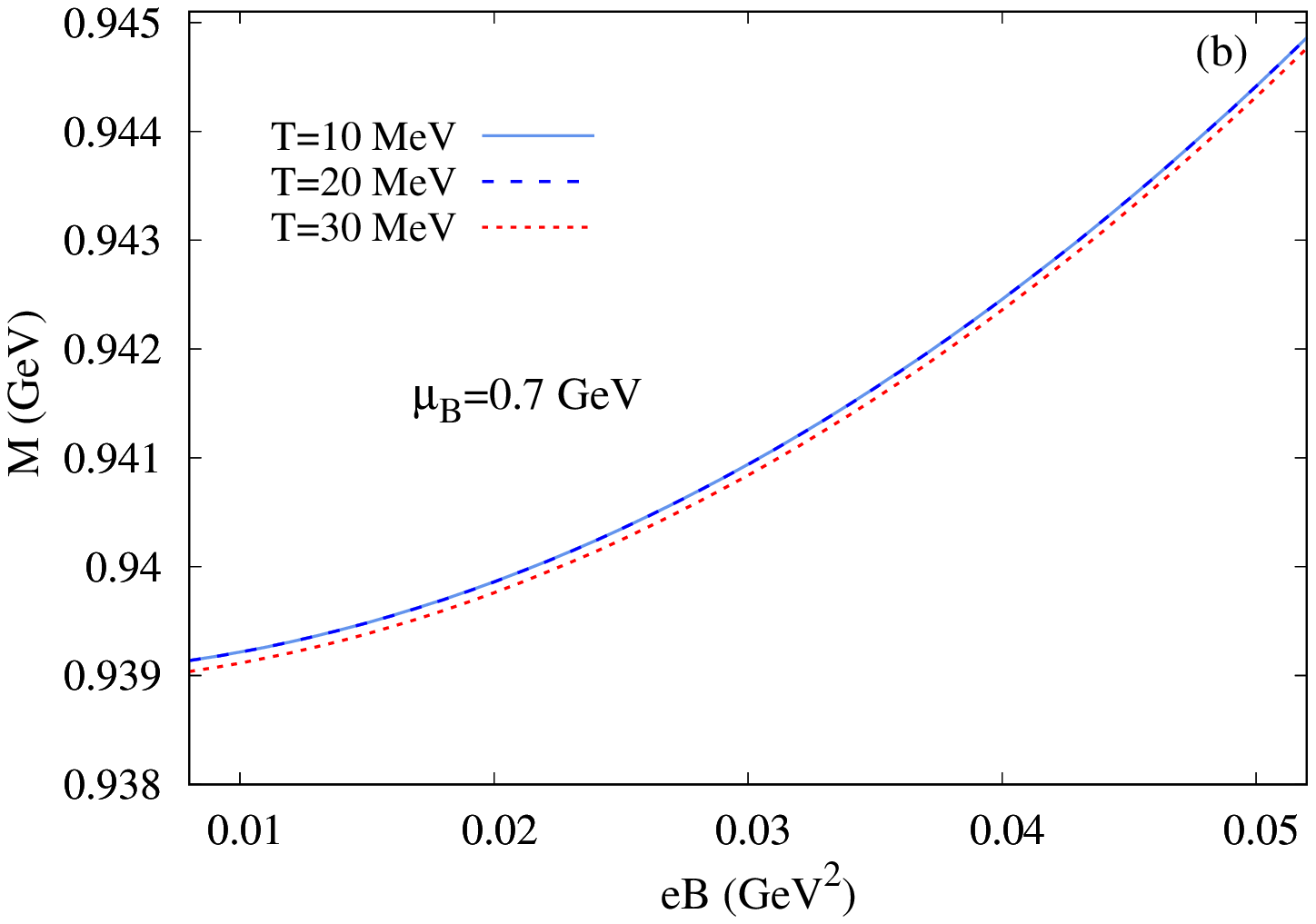}
	\includegraphics[angle = -90, scale=0.23]{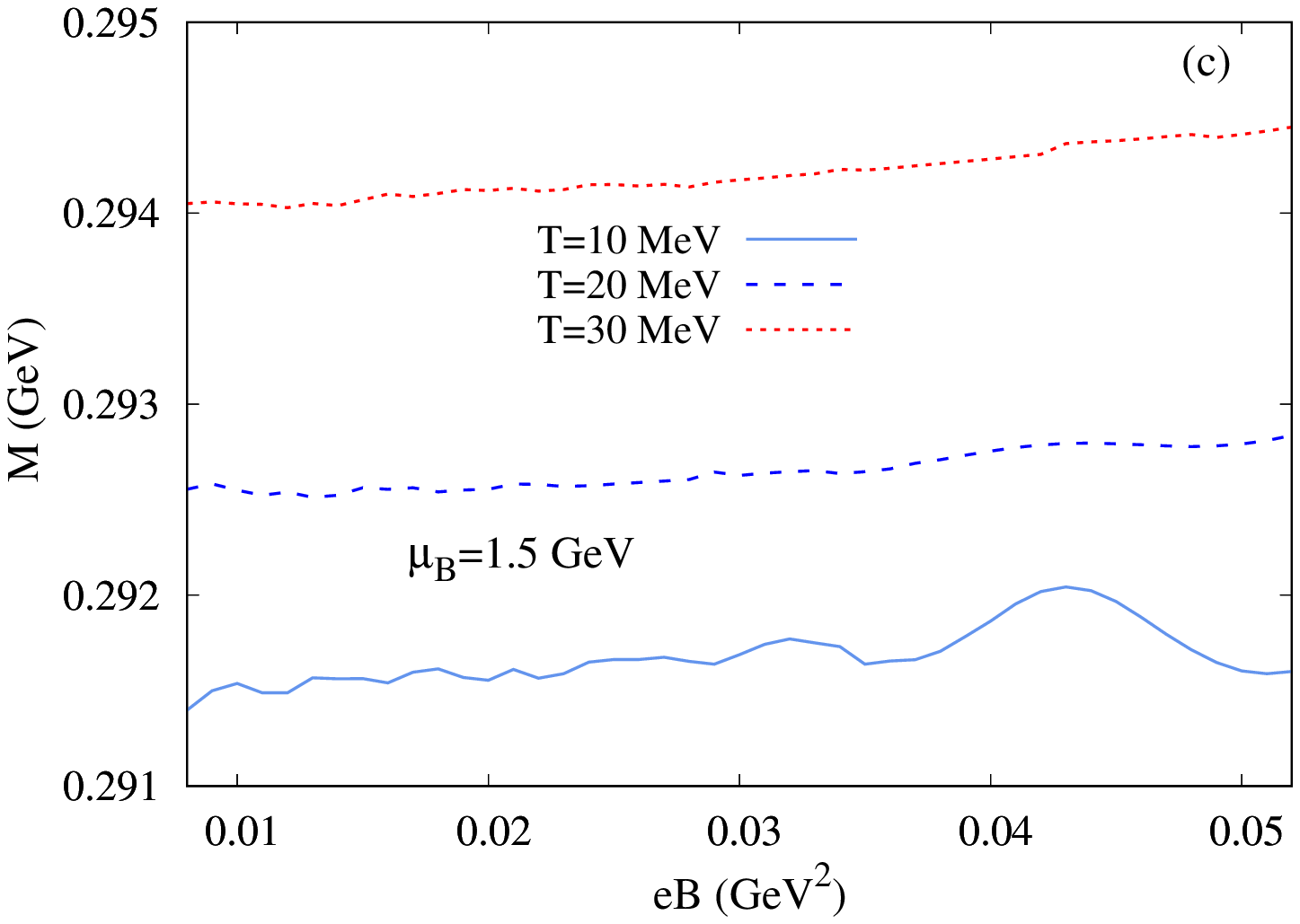}
	\caption{The effective nucleon mass $(M)$ as function of background magnetic field $eB$ for different values of temperature $T=10,~20,~30$ GeV at $(a)$ $\mu_B=0.3$ GeV, $(b)$ $\mu_B=0.7$ GeV, $(c)$ $\mu_B=1.5$ GeV.}\label{Fig.eBM}
\end{figure}
%

It is also interesting to observe the nucleon mass as a function magnetic field for various values of chemical potential ($\mu_B=0.3,~0.7,~1.5\rm~GeV$) and temperature ($T=10,~20,~30\rm~MeV$) as shown in Figs.~\ref{Fig.eBM}$(a)-(c)$. It is evident from the Figs.~\ref{Fig.eBM}$(a)-(b)$ that for a constant value of $\mu_B$ and $T$, the effective nucleon mass consistently rises with the magnetic field. This phenomenon is referred to as "Magnetic Catalysis". However, if the value of $\mu_B$ increases (e.g.$\mu_B=1.5\rm~GeV$ in Fig.~$\ref{Fig.eBM}(c)$), the variation of effective nucleon mass with $eB$ is marginal. 

%
%
\subsection{Nuclear Liquid-Gas Phase Transition}
%
\begin{figure}[h] 
	\includegraphics[angle = -90, scale=0.23]{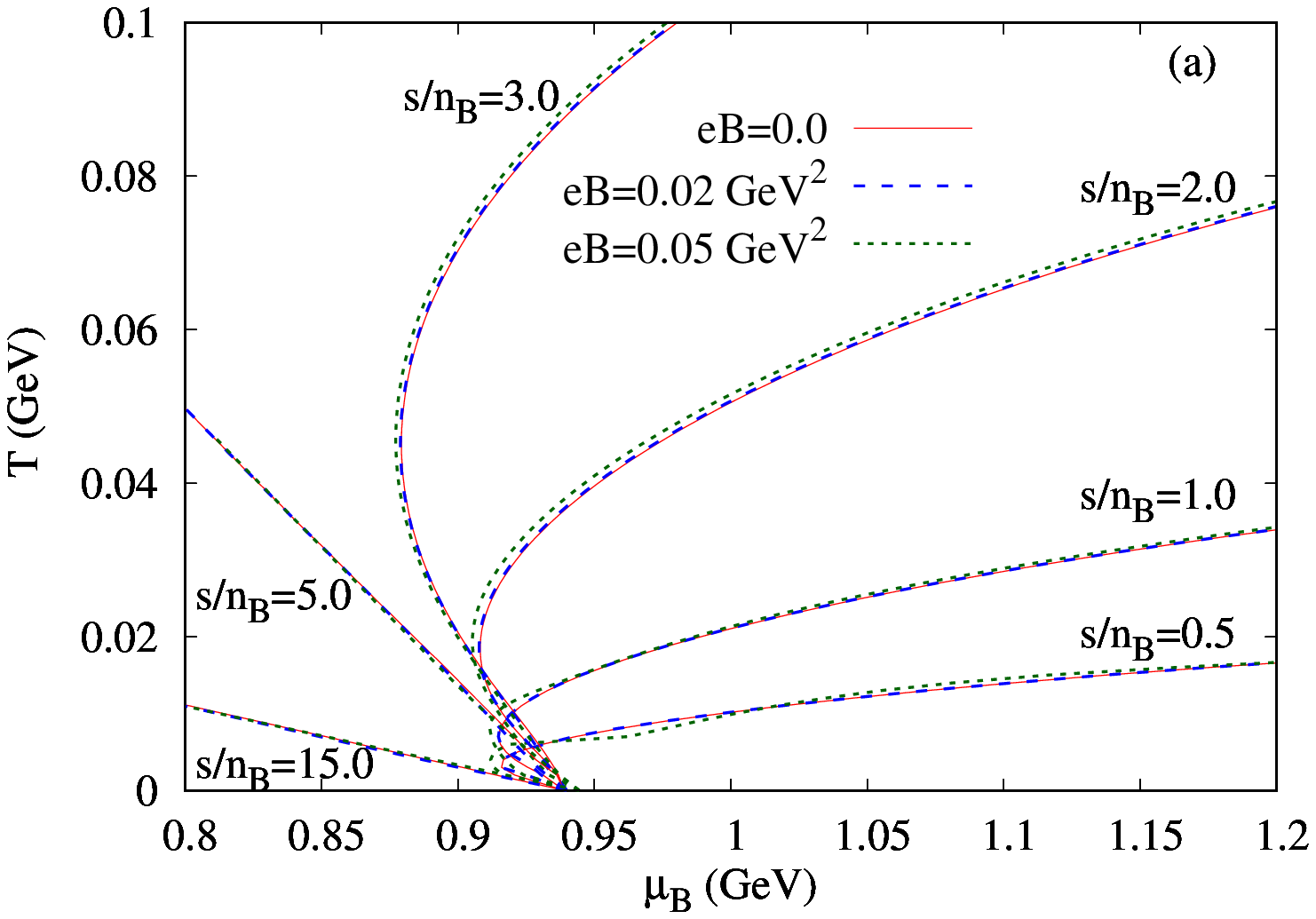}
	\includegraphics[angle = -90, scale=0.23]{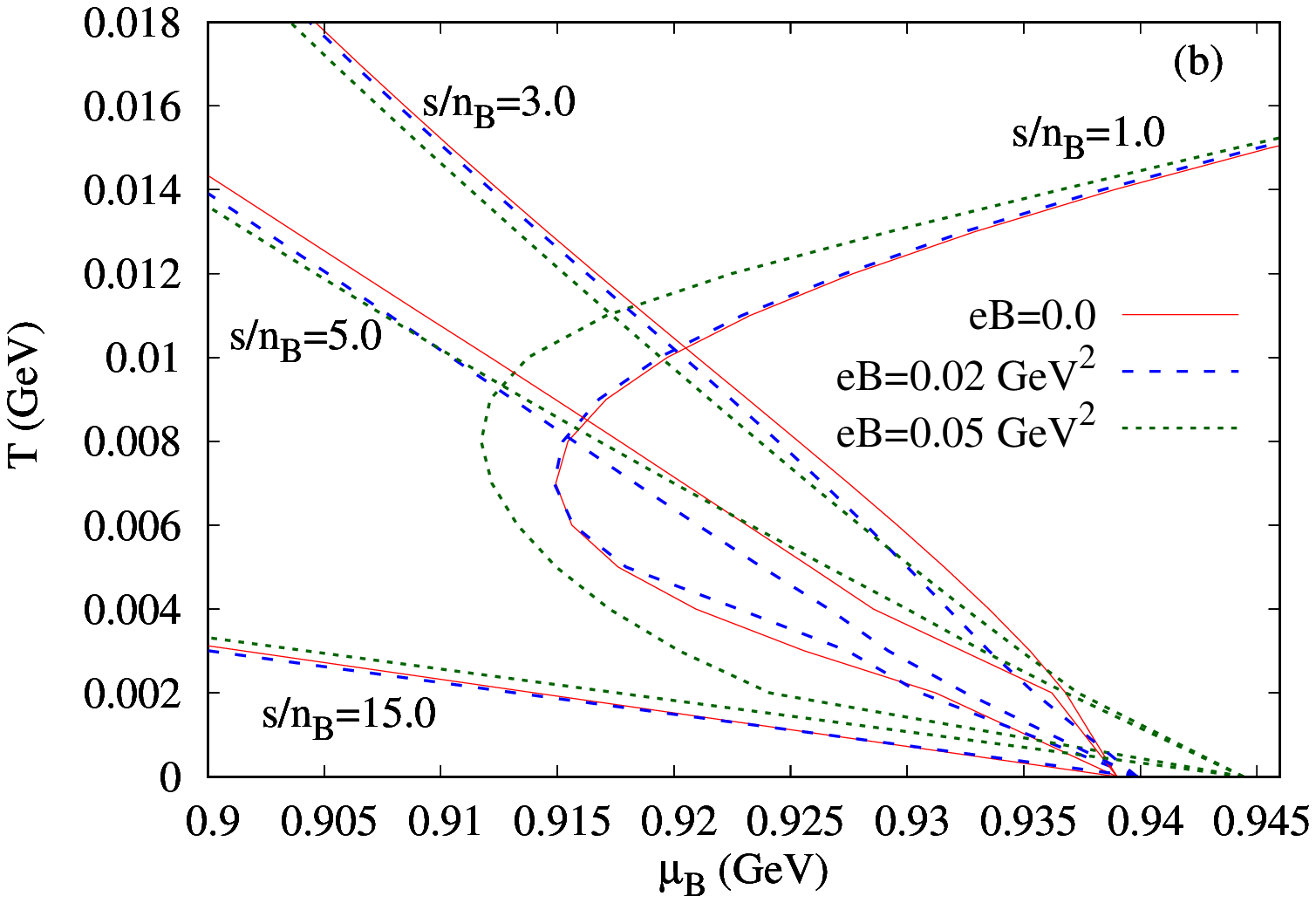}
	\includegraphics[angle = -90, scale=0.23]{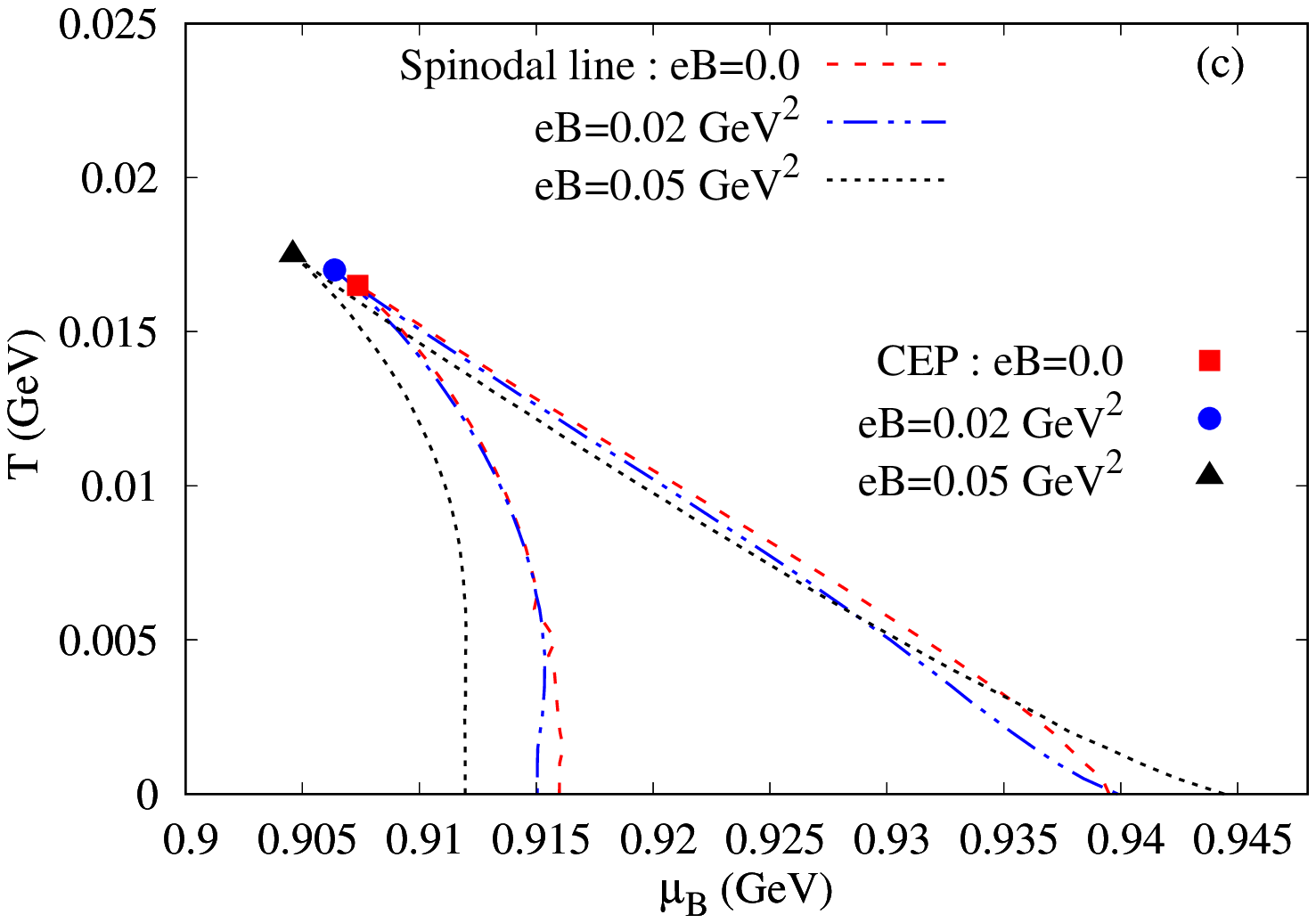}
	\caption{The isentropic curves for various values of magnetic field $eB~(=0,~0.02,~0.05~\rm GeV^2)$ and different values of $s/n_B$ $(a)$ in large $T-\mu_B$ plane, $(b)$ in small $T-\mu_B$ plane.  Different line-styles represent specific $s/n_B$ values while a uniform color scheme is used to denote a particular $eB$ value. $(c)$ The spinodial lines and critical end points (CEPs) for various values of $eB=0,~0.02,~0.05~\rm GeV^2$ in $T-\mu_B$ plane.}
	\label{MuTSbynB}
\end{figure}
Now, in Figs.~\ref{MuTSbynB}$(a)-(c)$, we illustrate the nuclear liquid-gas phase structure and isentropic curves for various $s/n_B$ values (ranging from 0.05 to 15.0), 
in the $T-\mu_B$ plane under background magnetic fields $eB=0,~0.02,~0.05~\rm GeV^2$. 
The isentropic curves depicted in Figs.~\ref{MuTSbynB}(a) and (b) illustrate the trajectories of an ideal fluid under adiabatic conditions. Notably, Fig.~\ref{MuTSbynB}$(b)$ is a similar plot as Fig.~\ref{MuTSbynB}$(a)$ in a smaller $T-\mu_B$ space. The graphs in Fig.~\ref{MuTSbynB}$(b)$ exhibit a discernible shift as the magnetic field strength varies for a specific value  of $s/n_B$. However, as $s/n_B$ increases, the distinctions between the graphs corresponding to the different magnetic field strengths decrease. Furthermore, it is evident from Fig.~\ref{MuTSbynB}$(b)$ that, at $T=0$, for a given value of $eB$, the plots for different values of $s/n_B$ meet at a single point on the $\mu_B$ axis. This point shifts towards higher $\mu_B$ with the increase of magnetic field strength. This is a reflection of the so called "Magnetic Catalysis" observed in Figs.~$\ref{Fig.eBM}(a)-(b)$. 
Fig.~\ref{MuTSbynB}(c) illustrates the CEP of a liquid-gas phase transition along with the corresponding spinodal lines in the $T-\mu_B$ plane for magnetic field strengths $eB=0.0,~0.02,~0.05~GeV^2$. The position of CEP shifts towards higher $T$ and lower $\mu_B$ as the magnetic field strength increases. It is to be noted that the spinodal lines 
are determined from the extrema of ${\partial M}/{\partial T}$. 
This gives rise to two distinct segments of spinodal lines which converge at the CEP, as illustrated in the figure \ref{MuTSbynB}(c). Notably, the presence of a magnetic field leads to significant changes in the spinodal lines.
%



\subsection{Magnetization of The Medium}
\begin{figure}[h] 
	\includegraphics[angle = -90, scale=0.23]{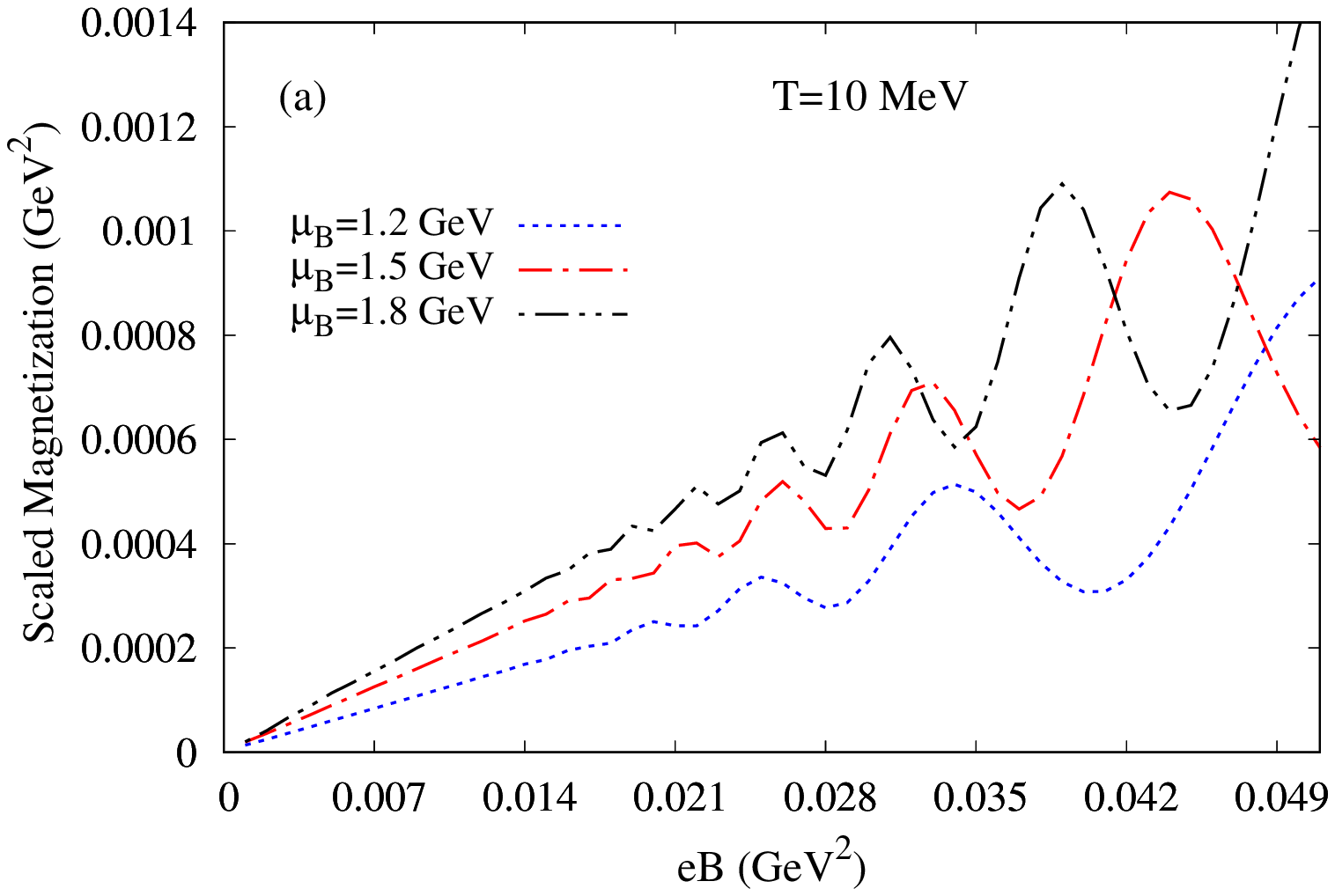}
	\includegraphics[angle = -90, scale=0.23]{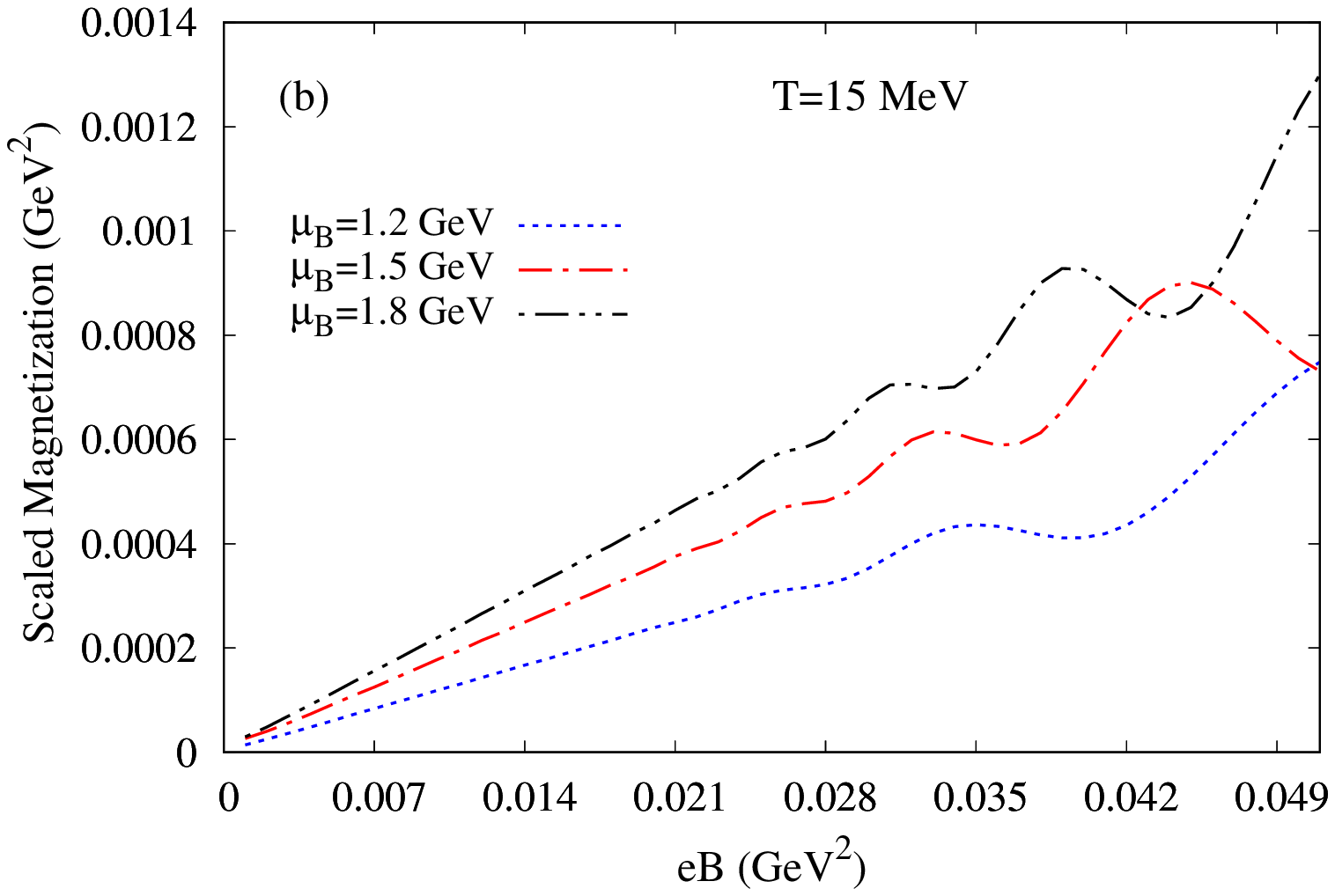}
	\includegraphics[angle = -90, scale=0.23]{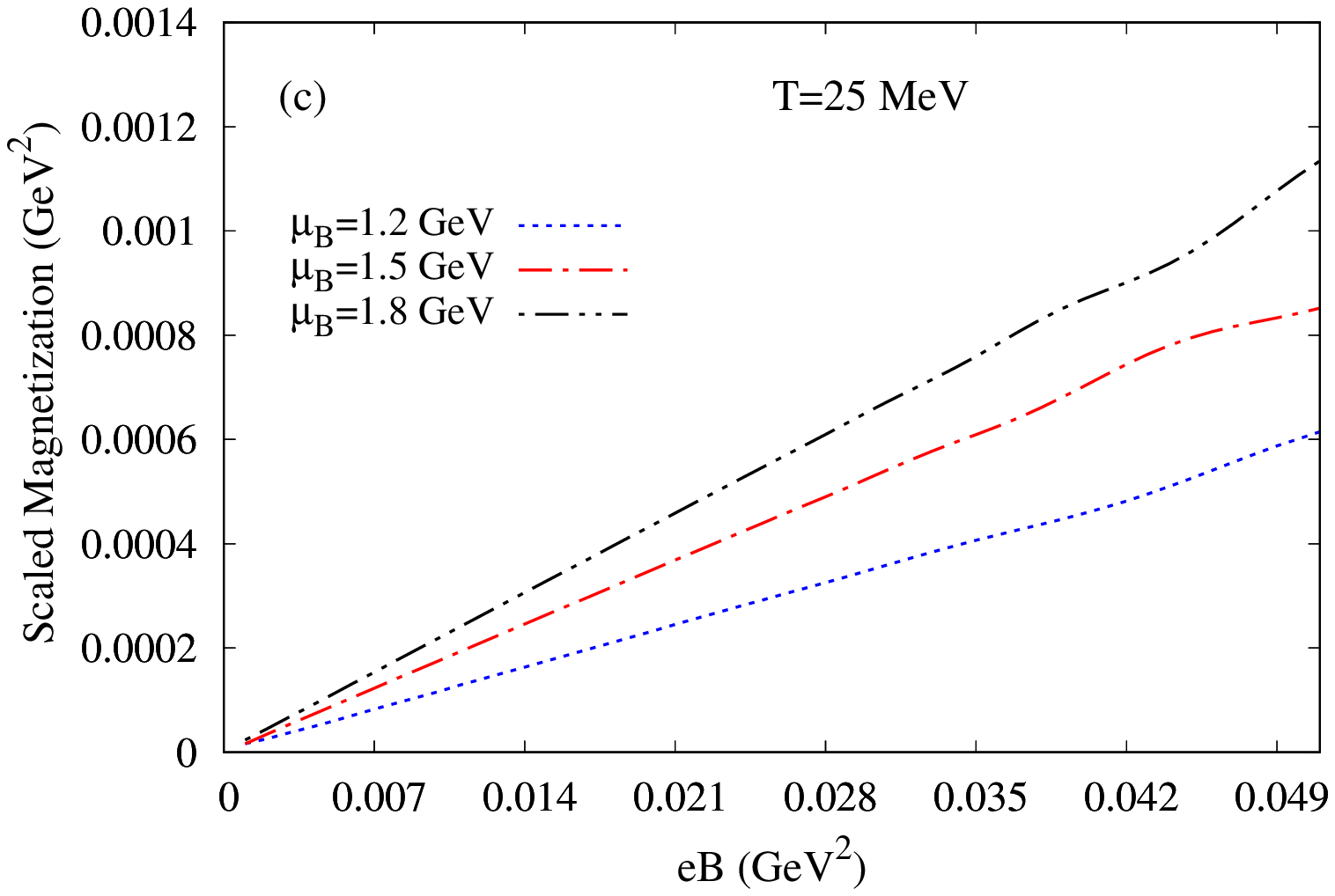}
	\caption{Scaled magnetization ($\mathcal{M}$) as a function of $eB$ for $\mu_B=1.2,~1.5,~1.8$ GeV at $(a)$ $T=5$ MeV, $(b)$ $T=10$ MeV, $(c)$ $T=25$ MeV.}
	\label{eBMagnetization}
\end{figure}
		In presence of background magnetic field the system will also be magnetized. The expression for the magnetization is given in the Eq.~\eqref{Magnetization}. We scale the magnetization of the system as :
		\begin{eqnarray}\label{S_Magne}
		\mathcal{M}_{\text{scaled}}=\mathcal{M}+B~.
		\end{eqnarray}
		In Figs.~$\ref{eBMagnetization}(a)-(c)$, the dependence of scaled magnetization on the background magnetic field $eB$ is illustrated for different values of the parameter $\mu_B=1.2,~1.5,~1.8$ GeV. Each subplot corresponds to distinct temperatures: Fig.~$\ref{eBMagnetization}(a)$ at $T=10$ MeV, Fig.~$\ref{eBMagnetization}(b)$ at $T=15$ MeV, Fig.~$\ref{eBMagnetization}(c)$ at $T=25$ MeV. In Fig.~$\ref{eBMagnetization}(a)$,  it is observed that the  positive scaled magnetization shows an oscillating trend with the magnetic field $eB$. 
		Additionally, as the temperature increases, there is a discernible reduction in the oscillating nature of the scaled magnetization, as depicted in Figs. ~$\ref{eBMagnetization}(b)-(c)$.

\subsection{Speed of Sound at Constant $s/n_B$}
\begin{figure}[h] 
	\includegraphics[angle = -90, scale=0.23]{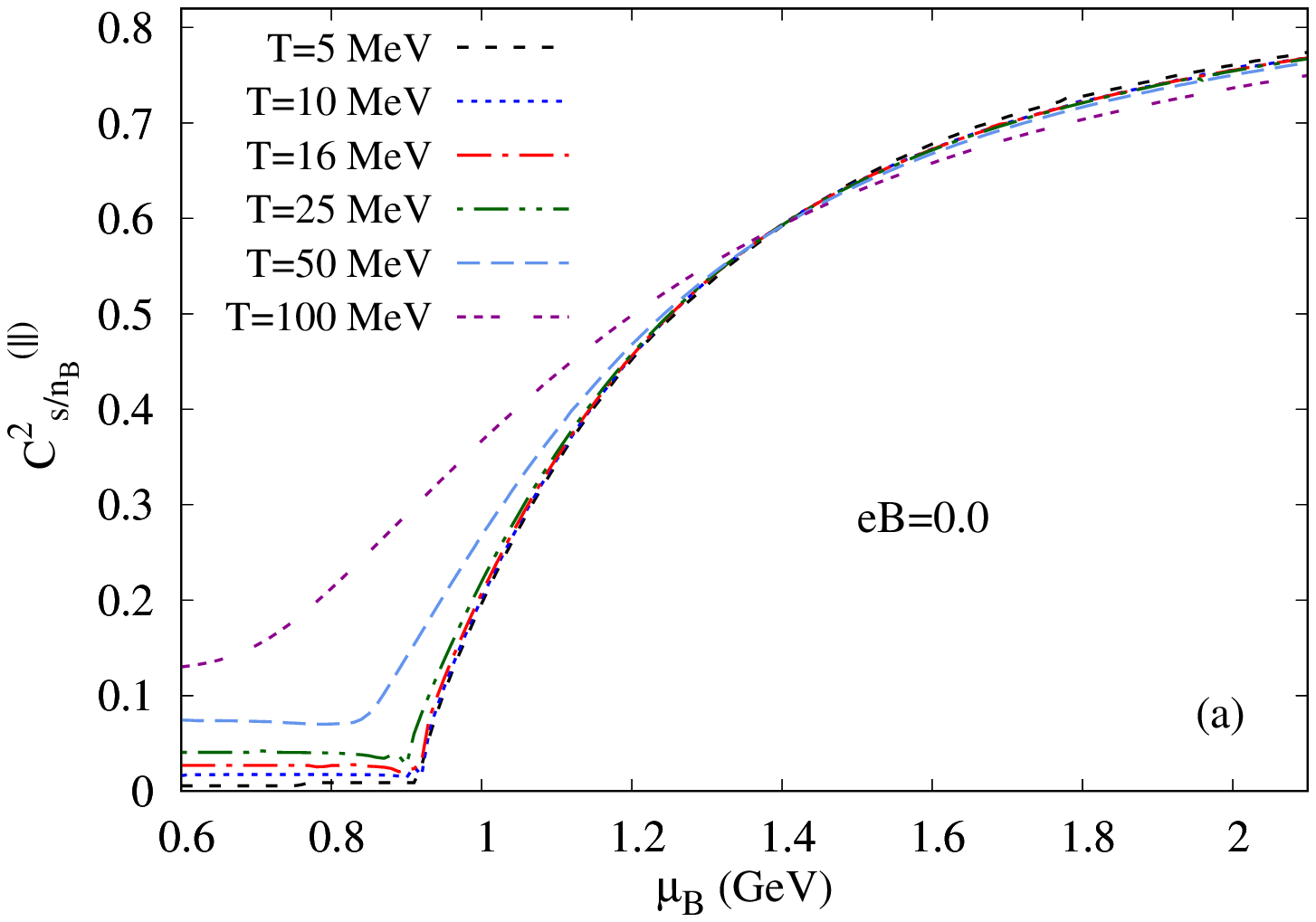}
	\includegraphics[angle = -90, scale=0.23]{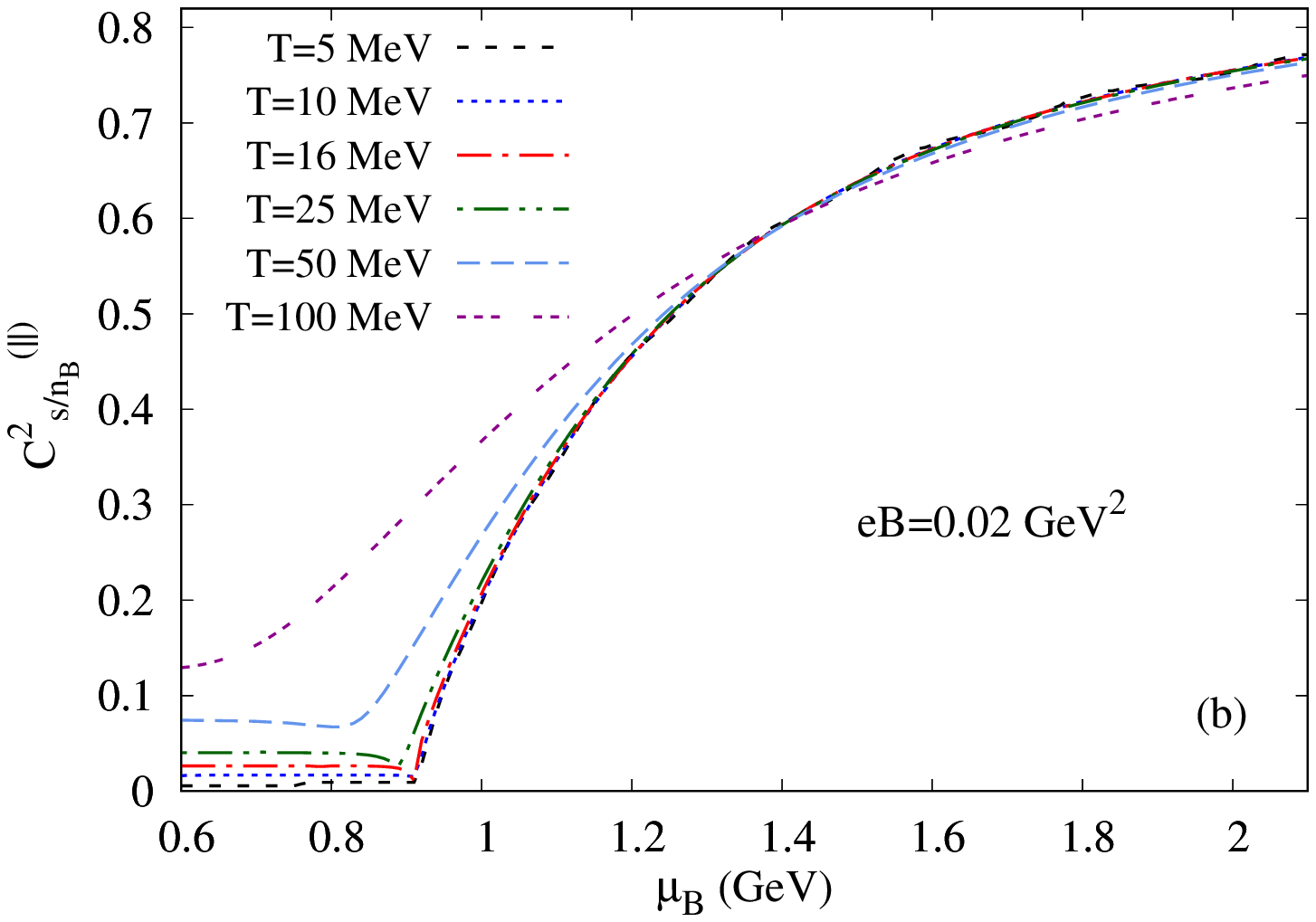}
	\includegraphics[angle = -90, scale=0.23]{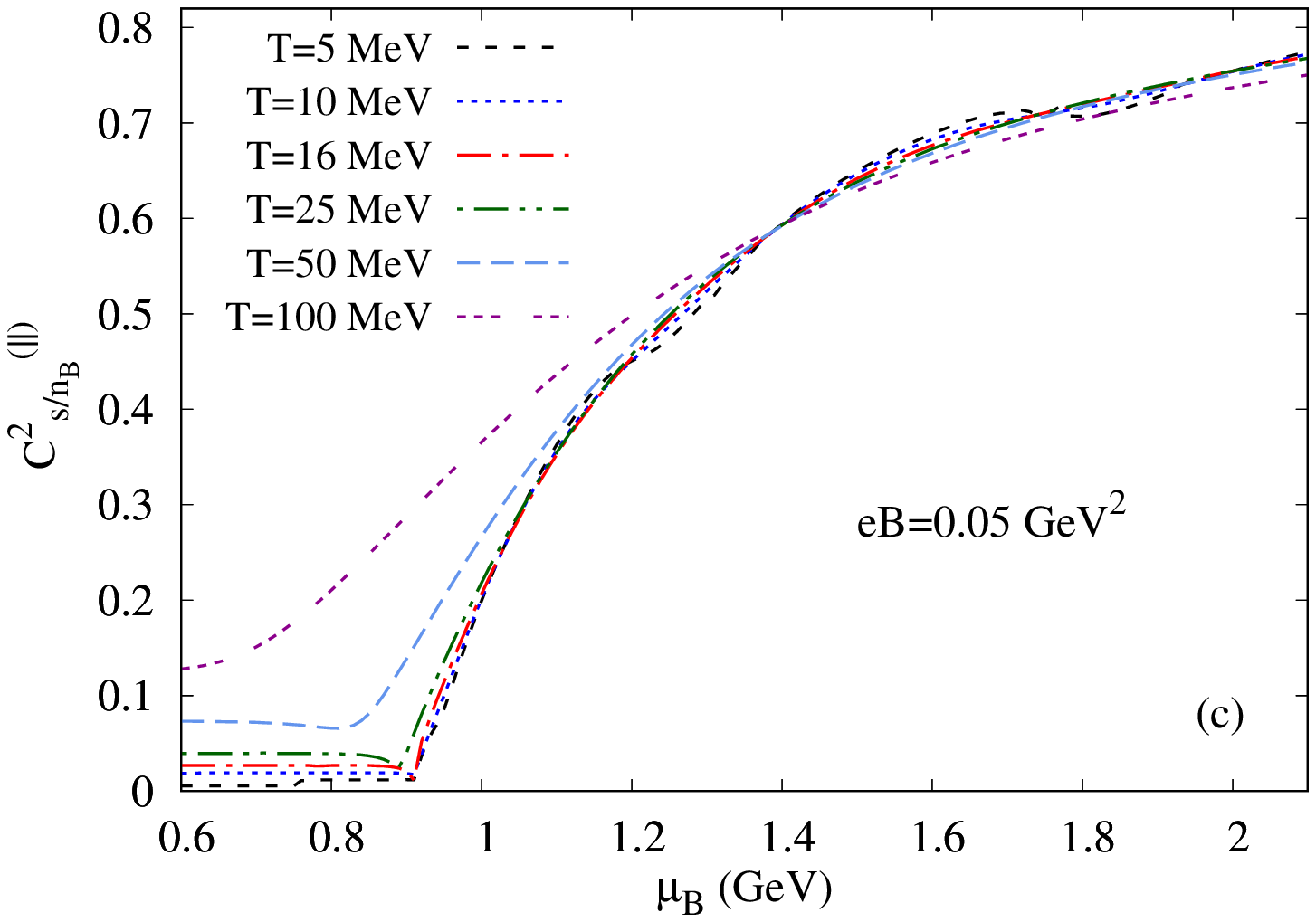}
	\caption{Parallel component of squared speed of sound $C_{s/n_B}^{2(\parallel)}$ as a function of chemical potential $\mu_B$ for a few fixed temperature ($T=5,~10,~16,~25,~50,~100\rm~MeV$) at (a) $eB=0$, (b) $eB=0.02\rm~GeV^2$, (c) $eB=0.05\rm~GeV^2$.}
	\label{Fig.MuC2sbynB}
\end{figure}
\begin{figure}[h] 
	\includegraphics[angle = -90, scale=0.23]{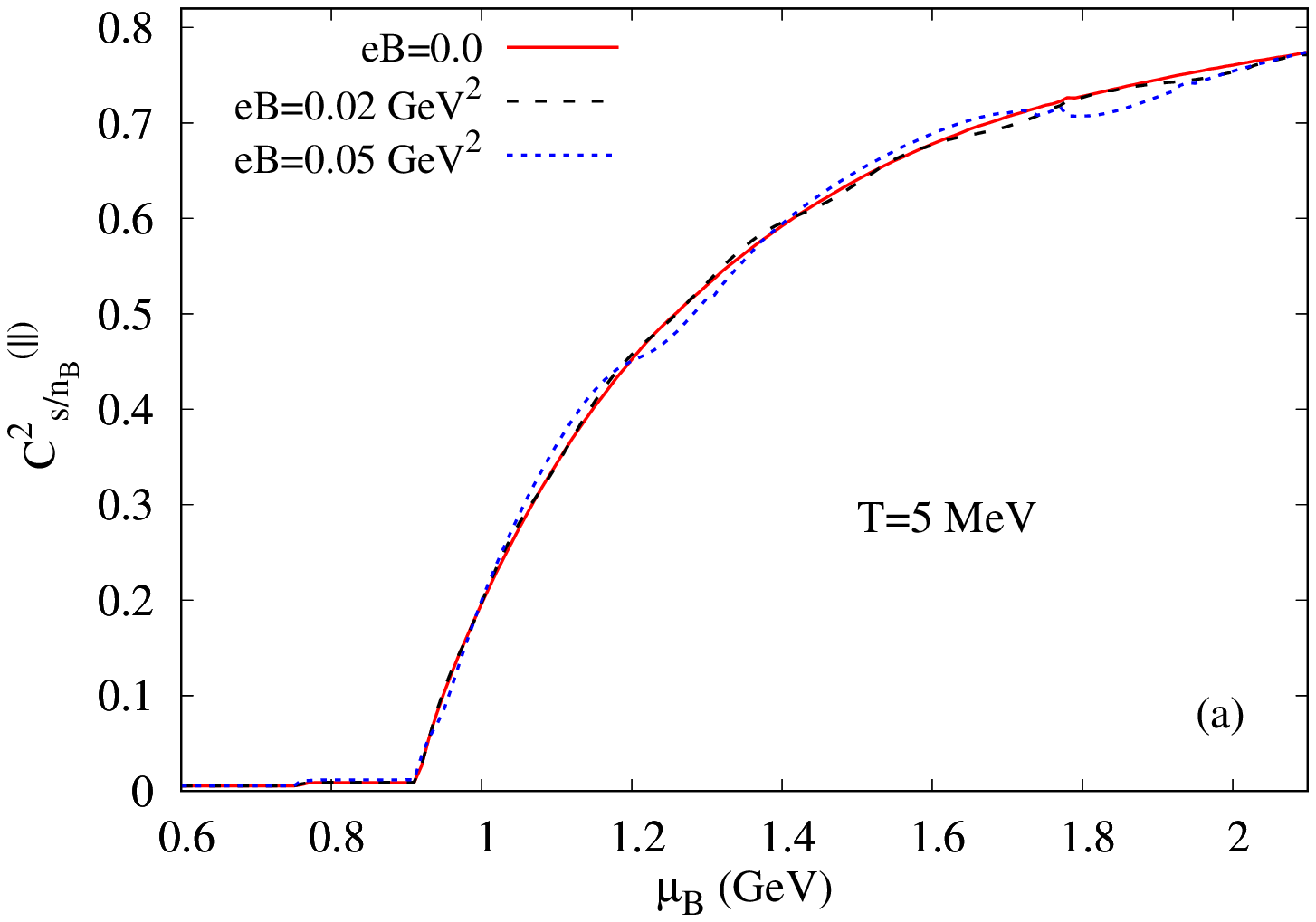}
	\includegraphics[angle = -90, scale=0.23]{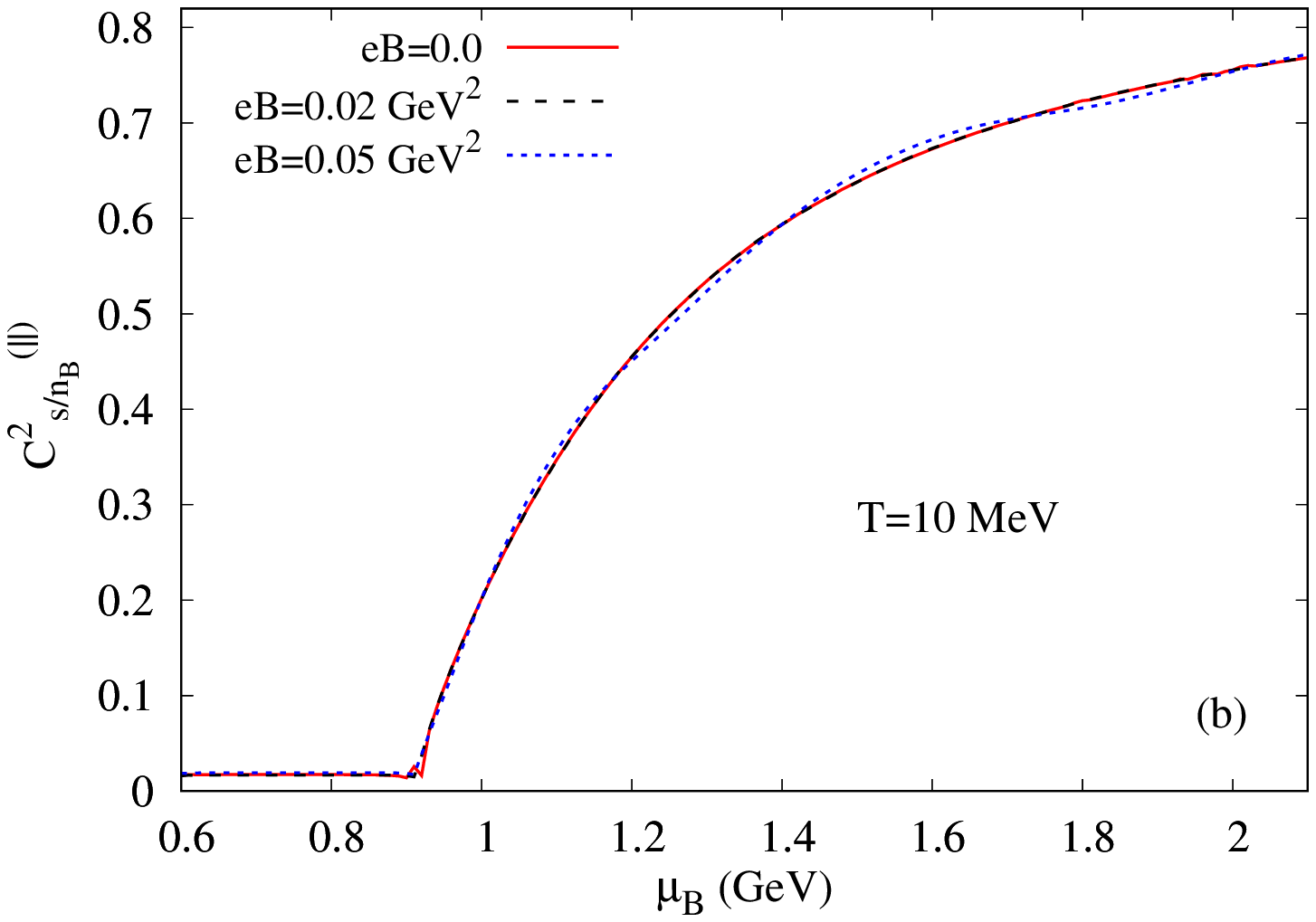}
	\includegraphics[angle = -90, scale=0.23]{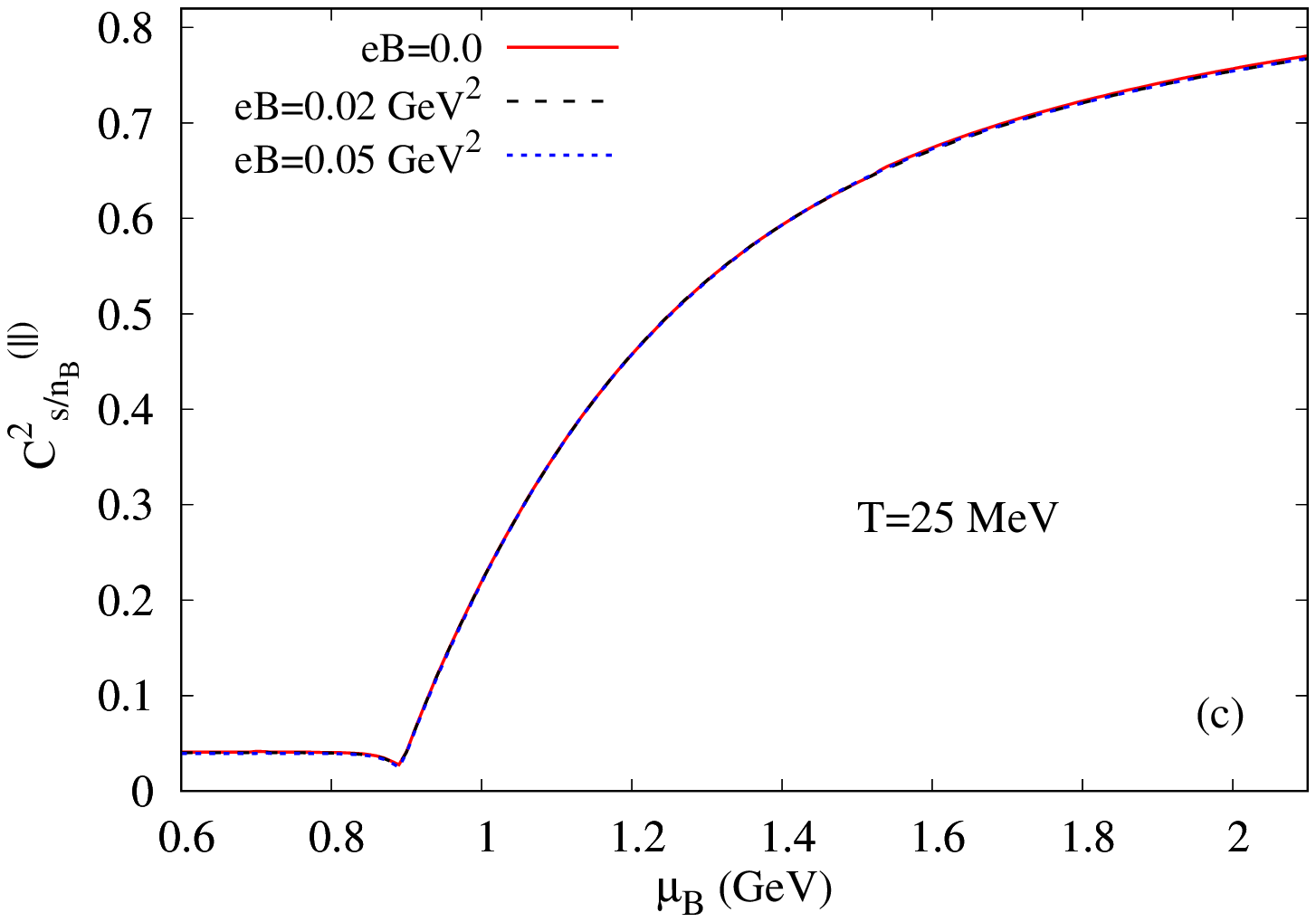}
	\caption{Parallel component of squared speed of sound  $C_{s/n_B}^{2(\parallel)}$ as a function of chemical potential $\mu_B$ for $eB=0.02,~0.05\rm~GeV^2$ at (a) $T=5\rm~MeV$, (b) $T=10\rm~MeV$, (c) $T=25\rm~MeV$.}
	\label{Fig.MuC2sbynB_T}
\end{figure}

In this subsection, we investigate the variations of the sound speed with chemical potential in presence of a background magnetic field in nuclear matter. As discussed in section \ref{SpdSound}, in presence of magnetic field $C^2_{s/n_B}$ splits into $C_{s/n_B}^{2(\parallel)}$ and $C_{s/n_B}^{2(\perp)}$ along and perpendicular to the magnetic field direction respectively. To find the sound speed, we use Eqs.~\eqref{C2sbynB}-\eqref{C2MuB} and  Eqs.~\eqref{C2sbynBP}-\eqref{C2MuBPeprp}.
We plot the parallel component of the squared speed of sound, denoted as $C_{s/n_B}^{2(\parallel)}$, as a function of baryon chemical potential $(\mu_B)$ for various temperatures $T=5,~10,~16,~25,~50,~100~\rm MeV$. These are shown in Fig.~$\ref{Fig.MuC2sbynB}(a)$ for the case of zero magnetic field strength $(eB=0)$, in Fig.~$\ref{Fig.MuC2sbynB}(b)$ for  $eB=0.02~\rm GeV^2$, in Fig.~$\ref{Fig.MuC2sbynB}(c)$ for $eB=0.05~\rm GeV^2$. The plot in Fig.~$\ref{Fig.MuC2sbynB}(a)$, under zero magnetic field conditions, illustrates that $C_{s/n_B}^{2(\parallel)}$ grows with the rising chemical potential for every temperature. This observation implies that, at $eB=0.0$ when $\mu_B > 1.32$ GeV, $C_{s/n_B}^{2(\parallel)}$ is larger at high temperature. This arises mainly from the temperature-driven influence on pressure and energy density at smaller chemical potentials. On the contrary, for $\mu_B <1.32~\rm GeV$, the inverse occurs, primarily due to the density-driven effect resulting from the reduction of dynamic nucleon mass. 
The parallel squared speed of sound changes when the magnetic field is switched on, as depicted in Figs.~$\ref{Fig.MuC2sbynB}(b)-(c)$. 
In the case of a non-zero magnetic field, it is also observed that $C_{s/n_B}^{2(\parallel)}$ is larger at high temperature due to the temperature-driven influence on pressure and energy density at smaller chemical potentials, as in the zero field case. However, in the domain of higher values of $\mu_B$, no monotonic behaviour is obeserved as in the zero field case.

The parallel squared speed of sound $C_{s/n_B}^{2(\parallel)}$ is presented as a function of $\mu_B$ for different magnetic field strengths $eB=0,~0.02,~0.05~\rm GeV^2$ in Fig.~\ref{Fig.MuC2sbynB_T}(a) for $T=5\rm~MeV$, in Fig.~\ref{Fig.MuC2sbynB_T}(b) for $T=10\rm~MeV$, and in Fig.~\ref{Fig.MuC2sbynB_T}(c) for $T=16\rm~MeV$ respectively. In Fig.~\ref{Fig.MuC2sbynB_T}(a), the plots corresponding to non-zero magnetic fields mildly oscillates around the $eB=0$ plot and the oscillations increase with the increase of magnetic field strength. As the temperature increases, in Figs.~$\ref{Fig.MuC2sbynB_T}(b)-(c)$, the oscillation decreases, i.e, the magnetic field effect decreases. This is also understandable from Fig.~$\ref{Fig.MuC2sbynB}(b)-(c)$. Therefore, the influence of magnetic field on $C_{s/n_B}^{2(\parallel)}$, which is particularly evident at lower temperatures, demonstrates a diminishing trend as temperature rises. 
%
\begin{figure}[h]
	\includegraphics[angle = -90, scale=0.23]{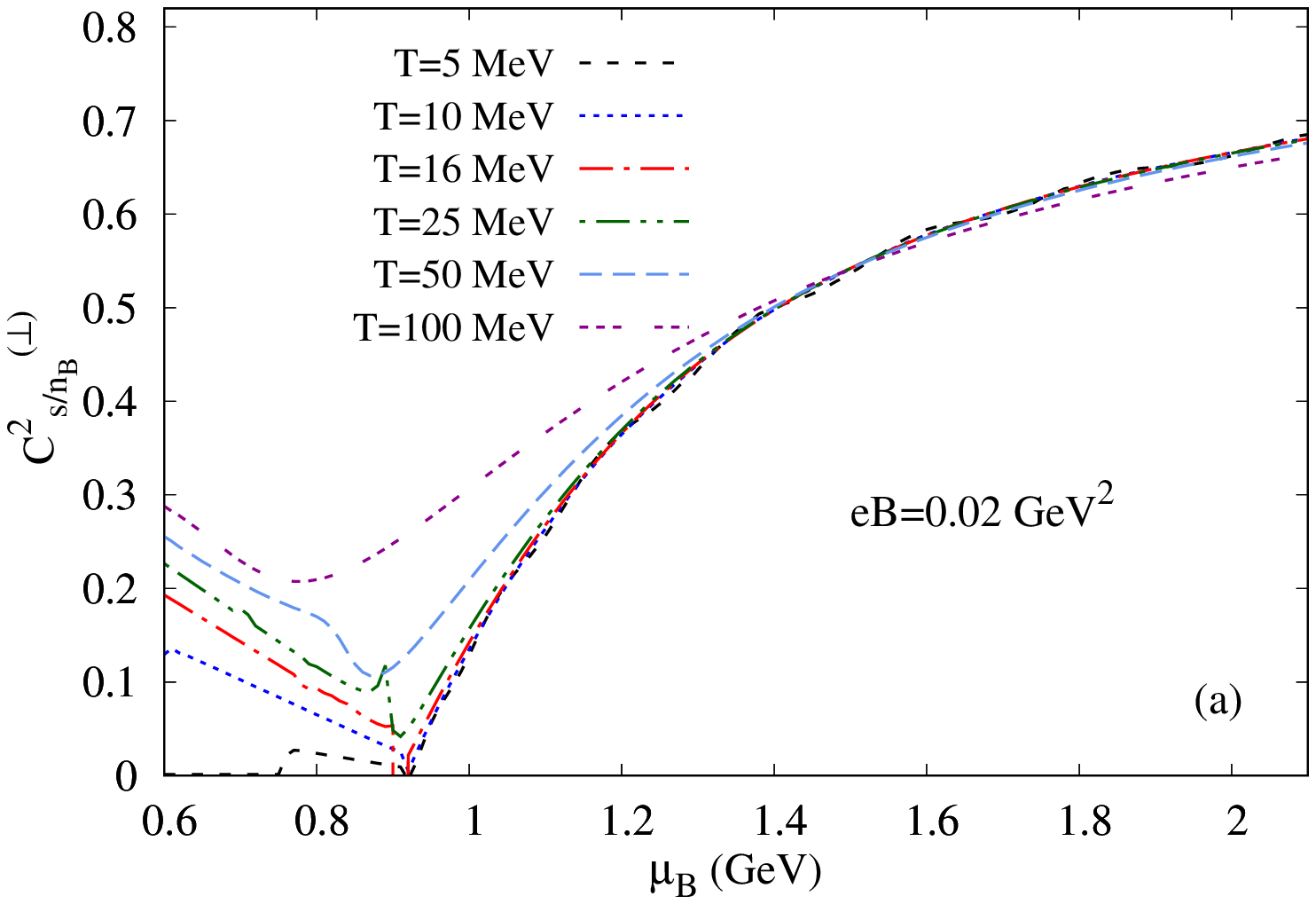}
	\includegraphics[angle = -90, scale=0.23]{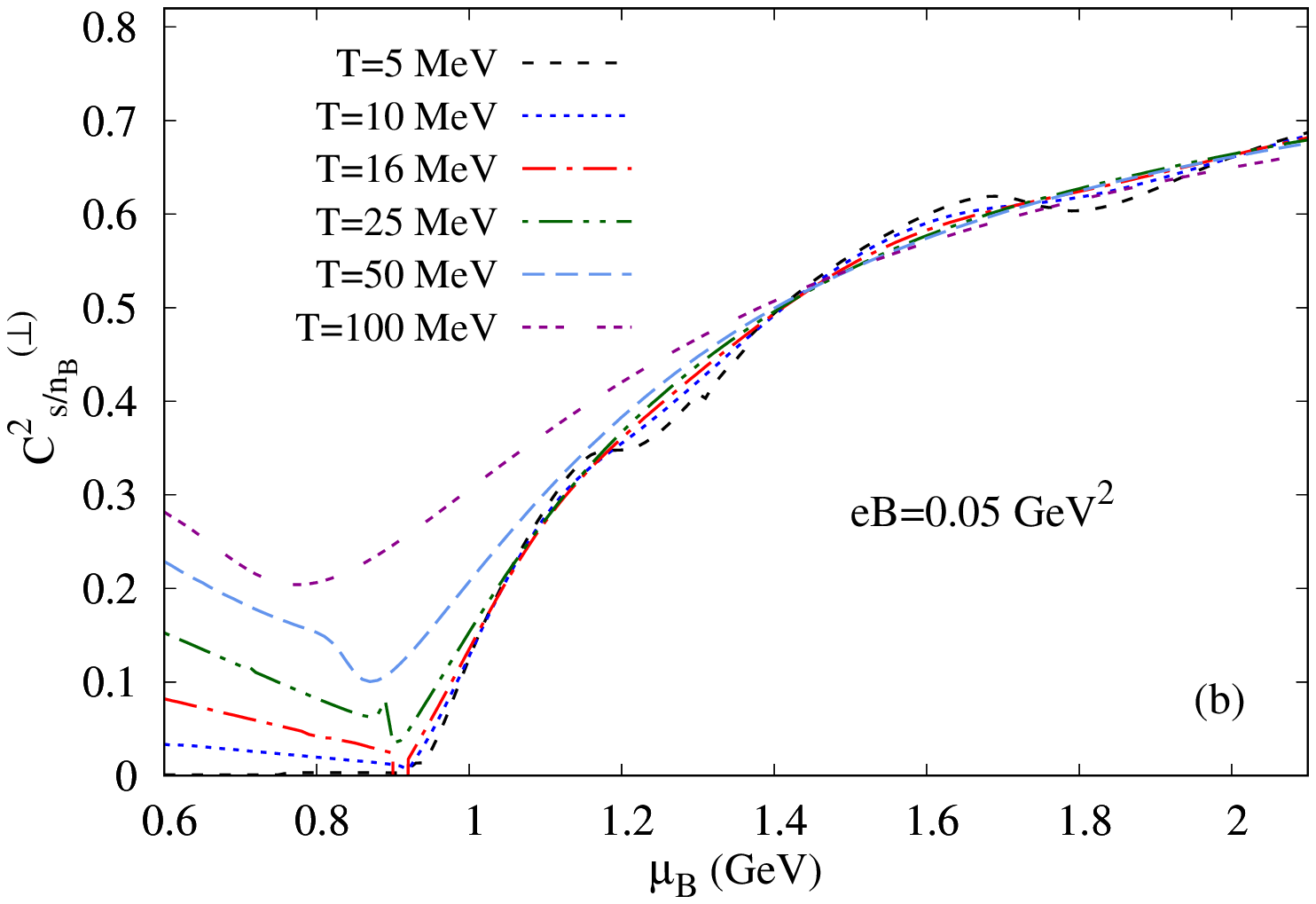}
	\includegraphics[angle = -90, scale=0.23]{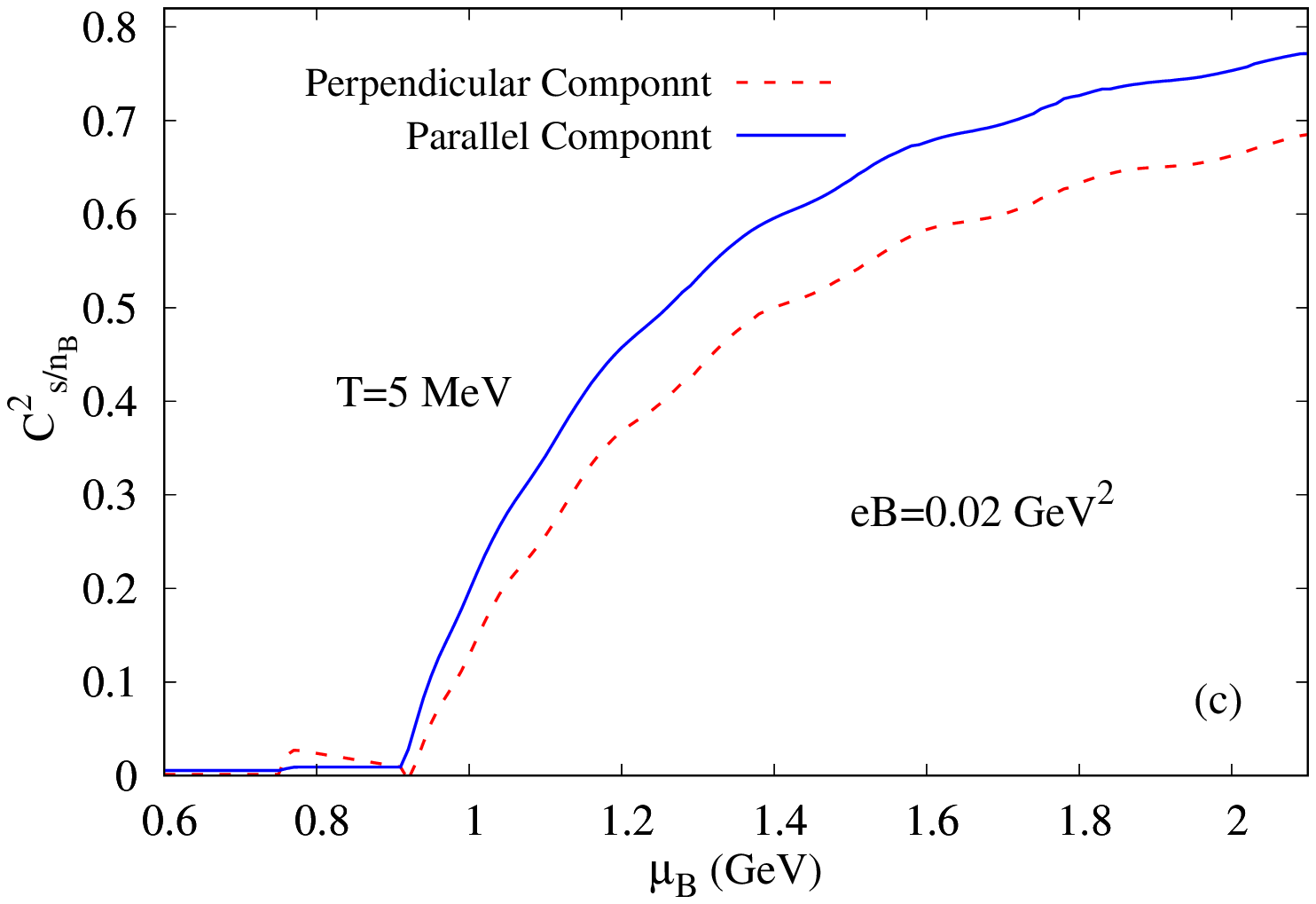} 
	\caption{$C_{s/n_B}^{2(\perp)}$ as a function of chemical potential $\mu_B$ for a few fixed temperature ($T=5,~10,~16,~25,~50,~100\rm~MeV$) at (a) $eB=0.02\rm~GeV^2$, (b) $eB=0.05\rm~GeV^2$. Parallel and perpendicular components of ${C^2_{s/n_B}}$ as a function of chemical potential $\mu_B$ at $T=5\rm~MeV$ and $eB=0.02\rm~GeV^2$ in (a).}
	\label{Fig.MuC2sbynB_Perp_B}
\end{figure}
\begin{figure}[h] 
	\includegraphics[angle = -90, scale=0.23]{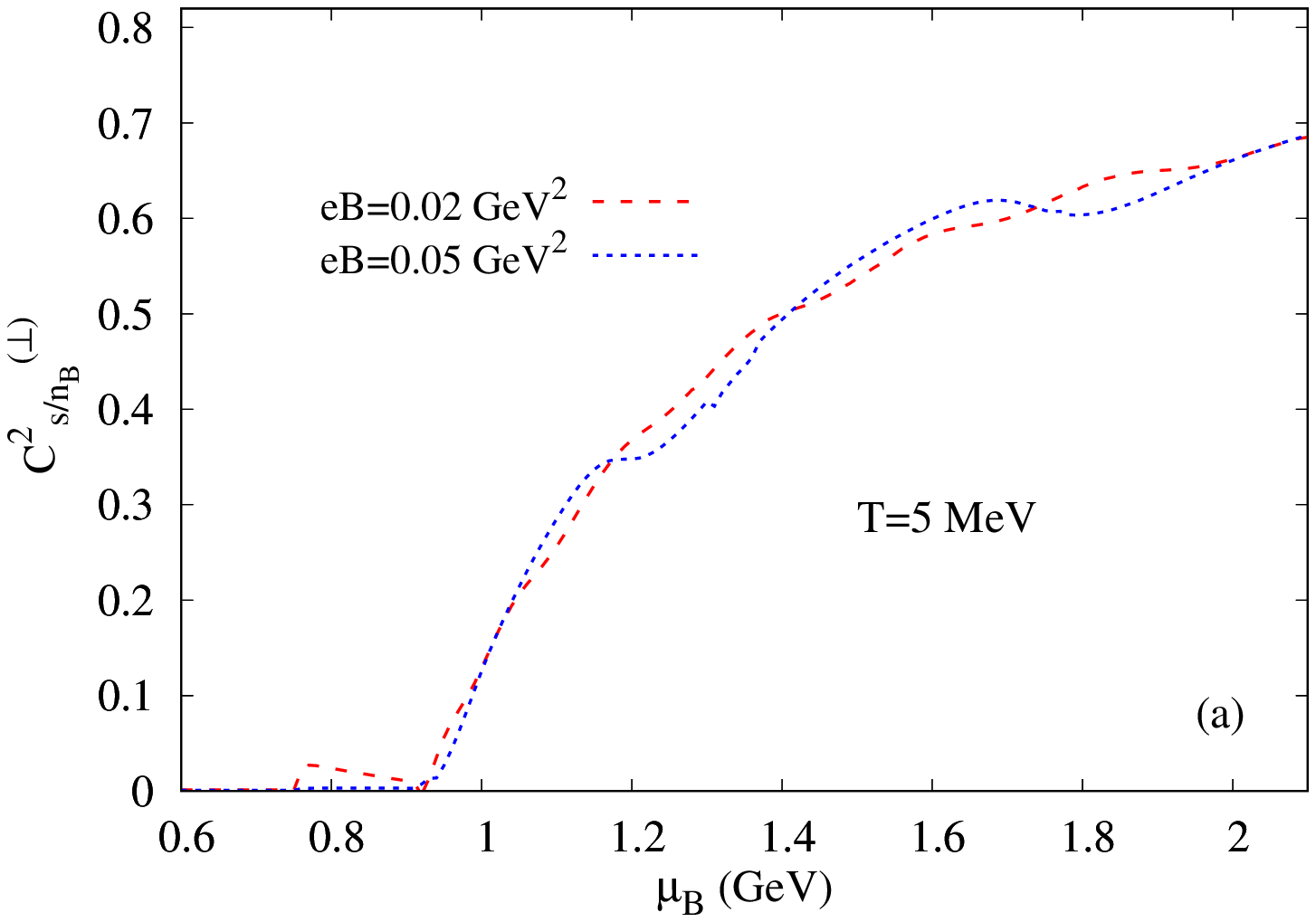}
	\includegraphics[angle = -90, scale=0.23]{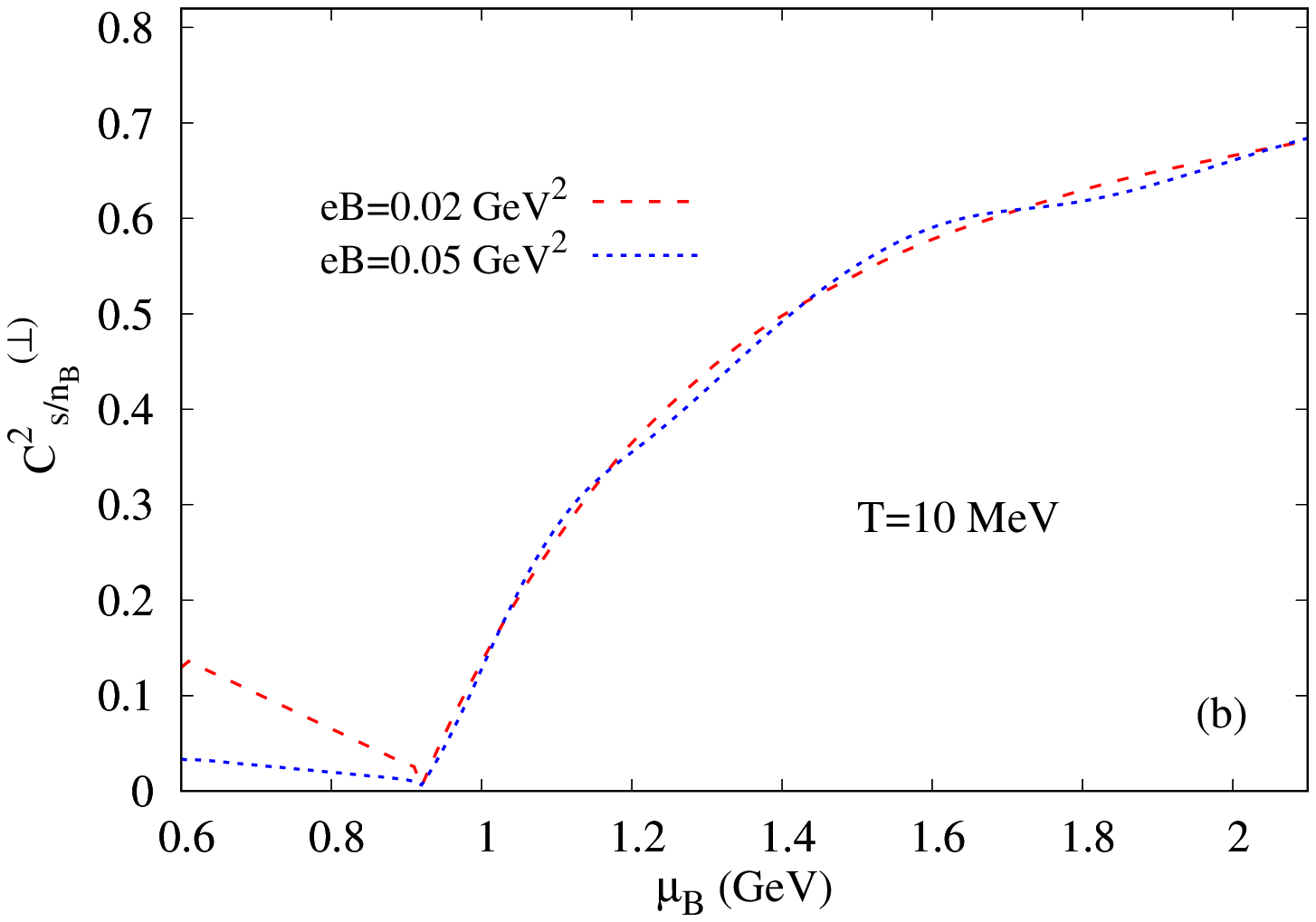}
	\includegraphics[angle = -90, scale=0.23]{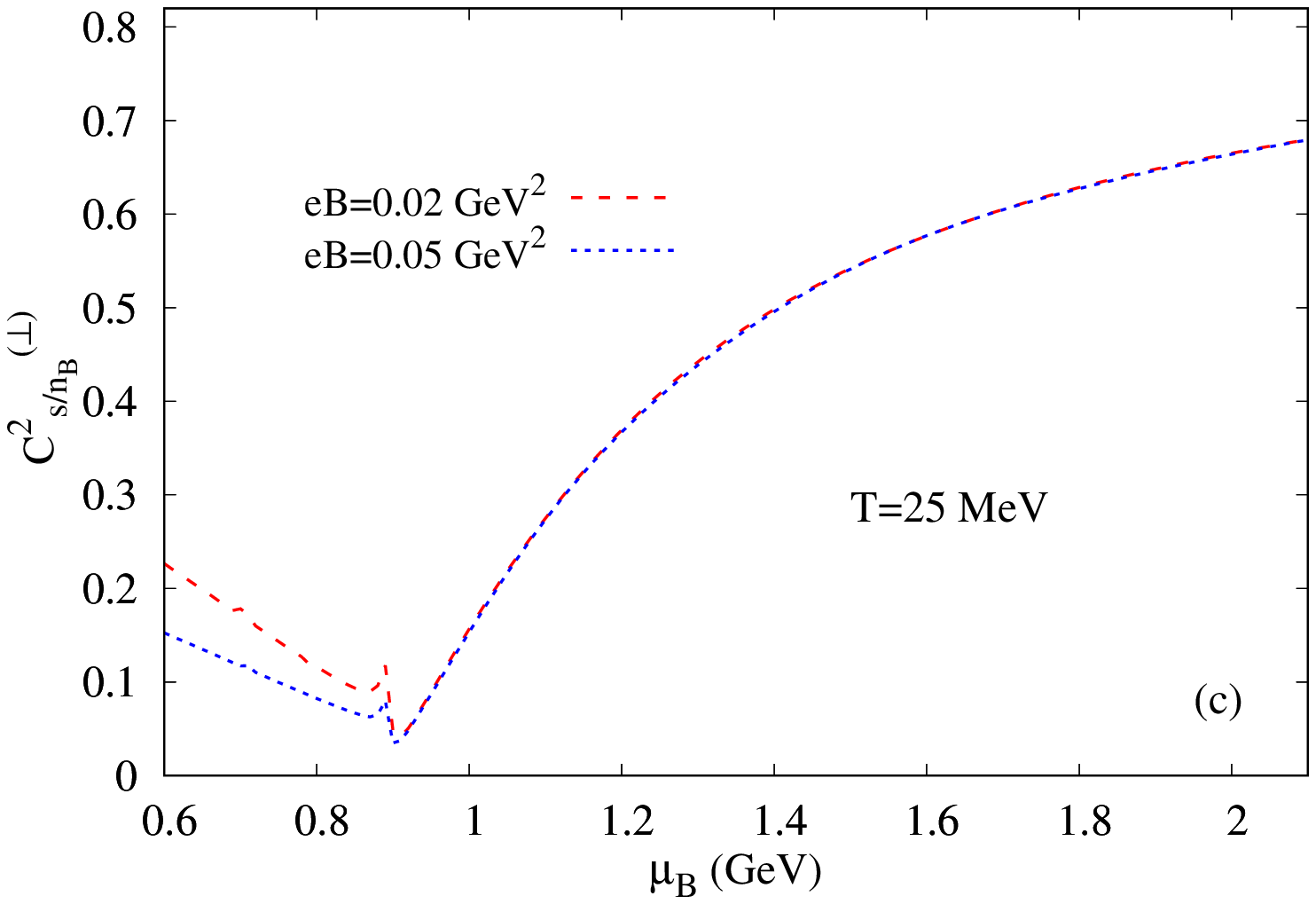}
	\caption{Perpendicular component of squared speed of sound $C_{s/n_B}^{2(\perp)}$ as a function of chemical potential $\mu_B$ for $eB=0.02,~0.05\rm~GeV^2$ at (a) $T=5\rm~MeV$, (b) $T=10\rm~MeV$, (c) $T=25\rm~MeV$.}
	\label{Fig.MuC2sbynB_Perp_T}
\end{figure}
%

The variation of $C_{s/n_B}^{2(\perp)}$ with respect to $\mu_B$ is plotted for various temperatures $T = 5, 10, 16, 25, 50, 100$ MeV in Fig.$~\ref{Fig.MuC2sbynB_Perp_B}(a)$ at $eB = 0.02$ GeV$^2$ and in Fig.~$\ref{Fig.MuC2sbynB_Perp_B}(b)$ at $eB = 0.05$ GeV$^2$ respectively. Fig.~$\ref{Fig.MuC2sbynB_Perp_B}(a)$ shows that the oscillating behavior of $C_{s/n_B}^{2(\perp)}$ diminishes with increasing temperature for larger $\mu_B$ values. However, in the lower $\mu_B$ domain, $C_{s/n_B}^{2(\perp)}$ rises with temperature. Fig.~$\ref{Fig.MuC2sbynB_Perp_B}(b)$ shows a similar trend as in Fig.~$\ref{Fig.MuC2sbynB_Perp_B}(a)$. The comparison between the parallel and perpendicular components of ${C^2_{s/n_B}}$ with respect to $\mu_B$ for a temperature of $T = 5$ MeV and a magnetic field of $eB = 0.02\rm~GeV^2$ is illustrated in Fig.~\ref{Fig.MuC2sbynB_Perp_B}(c). The figure shows a significant disparity between the parallel and perpendicular components of the sound speed. In the higher range of $\mu_B$ values, $C_{s/n_B}^{2(\perp)}$ surpasses $C_{s/n_B}^{2(\parallel)}$.

Fig.~$\ref{Fig.MuC2sbynB_Perp_T}(a)$ represents $C_{s/n_B}^{2(\perp)}$ as a function of $\mu_B$ for background fields $eB = 0.02, 0.05$ GeV at $T = 5$ MeV, while Fig.~$\ref{Fig.MuC2sbynB_Perp_T}(b)$ and Fig.~$\ref{Fig.MuC2sbynB_Perp_T}(c)$ do so for $T = 10$ MeV and $T = 25$ MeV respectively. In Figs.~$\ref{Fig.MuC2sbynB_Perp_B}(a)-(b)$ and $\ref{Fig.MuC2sbynB_Perp_T}(a)-(c)$, the minima of $C_{s/n_B}^{2(\perp)}$  around $\mu_B = 0.92$ GeV mark the occurrence of the liquid-gas phase transition which is consistent with $C_{s/n_B}^{2(\parallel)}$. Examination of Fig.~$\ref{Fig.MuC2sbynB_Perp_T}(a)$ at temperature $T=5$ MeV reveals that the oscillations in $C_{s/n_B}^{2(\perp)}$ intensify with an increase in the background magnetic field for larger $\mu_B$ values. As the temperature rises, the oscillations in $C_{s/n_B}^{2(\perp)}$ diminish (see Fig.~$\ref{Fig.MuC2sbynB_Perp_T}(b)$), achieving a smoother profile (see Fig.~$\ref{Fig.MuC2sbynB_Perp_T}(c)$) for $eB = 0.02, 0.05$ GeV in the domain of higher $\mu_B$ values. Conversely, in the domain of lower $\mu_B$ values, the values of $C_{s/n_B}^{2(\perp)}$ exhibit an increase with rising temperature. This trend is consistent with the patterns illustrated in Figs.~$\ref{Fig.MuC2sbynB_Perp_B}(a)-(b)$.

\begin{figure}[h] 
	\includegraphics[angle = -90, scale=0.23]{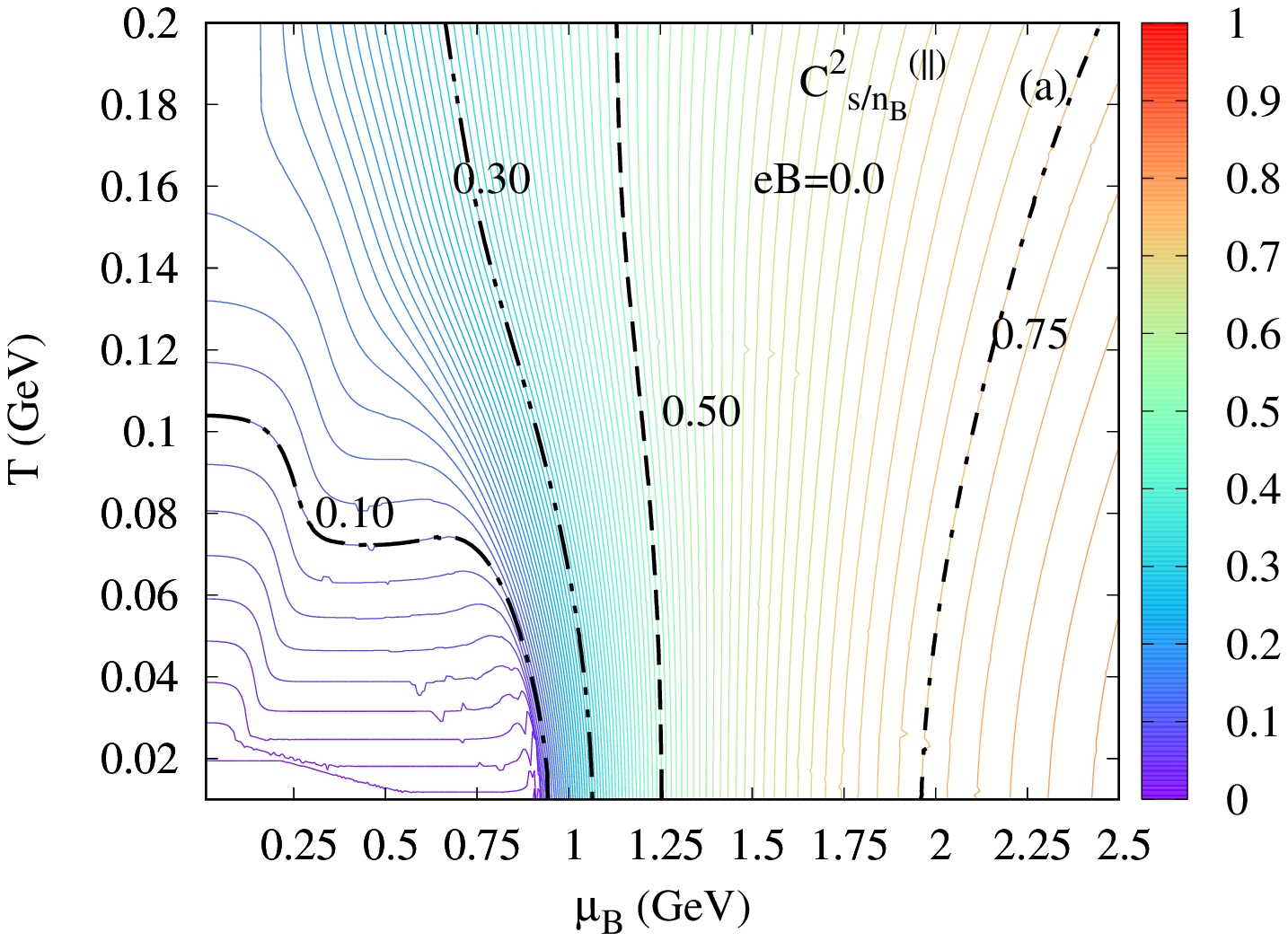}
	\includegraphics[angle = -90, scale=0.23]{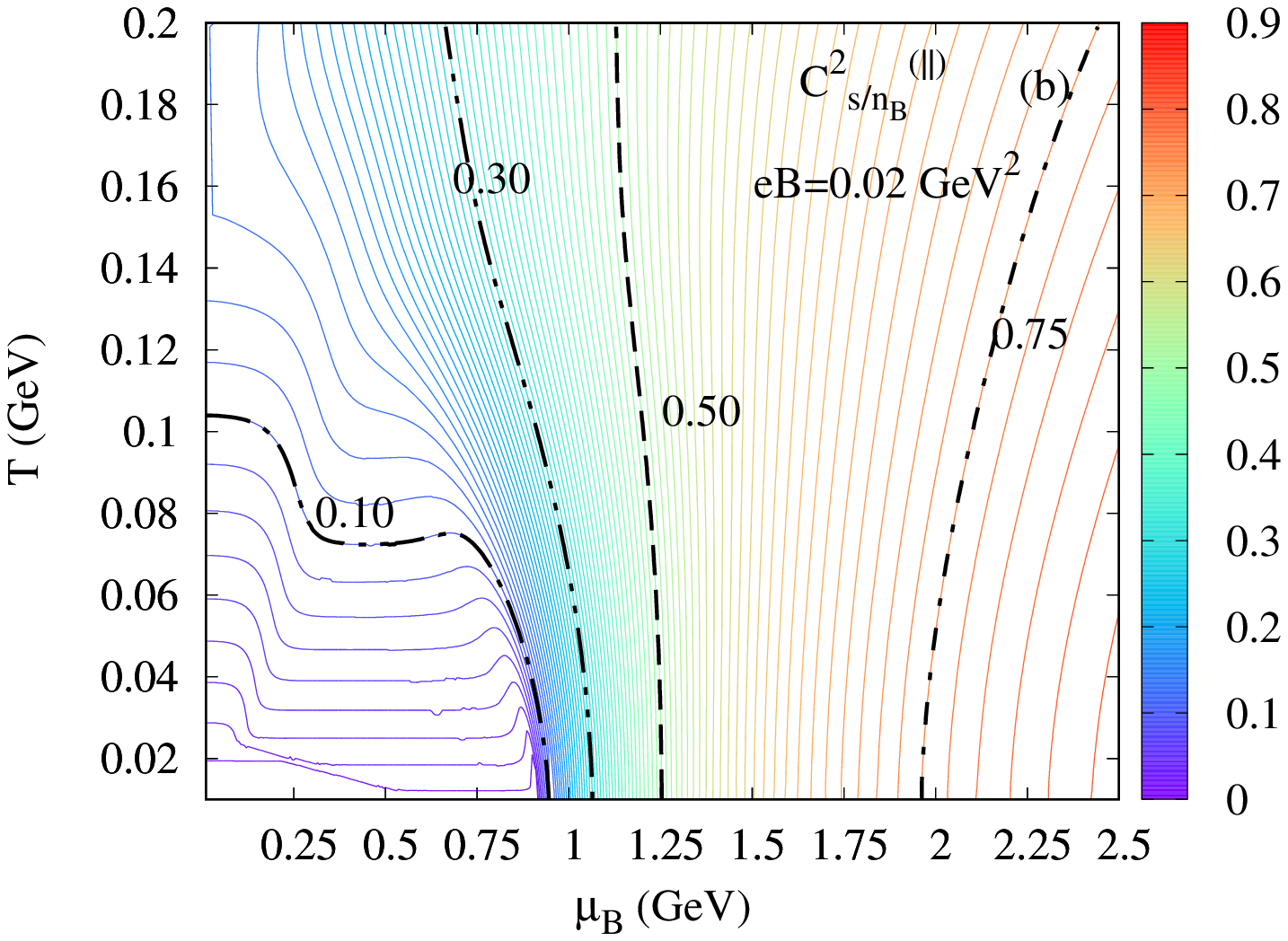}
	\includegraphics[angle = -90, scale=0.23]{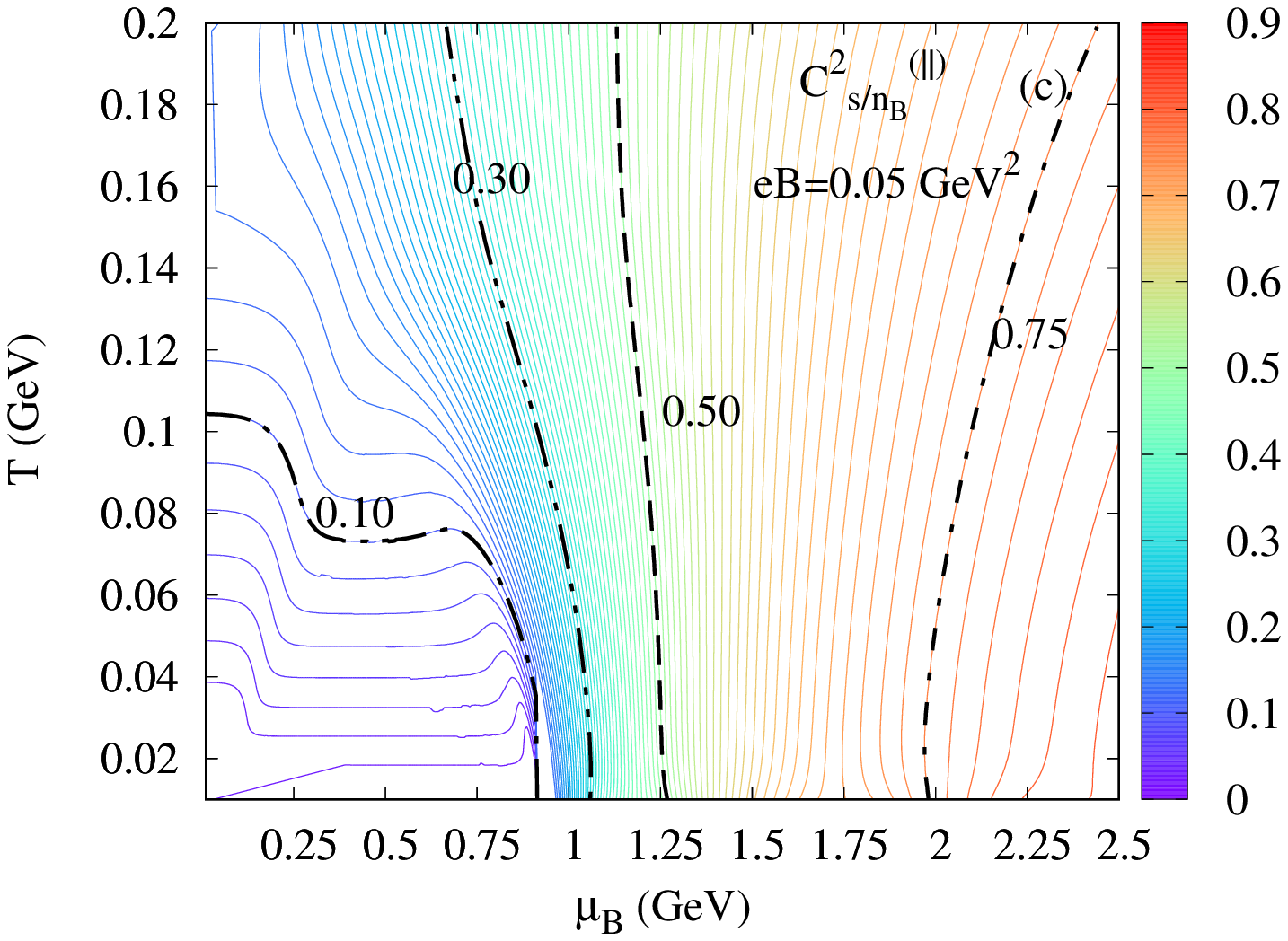}
	\caption{Contour plots of $C_{s/n_B}^{2(\parallel)}$ in the full $T-\mu_B$ plane at (a) $eB=0$, (b) $eB=0.02\rm~GeV^2$, (c) $eB=0.05\rm~GeV^2$.}
	\label{Fig.Con.C2sbynB}
\end{figure}

The contour plots in Figs.~$\ref{Fig.Con.C2sbynB}(a)-(c)$ demonstrate the profiles of $C_{s/n_B}^{2(\parallel)}$ in the full $T-\mu_B$ plane for $eB=0, 0.02,$ and $0.05~\rm GeV^2$ respectively. In each of the contour plots the profiles corresponding to $C_{s/n_B}^{2(\parallel)}=0.1,~0.30,~0.50,~0.75$ are identified separately. Recall that in Figs.~\ref{Fig.MuC2sbynB}, \ref{Fig.MuC2sbynB_T}, \ref{Fig.Con.C2sbynB} the squared speed of sound is larger than its conformal value , i.e, $\frac{1}{3}$ at high chemical potential or high density.  Importantly, causality is always preserved, i.e, $C_{s/n_B}^{2(\parallel)}<1$. Note that the value of $C_{s/n_B}^{2(\parallel)}$ approaches $\frac{1}{3}$ in quark matter at high chemical potential or high baryon density~\cite{PhysRevD.105.094024}.  

\subsection{Speed of Sound at Constant $n_B$ or $s$}
\begin{figure}[h] 
	\includegraphics[angle = -90, scale=0.23]{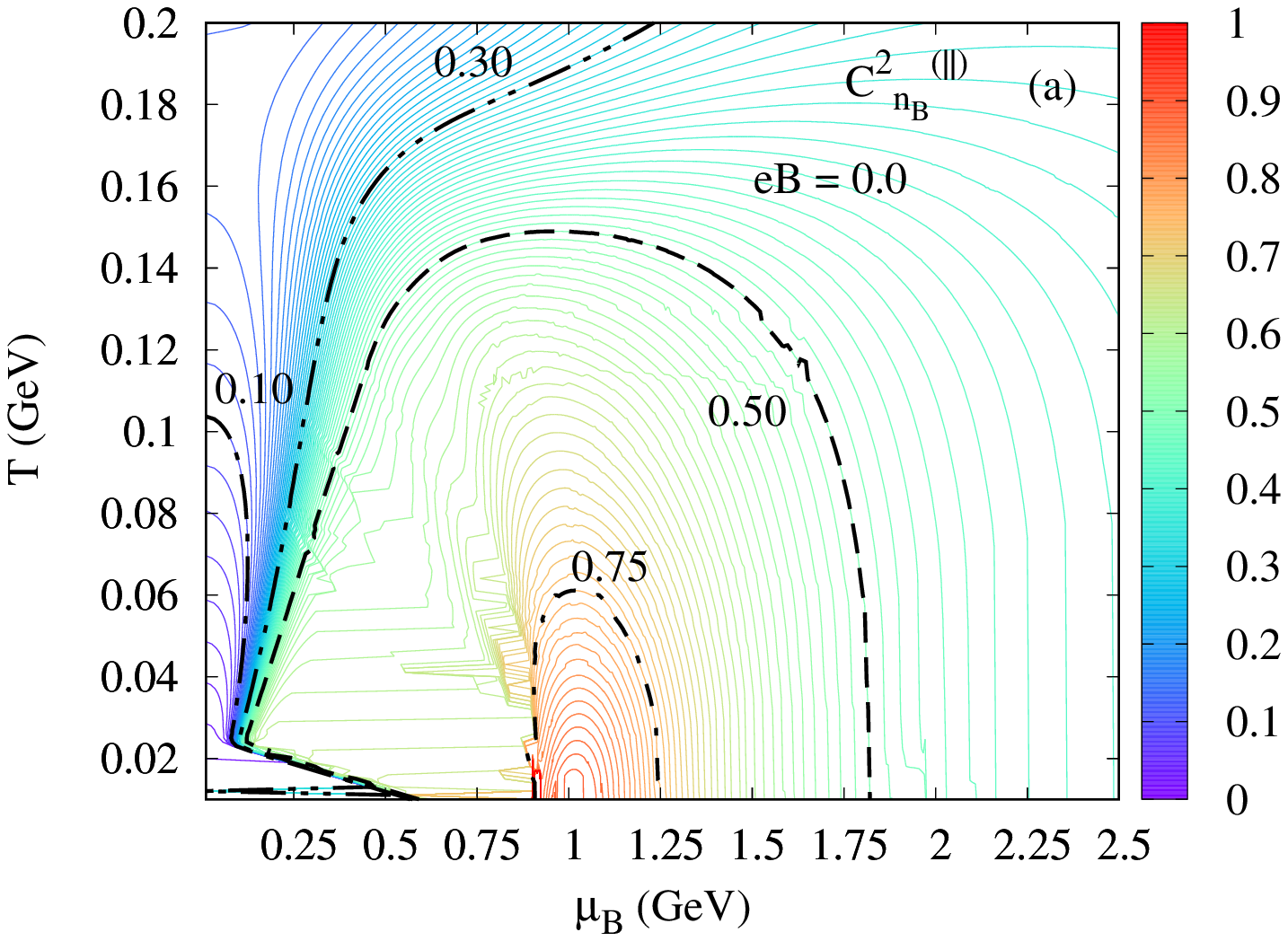}
	\includegraphics[angle = -90, scale=0.23]{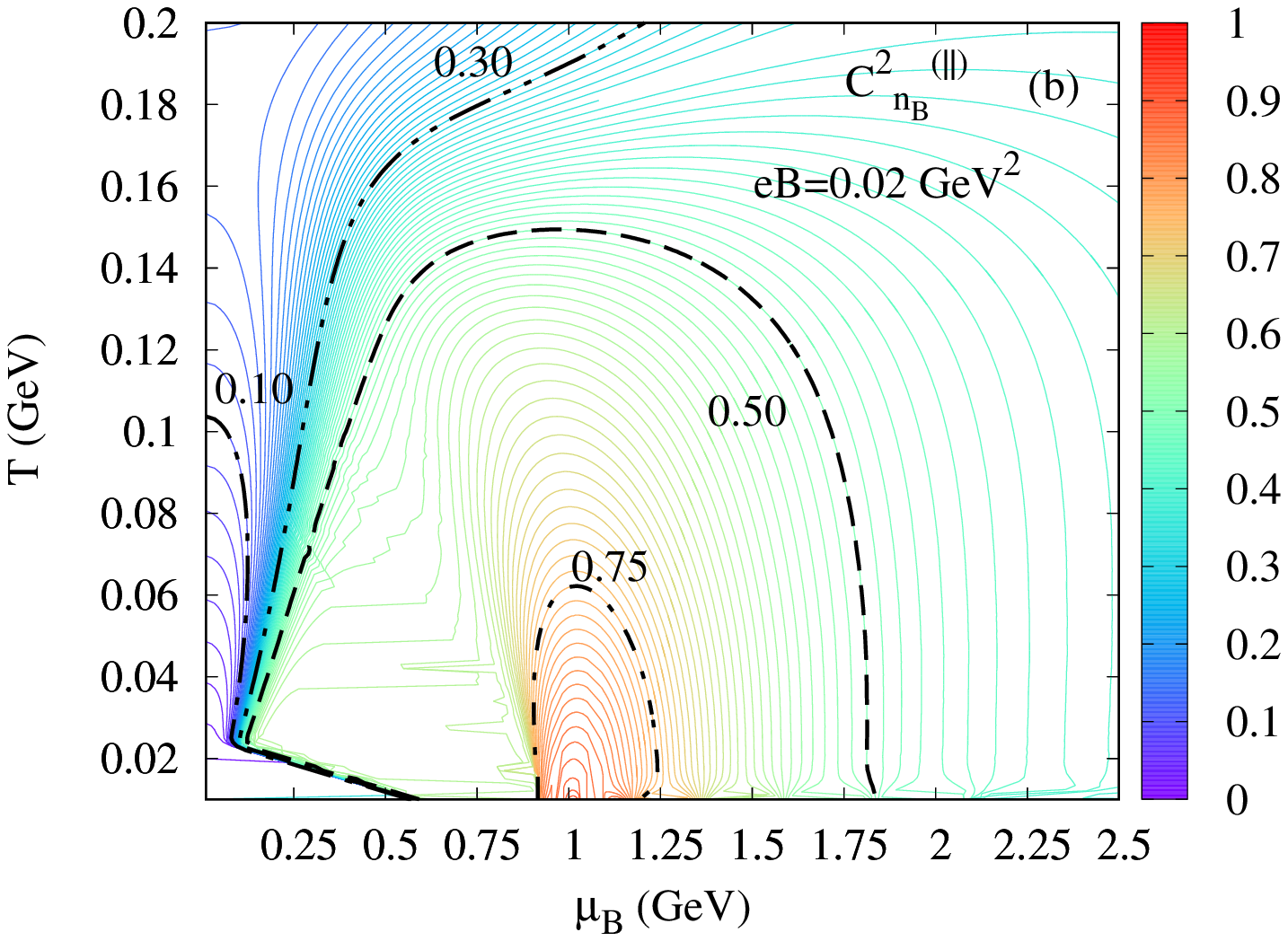}
	\includegraphics[angle = -90, scale=0.23]{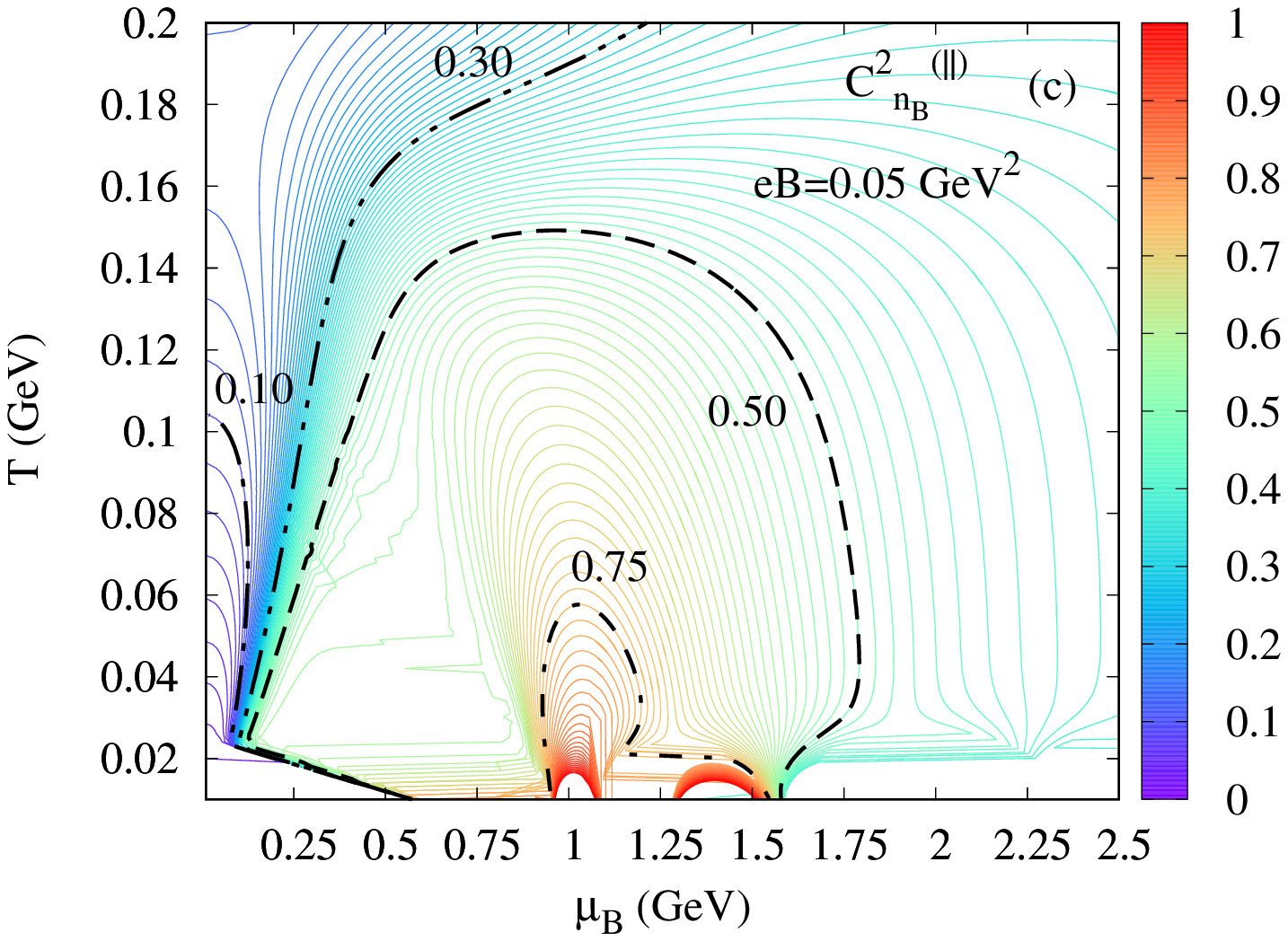}
	\caption{Contour plots of $C_{n_B}^{2(\parallel)}$ in the full $T-\mu_B$ plane at (a) $eB=0$, (b) $eB=0.02\rm~GeV^2$, (c) $eB=0.05\rm~GeV^2$.}
	\label{Fig.Con.C2nB}
\end{figure}

We present contour maps illustrating $C_{n_B}^{2(\parallel)}$ in the $T-\mu_B$ plane under various magnetic field strengths for $eB=0$ in Fig.~\ref{Fig.Con.C2nB}(a), $eB=0.02~\rm GeV^2$ in Fig.~\ref{Fig.Con.C2nB}(b), $eB=0.05~\rm GeV^2$ in Fig.~\ref{Fig.Con.C2nB}(c).
In the absence of a magnetic field, in Fig.~\ref{Fig.Con.C2nB}(a), $C_{n_B}^{2(\parallel)}$  exhibits a non-monotonic behaviour with chemical potential for almost all temperatures. Notably, there is a distinctive peak-like structure observed at intermediate chemical potential values. When the magnetic field is switched on, these characteristics manifest in a similar fashion. However, at lower temperatures $C_{n_B}^{2(\parallel)}$ exhibits a distinct variation which increases with the increasing magnetic field strength.  
%
%
%

\begin{figure}[h] 
	\includegraphics[angle = -90, scale=0.23]{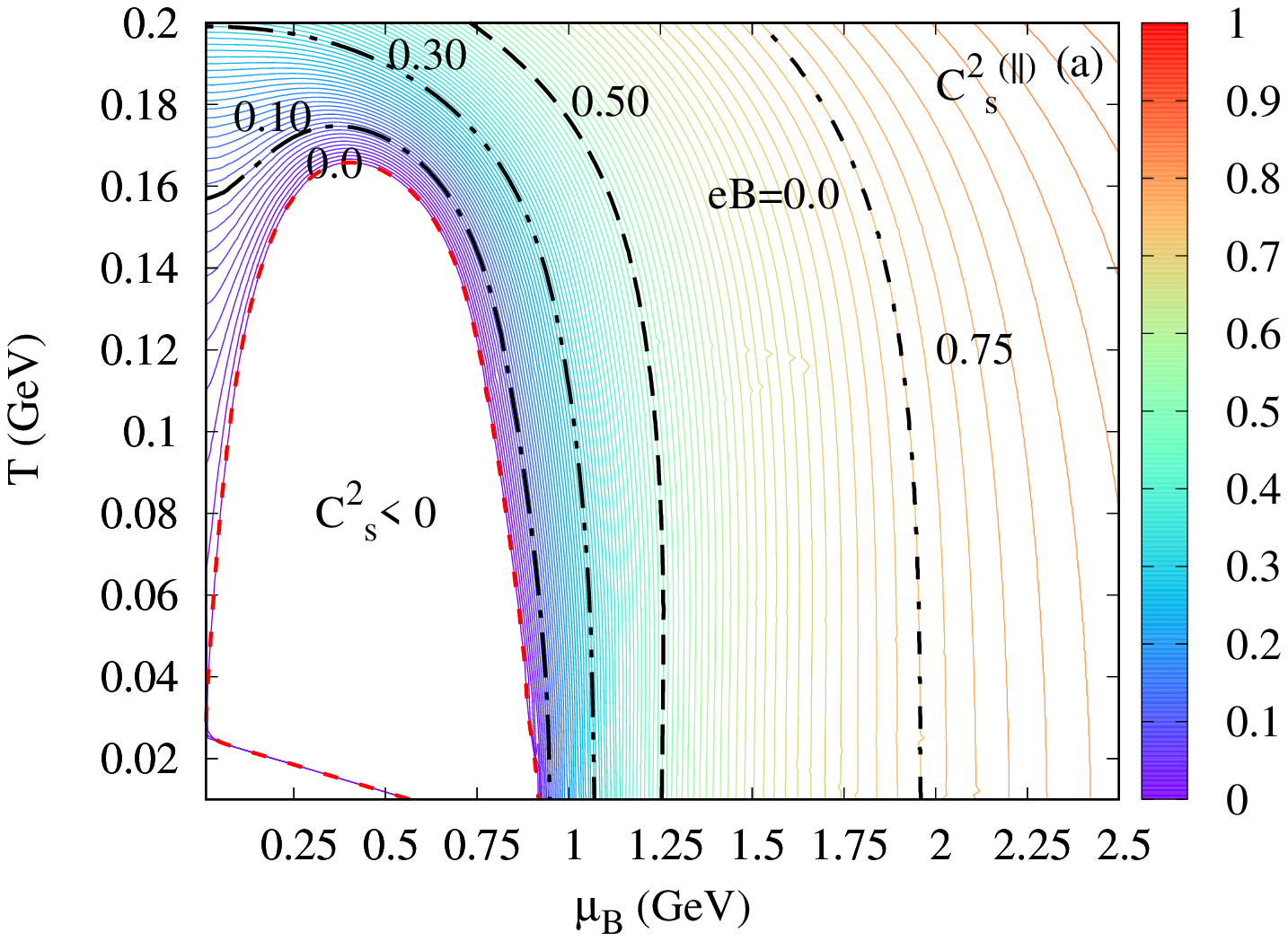}
	\includegraphics[angle = -90, scale=0.23]{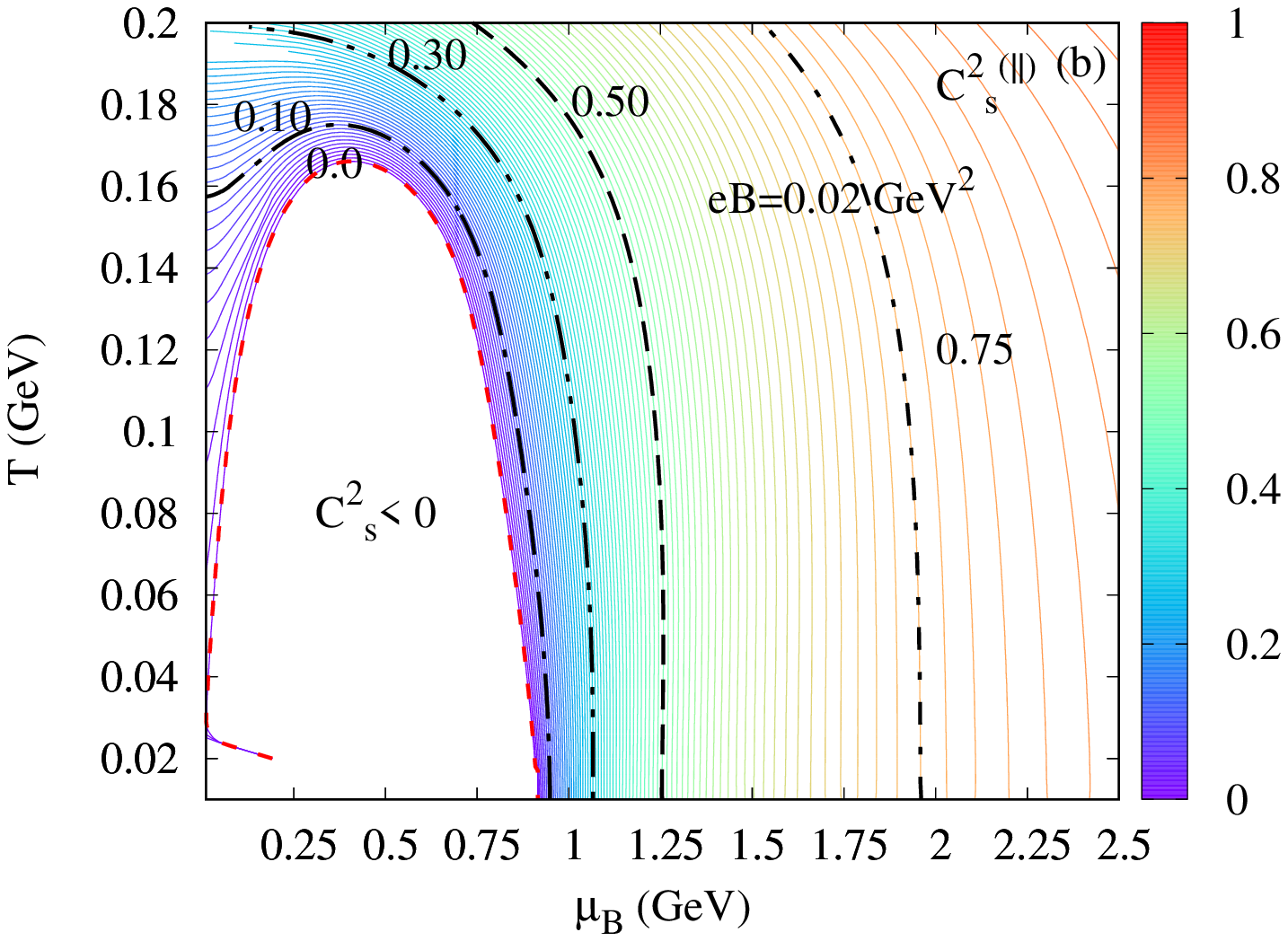}
	\includegraphics[angle = -90, scale=0.23]{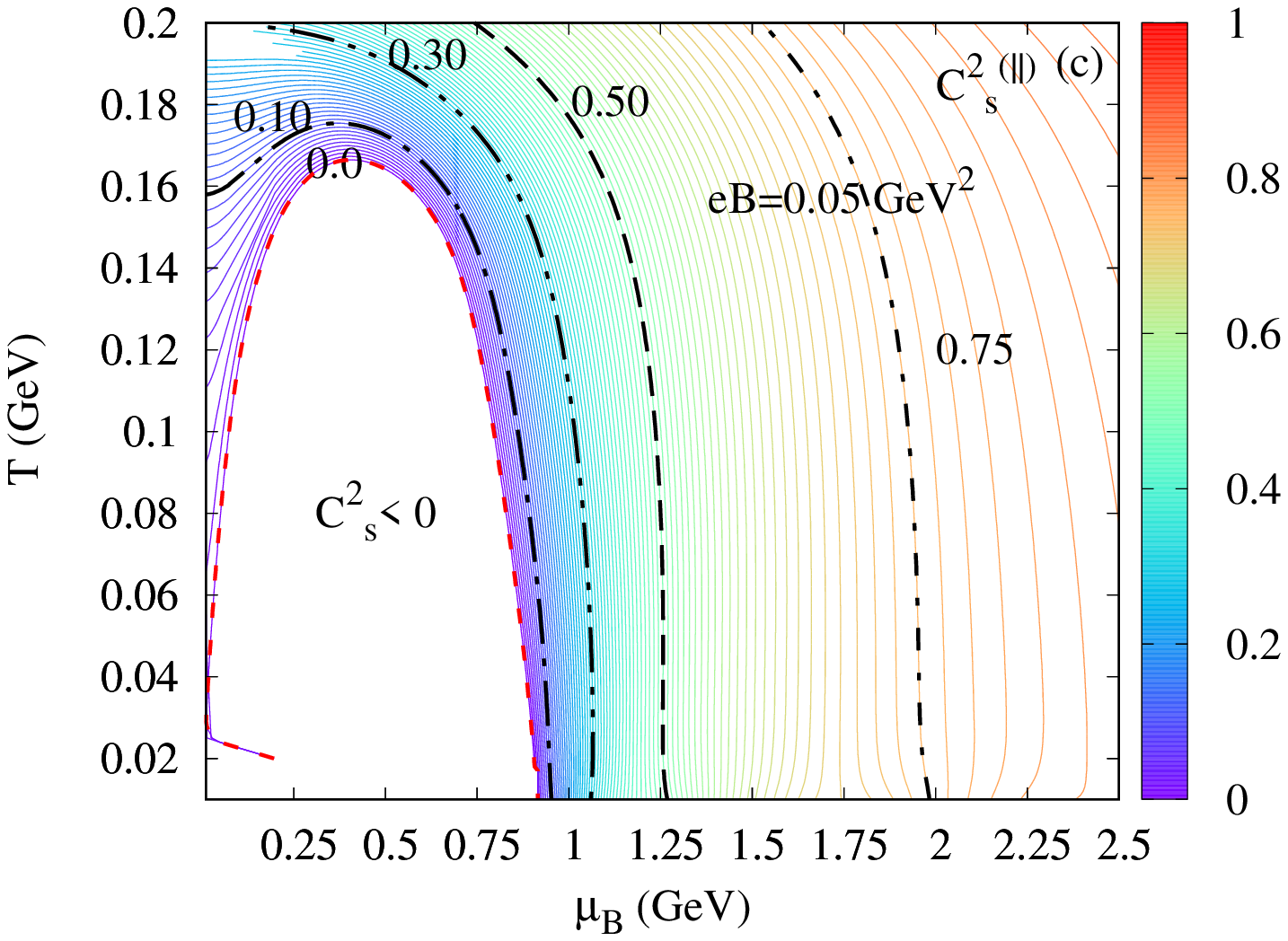}
	\caption{Contour plots of ${C^2_{s}}^{(\parallel)}$ in the full $T-\mu_B$ plane at (a) $eB=0$, (b) $eB=0.02\rm~GeV^2$, (c) $eB=0.05\rm~GeV^2$.}
	\label{Fig.Con.C2s}
\end{figure}

Next, we now show the contour maps of $C_{s}^{2(\parallel)}$ at constant entropy density under different magnetic field strengths, specifically, for $eB=0$ in Fig.~\ref{Fig.Con.C2s}(a), $eB=0.02~\rm GeV^2$ in Fig.~\ref{Fig.Con.C2s}(b), $eB=0.05~\rm GeV^2$ in Fig.~\ref{Fig.Con.C2s}(c) respectively. In the absence of magnetic field, as shown in Fig.~\ref{Fig.Con.C2s}(a), the contour plots of $C_{s}^{2(\parallel)}$ exhibit a complicated structure. The graphs demonstrate that $C_{s}^{2(\parallel)}$ has both negative and positive values separated by the red dashed  line in the Fig.~\ref{Fig.Con.C2s}(a). The value of $C_s^{2(\parallel)}$ vanishes on the boundary given by the red dashed line. This feature possibly represents a general phenomenon observed in first-order phase transitions within interacting systems where the fermion mass exhibits a dependency on both temperature and density. A similar behavior in the speed of sound within quark matter is observed in Ref.~\cite{PhysRevD.105.094024}. Indeed, the boundary denoted by the red dashed line in Figs.~$\ref{Fig.Con.C2s}(a)$ can be correlated with the thermodynamic formula
\begin{eqnarray}
	\FB{\frac{\partial \mu_B}{\partial T}}_{s/n_B}=\frac{\mu_B\FB{\frac{\partial p^{(\parallel)}}{\partial\epsilon}}_s}{T\FB{\frac{\partial p^{(\parallel)}}{\partial\epsilon}}_{n_B}}=\FB{\mu_B/T}\frac{{C_s}^{2(\parallel)}}{C_{n_B}^{2(\parallel)}}~~.
	\label{Eq.dMubydT}
\end{eqnarray} 
Therefore, one can obtain the boundary of $C_s^{2(\parallel)}=0$ using Eq.~\eqref{Eq.dMubydT} and by taking the condition $\FB{\frac{\partial \mu_B}{\partial T}}_{s/n_B}=0$. Moreover, one of the two physical quantities $C_s^{2(\parallel)}$ and  $C_{n_B}^{2(\parallel)}$ takes negative value when $\FB{\frac{\partial \mu_B}{\partial T}}_{s/n_B}<0$. Since  $C_{n_B}^{2(\parallel)}$ is always positive, $C_s^{2(\parallel)}$ is negative in this situation. The region enclosed by the dashed red line in Figs.~$\ref{Fig.Con.C2s}(a)$ represents this region. With the introduction of a magnetic field, the contour plots exhibit a similar behavior except at lower temperature and high chemical potential as demonstrated in Figs.~$\ref{Fig.Con.C2s}(b)-(c)$.

\subsection{Speed of Sound at Constant $T$}
%

\begin{figure}[h] 
	\includegraphics[angle = -90, scale=0.23]{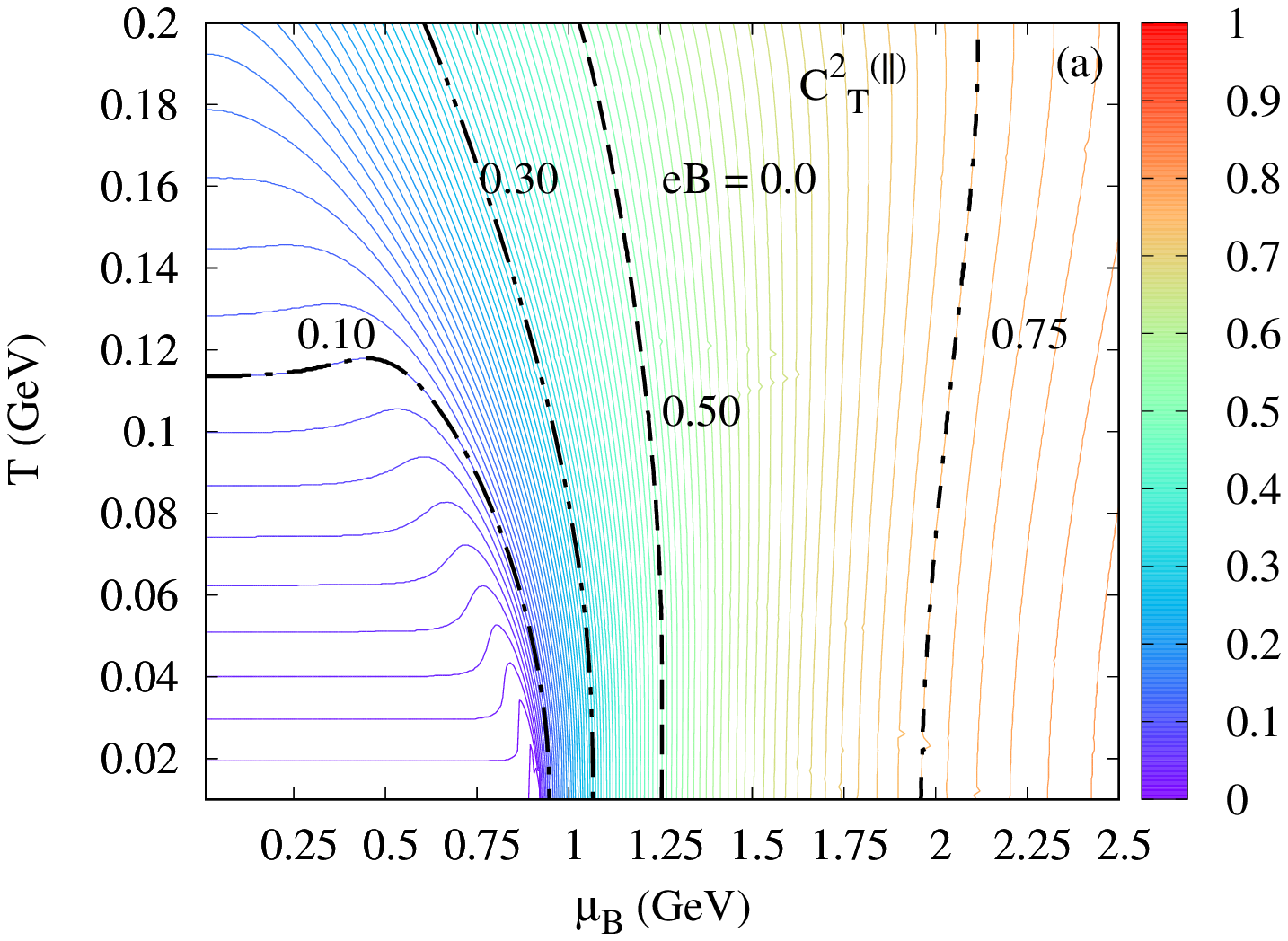}
	\includegraphics[angle = -90, scale=0.23]{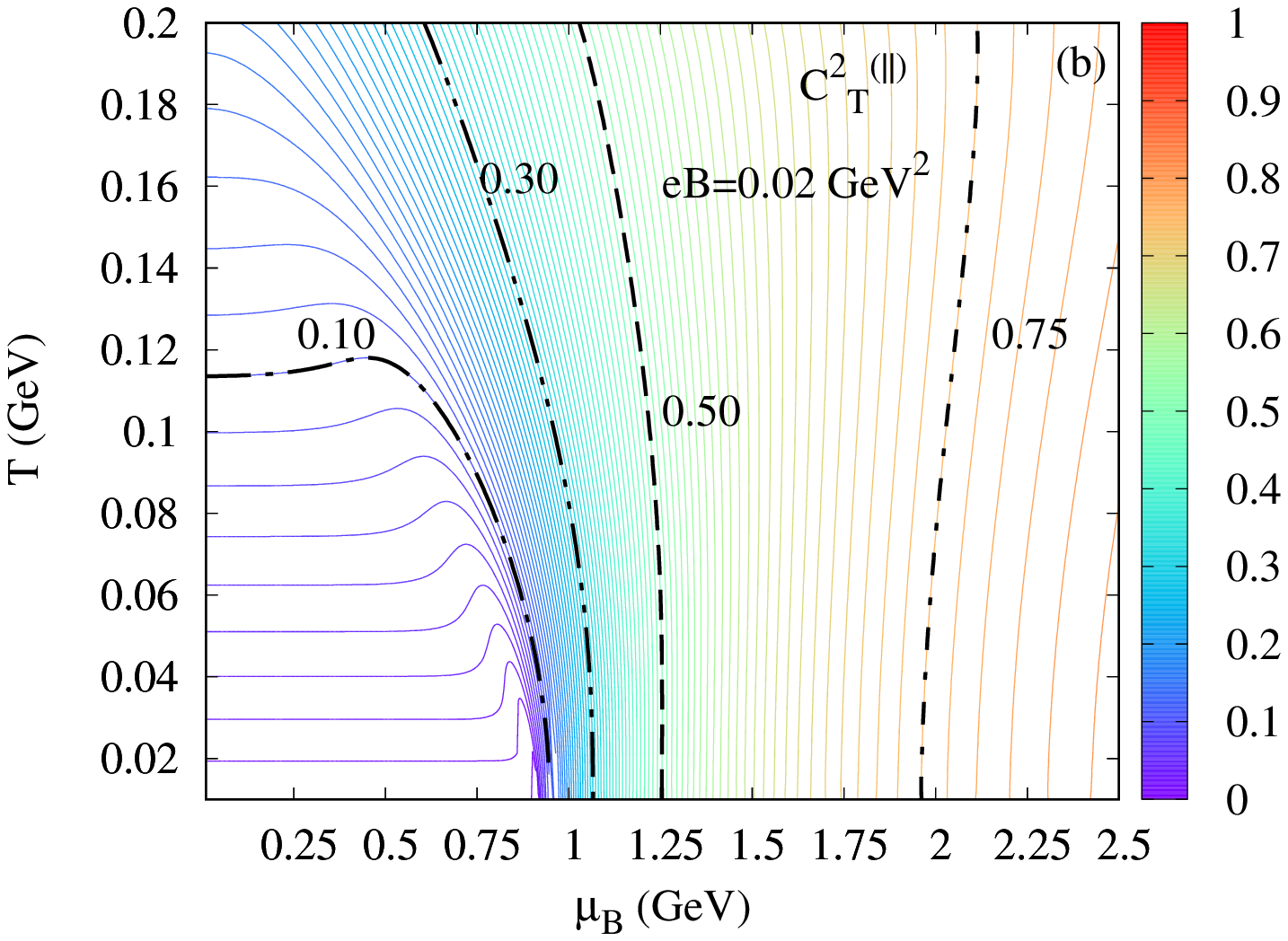}
	\includegraphics[angle = -90, scale=0.23]{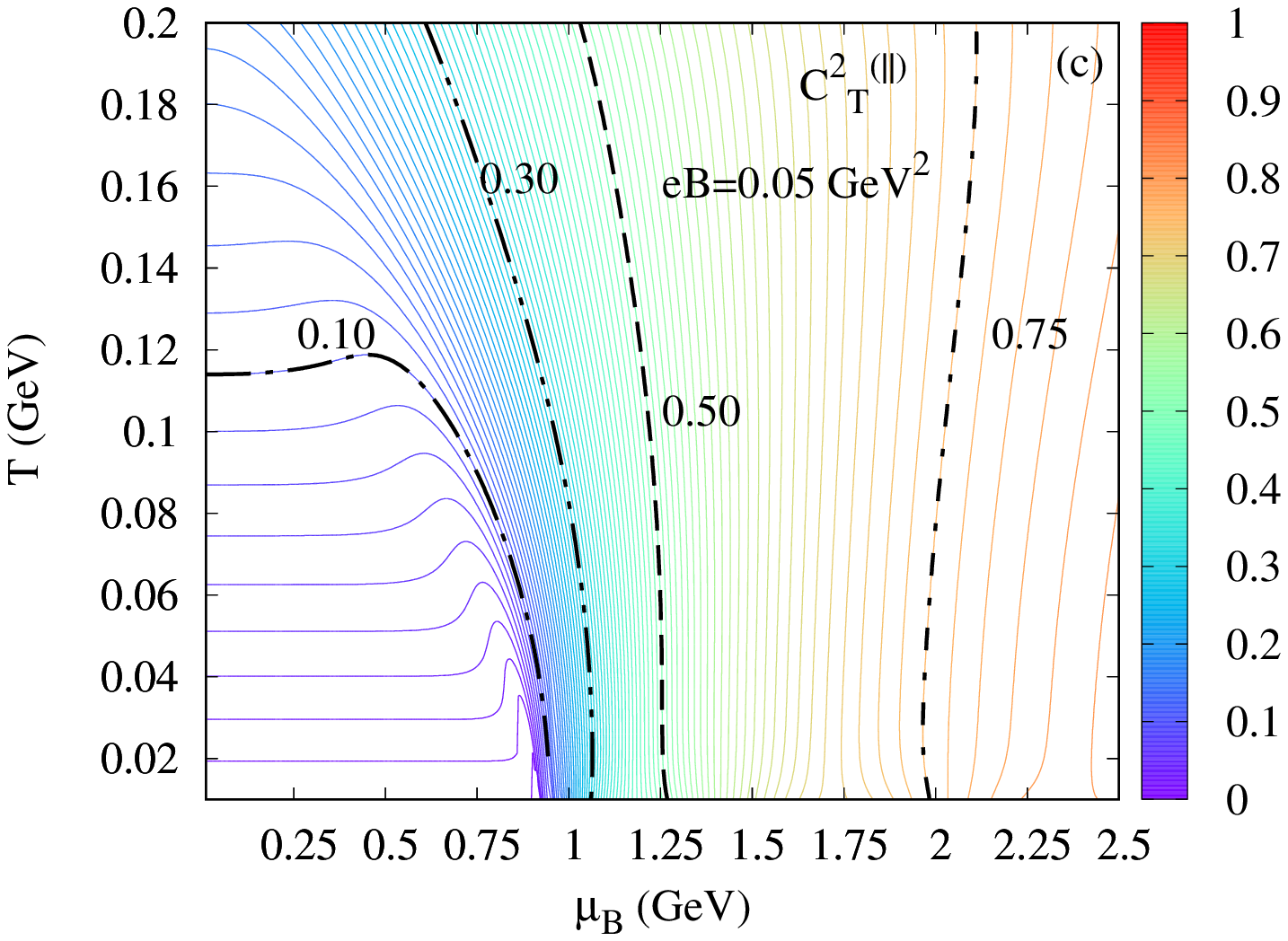}
	\caption{Contour plots of ${C^2_{T}}^{(\parallel)}$ in the full $T-\mu_B$ plane at (a) $eB=0$, (b) $eB=0.02\rm~GeV^2$, (c) $eB=0.05\rm~GeV^2$.}
	\label{Fig.Con.C2T}
\end{figure}

In the following, we explore the speed of sound at constant temperature $T$. 
The estimation of $C_T^{2(\parallel)}$ as a function of the baryon number density during the chemical freeze-out of quark-gluon plasma created in relativistic heavy-ion collisions has been carried out in Ref.~\cite{PhysRevLett.127.042303}. In recent times, there has been a lot of discussion regarding the density-dependent $C_T^{2(\parallel)}$ in the context of neutron star matter. The observational data indicates a substantial value of $C_T^{2(\parallel)}$ (greater than 1/3) at densities several times that of nuclear saturation. Here, we see the behavior of $C_T^{2(\parallel)}$ in nuclear matter in the full $T-\mu_B$ plane under different magnetic field strengths, specifically, for $eB=0$ in Fig.~\ref{Fig.Con.C2T}(a), $eB=0.02~\rm GeV^2$ in Fig.~\ref{Fig.Con.C2T}(b), $eB=0.05~\rm GeV^2$ in Fig.~\ref{Fig.Con.C2T}(c). {In the case of zero magnetic field strength in Fig.~\ref{Fig.Con.C2T}(a), the value of $C_T^{2(\parallel)}$ always increases with the rising chemical potential.}  At low temperature the behaviour of $C_T^{2(\parallel)}$ closely resembles $C_{s/n_B}^{2(\parallel)}$.
The effects of the magnetic field are shown in Figs.~$\ref{Fig.Con.C2T}(b)-(c)$. 

\subsection{Isothermal Compressibility}
\begin{figure}[h] 
	\includegraphics[angle = -90, scale=0.23]{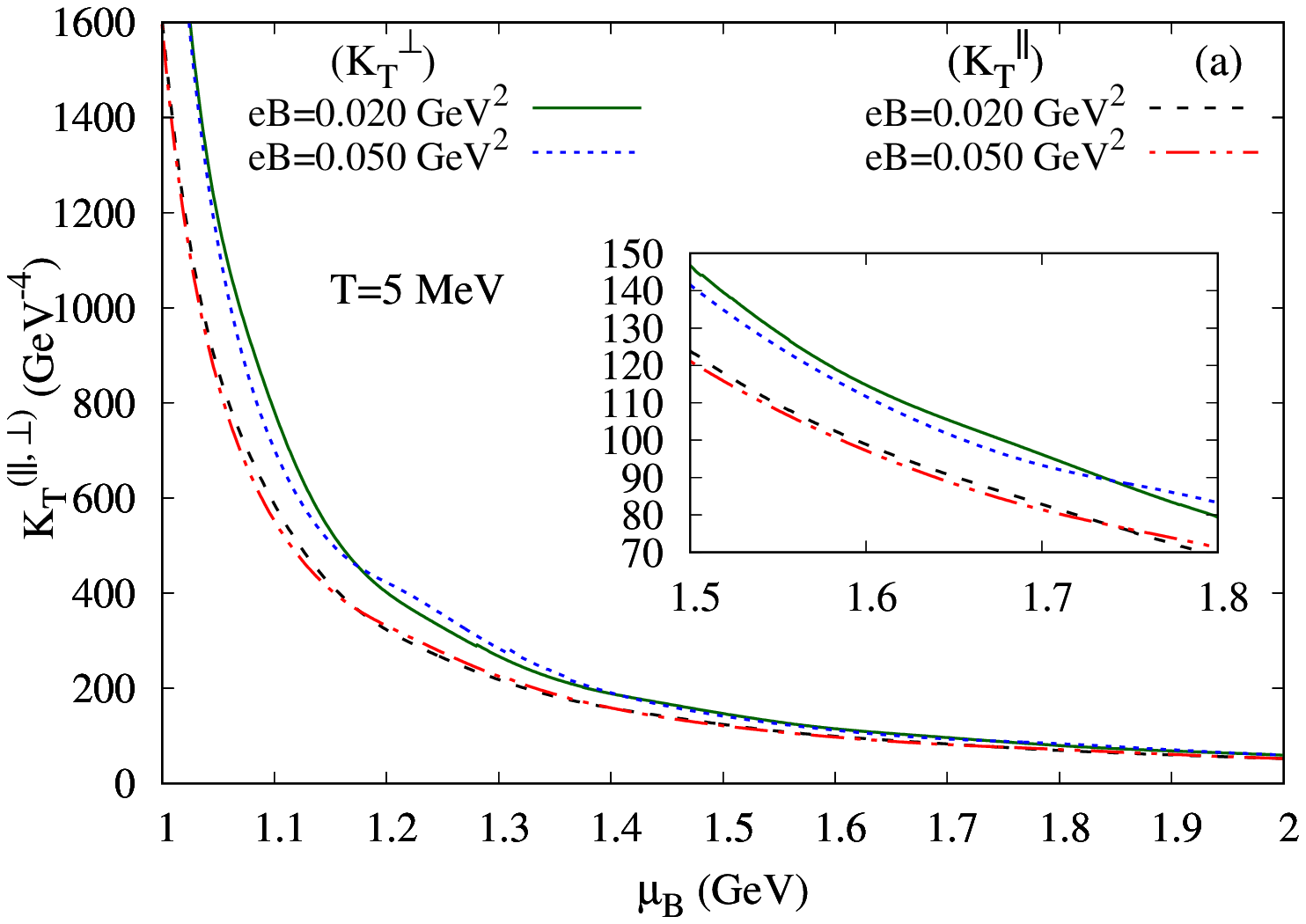}
	\includegraphics[angle = -90, scale=0.23]{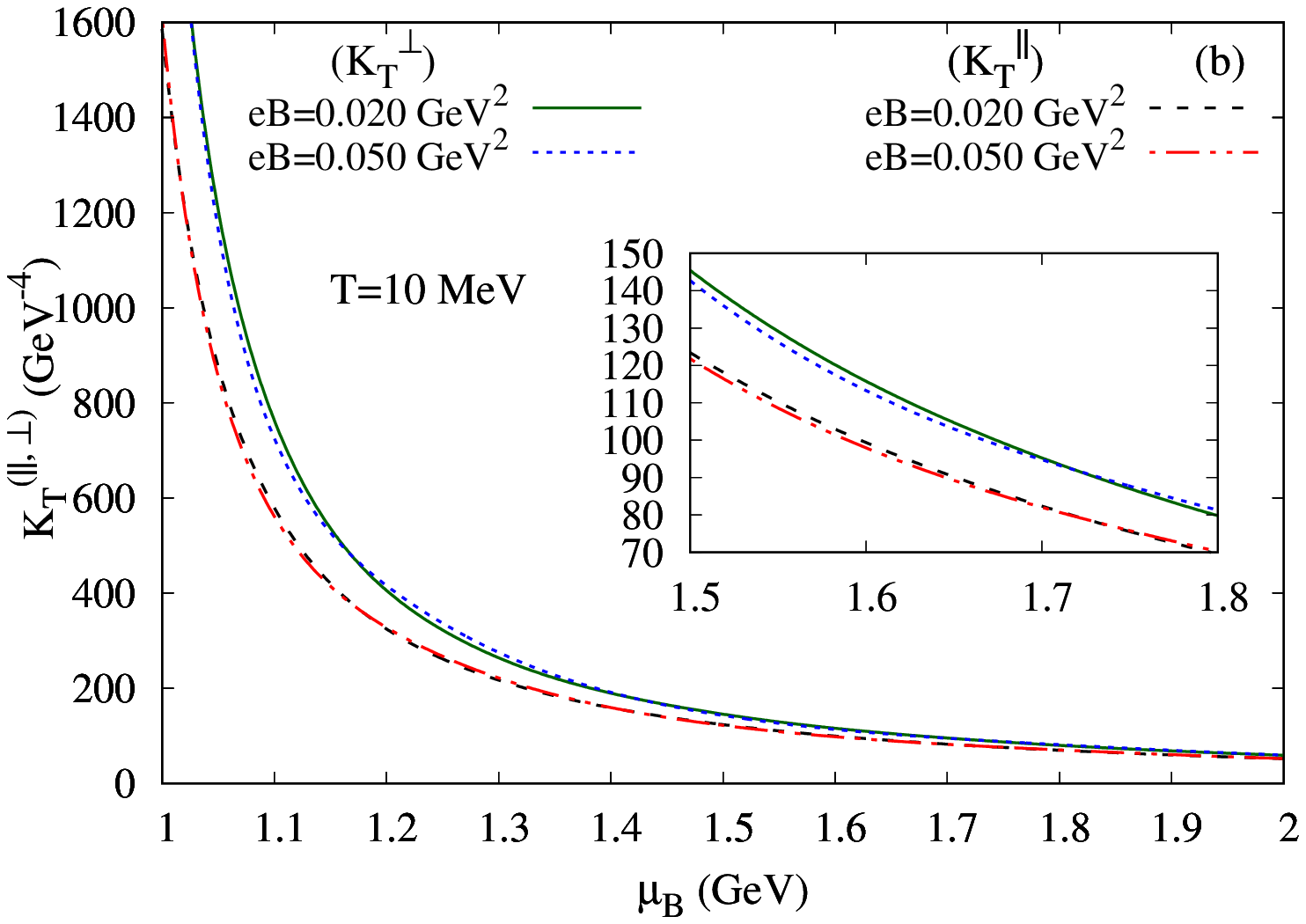}
	\includegraphics[angle = -90, scale=0.23]{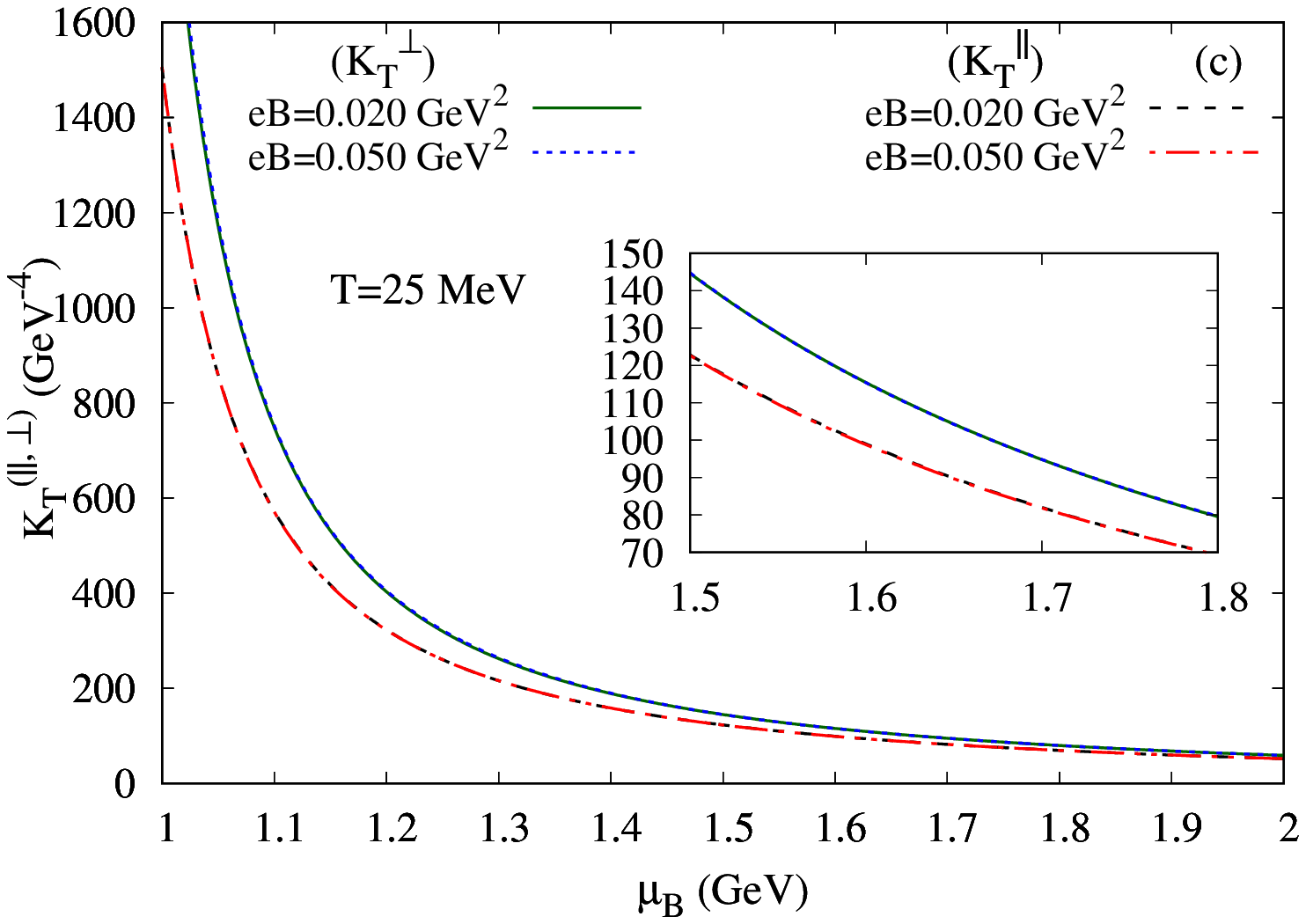}
	\caption{Parallel and perpendicular components of isothermal compressibility $K_T^{(\parallel,~\perp)}$ as function of $\mu_B$ for $eB=0.02,~0.05\rm~GeV^2$ at (a) $T=5$ MeV, (b) $T=10$ MeV, (c) $T=25$ MeV.}
	\label{KT}
\end{figure}
In the absence of magnetic fields, the isothermal compressibility ($K_T$) exhibits isotropy. 
However, in the presence of magnetic field $K_T$ becomes anisotropic and splits into $K_T^{(\parallel)}$ (along the magnetic field direction) and $K_T^{(\perp)}$ (perpendicular to the magnetic field direction). To estimate isothermal compressibility, we use Eqs.~\eqref{KT_para}-\eqref{KT_per}.  $K_T^{(\parallel,~\perp)}$ are plotted as a function of chemical potential $\mu_B$ for magnetic field $eB=0.02,~0.05\rm~GeV^2$. The plots are depicted in Fig.~$\ref{KT}(a)$ at $T=5$ MeV, Fig.~$\ref{KT}(b)$ at $T=10$ MeV and Fig.~$\ref{KT}(c)$ at $T=25$ MeV. Fig.~$\ref{KT}(a)$ shows that $K_T^{(\parallel)}$ is smaller than $K_T^{(\perp)}$ for given values of $T,~\mu_B,~eB$. Therefore, the equation of state is stiffer along the magnetic field direction. Figs.~$\ref{KT}(b)-(c)$ show a similar behaviour at higher temperatures. In all the graphs both the compressibilities become smaller at higher $\mu_B$ indicating a stiffer EoS in this region.

\section{Summary \& Conclusion}\label{SC}
In summary, we have investigated the  modifications of nucleon mass, the nuclear liquid-gas phase transition, squared speed of sound  and isothermal compressibility in nuclear matter subjected to a background magnetic field at finite temperature and chemical potential (baryon density) within the framework of the nonlinear Walecka model.
Our findings reveal that the effective mass of the nucleon increases with the growing background magnetic field, a phenomenon known as Magnetic Catalysis. Additionally, the presence of a magnetic field is found to influence the positions of the critical end point (CEP) and spinodal lines in the $T-\mu_B$ plane. Furthermore, our study demonstrates that the presence of magnetic field induces anisotropy in the sound speed, showing variations between ${C^2_x}^{(\parallel, \perp)}$ components. Our calculations support the assertion that the sound speed in nuclear matter can exceed $\sqrt[]{1/3}$ at high chemical potential even in the presence of the magnetic field. However, it is important to note that causality is always upheld, ensuring $C^2_x<1$. Notably, we also observed that the influence of the magnetic field on the sound speed is most pronounced at high chemical potential  and low temperature. Moreover, our investigation revealed that the magnetic field can induce anisotropy in the isothermal compressibility of nuclear matter in a similar manner. It is found that $K_T^{(\parallel)}$ is smaller than $K_T^{(\perp)}$ for given values of $T,~\mu_B,~eB$ indicating that the equation of state is stiffer along the magnetic field direction.
\section*{Acknowledgments}
The authors thank Snigdha Ghosh for valuable discussions at various stages of the work.
\appendix
\section{$eB-$Dependent Vacuum Contribution}\label{A1}
The vacuum contribution to the free energy is 
\begin{eqnarray}
\Omega_{\rm sea}&=&-2\int\frac{d^3k}{\FB{2\pi}^3}E-\frac{qB}{2\pi}\sum_{n=0}^{\infty}\alpha_n\int\frac{dk_z}{2\pi}E_n\\
&=&-I_0-I_1=-2I_0-\FB{I_1-I_0}
\end{eqnarray}
where
\begin{eqnarray}
I_1&=&\frac{qB}{2\pi}\sum_{n=0}^{\infty}\alpha_n\int\frac{dk_z}{2\pi}E_n~~,\label{I1}\\
I_0&=&2\int\frac{d^3k}{\FB{2\pi}^3}E\label{I0}~.
\end{eqnarray}
The momentum integration in $d-$dimension is
\begin{eqnarray}
	\int\frac{d^np}{\FB{2\pi}^n}\FB{p^2+M^2}^{-A}=\frac{\Gamma\FB{A-d/2}}{\FB{4\pi}^{d/2}\Gamma(A)}\FB{\frac{1}{M^2}}^{A-d/2}\label{IntIden}~.
\end{eqnarray}
We first convert the integration in Eq.~\ref{I1} into $d-$dimension and then use the standard dimensional regularization formula Eq.~\ref{IntIden} with $A=-\frac{1}{2}$ and $d=1-\epsilon(<<1)$. Thus,

\begin{eqnarray}
	I_1&=&\frac{\FB{eB}}{2\pi}\frac{\Gamma(-1+\epsilon/2)}{\FB{4\pi}^{(1-\epsilon)/2}\Gamma(-1/2)}\TB{2\sum_{n=0}^{\infty}\FB{\frac{1}{M^2+2neB}}^{-1+\frac{\epsilon}{2}}-\FB{\frac{1}{M^2}}^{-1+\frac{\epsilon}{2}}}~~.\label{I1_1}
\end{eqnarray}
Now defining $x=\frac{M^2}{2eB}$ and using the formula of Hurwitz zeta function $\zeta(z,x)=\sum_{n=0}^{\infty}\frac{1}{\FB{n+x}^z}$ Eq.~\ref{I1_1} can be written as
\begin{eqnarray}
I_1&=&-\frac{\FB{eB}^2}{2\pi^2}\FB{{\frac{eB}{2\pi}}}^{-\frac{\epsilon}{2}}\Gamma\FB{-1+\frac{\epsilon}{2}}\TB{\zeta\FB{-1+\frac{\epsilon}{2},x}-\frac{1}{2x^{-1+\frac{\epsilon}{2}}}}~.\label{I1_2}
\end{eqnarray}
Eq.~\ref{I1_2} can be further simplified to obtain
\begin{eqnarray}
 I_1&=&\frac{\FB{eB}^2}{2\pi^2}\TB{-\frac{x^2}{\epsilon}-\frac{1}{2}x^2+\frac{1}{2}\gamma x^2+\frac{1}{2}x^2\text{ln}\frac{eB}{2\pi}+\frac{1}{2}x\text{ln}x+\zeta '(-1,x)}+x\text{-independent terms}~.
\end{eqnarray}
Similarly, we convert the integration in Eq.~\ref{I0} into $d$-dimension using the standard dimensional regularization formula  Eq.~\ref{IntIden} with $A=-\frac{1}{2}$ and $d=3-\epsilon(<<1)$. Then the simplified equation can be written as:
\begin{eqnarray}
I_0&=&-\frac{1}{8\pi}M^4\FB{\frac{1}{\epsilon}+\frac{3-2\gamma}{4}-\frac{1}{2}\text{ln}M^2+\frac{1}{2}\text{ln}4\pi}~.\label{I0_1}
\end{eqnarray}
Inserting $M^2=2eBx$ and simplifying, Eq.~\ref{I0_1} can be expressed as 

\begin{eqnarray}
	I_0=-\frac{(eB)^2}{2\pi^2}\TB{\frac{x^2}{\epsilon}+\frac{3-2\gamma}{4}x^2-\frac{1}{2}x^2\text{ln}x-\frac{1}{2}x^2\text{ln}\frac{eB}{2\pi}}~.
\end{eqnarray}
Now
\begin{eqnarray}
\Omega_{\rm sea}&=&-2I_0-\FB{I_1-I_0}\\
&=&\FB{2I_0+x\text{-independent terms}}+\Omega_{\rm vac}^B
\end{eqnarray}
where $eB-$dependent vacuum part is
\begin{eqnarray}
\Omega_{\rm vac}^B&=&-\frac{(eB)^2}{2\pi^2}\SB{\zeta'\FB{-1,x}+\frac{1}{4}x^2+\frac{1}{2}x(1-x)~\text{ln}~x}~.
\end{eqnarray}

\section{Important Thermodynamic Relation}\label{A2}
\begin{eqnarray}
\FB{\frac{\partial\epsilon}{\partial T}}_{\mu_B}&=&T\FB{\frac{\partial s}{\partial T}}_{\mu_B}+\mu_B\FB{\frac{\partial n_B}{\partial T}}_{\mu_B}~~,~~~~~
\FB{\frac{\partial\epsilon}{\partial\mu_B}}_T=T\FB{\frac{\partial s}{\partial\mu_B}}_T+\mu_B\FB{\frac{\partial n_B}{\partial\mu_B}}_T~~,
\end{eqnarray}
\begin{eqnarray}
\FB{\frac{\partial(s/n_B)}{\partial\mu_B}}_{T}&=&\frac{1}{n_B}\FB{\frac{\partial s}{\partial\mu_B}}_T-\frac{s}{n_B^2}\FB{\frac{\partial n_B}{\partial\mu_B}}_T~~,~~~~~\FB{\frac{\partial(s/n_B)}{\partial T}}_{\mu_B}=\frac{1}{n_B}\FB{\frac{\partial s}{\partial T}}_T-\frac{s}{n_B^2}\FB{\frac{\partial n_B}{\partial T}}_{\mu_B}.
\end{eqnarray}

\section{Susceptibilities}\label{Appendix.ES}

In section \ref{SpdSound}, we have seen that the speed of sound contain $\FB{\frac{\partial\mathcal{M}}{\partial T}}_{\mu_B}$, $\FB{\frac{\partial\mathcal{M}}{\partial\mu_B}}_T$, $\FB{\frac{\partial s}{\partial \mu_B}}_T$, $\FB{\frac{\partial s}{\partial T}}_{\mu_B}$, $\FB{\frac{\partial n_B}{\partial \mu_B}}_T$, $\FB{\frac{\partial n_B}{\partial T}}_{\mu_B}$ and can be obtained from the free energy  $\Omega$. The expressions are given below:

\begin{eqnarray}
	\FB{\frac{\partial\mathcal{M}}{\partial T}}_{\mu_B}&=&-\frac{M}{2\pi^2}\TB{2x(1-\text{ln}x)+\text{ln}\Gamma(x)+x\FB{\Psi(x)-1}+\frac{1}{2}\FB{1+\text{ln}\frac{x}{2\pi}}}\frac{\partial M}{\partial T}-\sum_{n=0}^{\infty}\alpha_n\int\frac{dp_z}{4\pi^2}\TB{\text{ln}(1-f_n^+)+\text{ln}(1-f_n^-)}\nn\\
	&&-T\sum_{n=0}^{\infty}\alpha_n\int\frac{dp_z}{4\pi^2}\TB{-\frac{E_n}{T^2}(f_n^++f_n^-)-\frac{\mu^\star}{T^2}(f_n^--f_n^+)+\frac{1}{T}\frac{M}{E_n}\frac{\partial M}{\partial T}(f_n^++f_n^-)+\frac{1}{T}\frac{\partial\mu^\star}{\partial T}(f_n^--f_n^+)}\nn\\
	&&-eB\sum_{n=0}^{\infty}\alpha_n\int\frac{dp_z}{4\pi^2}\frac{n}{E_n}\left[-\frac{M}{E_n^2}\frac{\partial M}{\partial T}(f^+_n+f^-_n)+\frac{1}{T^2}\SB{(E_n+\mu^\star)f^-(1-f^-)+(E_n-\mu^\star)f^+_n(1-f^+_n)}\right. \nn\\&&\hspace{3.5cm}\left. -\frac{1}{T}\frac{M}{E_n}\frac{\partial M}{\partial T}\SB{f^+_n(1-f^+_n)+f^-_n(1-f^-_n)}-\frac{1}{T}\frac{\partial\mu^\star}{\partial T}\SB{f^-_n(1-f^-_n)-f^+_n(1-f^+_n)}\right]~~,\\
	%
	\FB{\frac{\partial\mathcal{M}}{\partial\mu_B}}_T&=&-\frac{M}{2\pi^2}\TB{2x(1-\text{ln}x)+\text{ln}\Gamma(x)+x\FB{\Psi(x)-1}+\frac{1}{2}\FB{1+\text{ln}\frac{x}{2\pi}}}\frac{\partial M}{\partial\mu_B}\nn\\
	&&-\sum_{n=0}^{\infty}\alpha_n\int\frac{dp_z}{4\pi^2}\TB{\frac{M}{E_n}\frac{\partial M}{\partial\mu_B}(f_n^++f^-_n)+\frac{\partial\mu^\star}{\partial\mu_B}(f_n^--f^+_n)}+qB\sum_{n=0}^{\infty}\alpha_n\int\frac{dp_z}{4\pi^2}\frac{n}{E_n^3}M(f^+_n+f^-_n)\frac{\partial M}{\partial\mu_B}\nn\\
	&&-\frac{eB}{T}\sum_{n=0}^{\infty}\alpha_n\int\frac{dp_z}{4\pi^2}\frac{n}{E_n}\TB{\frac{M}{E_n}\frac{\partial M}{\partial\mu_B}\SB{f^+_n(1-f^+_n)+f^-_n(1-f^-_n)}+\frac{\partial\mu^\star}{\partial\mu}\SB{f^-_n(1-f^-_n)-f^+_n(1-f^+_n)}}
\end{eqnarray}	
where $\Psi(x)=\frac{\partial}{\partial x}(\text{ln}~\Gamma(x))$ is digamma function and
\begin{eqnarray}
\FB{\frac{\partial s}{\partial T}}_{\mu_B}&=&\int\frac{d^3k}{(2\pi)^3}\frac{1}{T^2}\TB{\SB{\frac{E-\mu^\star}{T}-\frac{M}{E}\frac{\partial M}{\partial T}+\frac{\partial \mu^\star}{\partial T}}\FB{E-\mu^\star}f^+(1-f^+)+\SB{\frac{E+\mu^\star}{T}-\frac{M}{E}\frac{\partial M}{\partial T}-\frac{\partial \mu^\star}{\partial T}}\FB{E+\mu^\star}f^-(1-f^-)}\nn\\
&+&\frac{qB}{2\pi}\sum_{n=0}^{\infty}\alpha_n\int\frac{dk_z}{2\pi}\frac{1}{T^2}\left[ \SB{\frac{E_n-\mu^\star}{T}-\frac{M}{E_n}\frac{\partial M}{\partial T}+\frac{\partial \mu^\star}{\partial T}}\FB{E_n-\mu^\star}f_n^+(1-f_n^+)+\SB{\frac{E_n+\mu^\star}{T}-\frac{M}{E_n}\frac{\partial M}{\partial T}-\frac{\partial \mu^\star}{\partial T}}\right. \nn\\&&\left. \hspace{12cm}\FB{E_n+\mu^\star}f_n^-(1-f_n^-)\right]~~,\label{dsbydT}
%
\\\FB{\frac{\partial s}{\partial\mu_B}}_{T}&=&-\frac{qB}{2\pi}\frac{1}{T^2}\sum_{n=0}^{\infty}\alpha_n\int\frac{dk_z}{2\pi}\TB{\SB{\frac{M}{E_n}\frac{\partial M}{\partial \mu_B}-\frac{\partial \mu^\star}{\partial \mu_B}}\FB{E_n-\mu^\star}f_n^+(1-f_n^+)+\SB{\frac{M}{E_n}\frac{\partial M}{\partial\mu_B}+\frac{\partial\mu^\star}{\partial\mu_B}}\FB{E_n+\mu^\star}f_n^-(1-f_n^-)}\nn\\
&-&\frac{2}{T^2}\int\frac{d^3k}{(2\pi)^3}\TB{\SB{\frac{M}{E}\frac{\partial M}{\partial \mu_B}-\frac{\partial \mu^\star}{\partial \mu_B}}\FB{E-\mu^\star}f^+(1-f^+)+\SB{\frac{M}{E}\frac{\partial M}{\partial\mu_B}+\frac{\partial\mu^\star}{\partial\mu_B}}\FB{E+\mu^\star}f^-(1-f^-)}~~,\label{dsbydMu}
\end{eqnarray}
\begin{eqnarray}
\FB{\frac{\partial n_B}{\partial T}}_{\mu_B}&=&\frac{qB}{2\pi}\sum_{n=0}^{\infty}\alpha_n\int\frac{dk_z}{2\pi}\TB{\SB{\frac{E_n-\mu^\star}{T^2}-\frac{1}{T}\FB{\frac{M}{E_n}\frac{\partial M}{\partial T}-\frac{\partial\mu^\star}{\partial T}}}f_n^+(1-f_n^+)-\SB{\frac{E_n+\mu^\star}{T^2}-\frac{1}{T}\FB{\frac{M}{E_n}\frac{\partial M}{\partial T}+\frac{\partial\mu^\star}{\partial T}}}f_n^-(1-f_n^-)}\nn\\
&+&2\int\frac{d^3k}{(2\pi)^3}\TB{\SB{\frac{E-\mu^\star}{T^2}-\frac{1}{T}\FB{\frac{M}{E}\frac{\partial M}{\partial T}-\frac{\partial \mu^\star}{\partial T}}}f^+(1-f^+)-\SB{\frac{E+\mu^\star}{T^2}-\frac{1}{T}\FB{\frac{M}{E}\frac{\partial M}{\partial T}+\frac{\partial\mu^\star}{\partial T}}}f^-(1-f^-)}\label{dnBbydT}\\
&=&Y_T+Y_M\frac{\partial M}{\partial T}+Y_{\mu^\star}\frac{\partial \mu^\star}{\partial T}~~,
\\
\FB{\frac{\partial n_B}{\partial\mu_B}}_{T}&=&\frac{qB}{2\pi}\sum_{n=0}^{\infty}\alpha_n\int\frac{dk_z}{2\pi}\TB{-\frac{1}{T}\FB{\frac{M}{E_n}\frac{\partial M}{\partial \mu_B}-\frac{\partial\mu^\star}{\partial\mu_B}}f_n^+(1-f_n^+)+\frac{1}{T}\FB{\frac{M}{E_n}\frac{\partial M}{\partial\mu_B}+\frac{\partial\mu^\star}{\partial\mu_B}}f_n^-(1-f_n^-)}\nn\\
&&\hspace{3cm}+~2\int\frac{d^3k}{(2\pi)^3}\TB{-\frac{1}{T}\FB{\frac{M}{E}\frac{\partial M}{\partial \mu_B}-\frac{\partial \mu^\star}{\partial\mu_B}}f^+(1-f^+)+\frac{1}{T}\FB{\frac{M}{E}\frac{\partial M}{\partial\mu_B}+\frac{\partial\mu^\star}{\partial\mu_B}}f^-(1-f^-)}\label{dnBbydMu}\\
&=& Y_M\frac{\partial M}{\partial\mu_B}+Y_{\mu^\star}\frac{\partial \mu^\star}{\partial \mu_B}~~,
\end{eqnarray}
\begin{eqnarray}
\FB{\frac{\partial n_s}{\partial T}}_{\mu_B}&=&2\int\frac{d^3k}{(2\pi)^3}\FB{\frac{1}{E}-\frac{M^2}{E^3}}\frac{\partial M}{\partial T}(f^++f^-)+\frac{eB}{2\pi}\sum_{n=0}^{\infty}\alpha_n\int\frac{dk_z}{2\pi}\FB{\frac{1}{E_n}-\frac{M^2}{E_n^3}}\frac{\partial M}{\partial T}(f_n^++f_n^-)\nn\\
&&+2\int\frac{d^3k}{(2\pi)^3}\frac{M}{E}\TB{\SB{\frac{E+\mu^\star}{T^2}-\frac{1}{T}\FB{\frac{M}{E}\frac{\partial M}{\partial T}+\frac{\partial \mu^\star}{\partial T}}}f^-(1-f^-)+\SB{\frac{E-\mu^\star}{T^2}-\frac{1}{T}\FB{\frac{M}{E}\frac{\partial M}{\partial T}-\frac{\partial \mu^\star}{\partial T}}}f^+(1-f^+)}\nn\\
&&+\frac{eB}{2\pi}\sum_{n=0}^{\infty}\alpha_n\int\frac{dk_z}{2\pi}\frac{M}{E_n}\TB{\SB{\frac{E_n+\mu^\star}{T^2}-\frac{1}{T}\FB{\frac{M}{E_n}\frac{\partial M}{\partial T}+\frac{\partial \mu^\star}{\partial T}}}f_n^-(1-f_n^-)+\SB{\frac{E_n-\mu^\star}{T^2}-\frac{1}{T}\FB{\frac{M}{E_n}\frac{\partial M}{\partial T}-\frac{\partial \mu^\star}{\partial T}}}f_n^+(1-f_n^+)}\nn\\
&=&X_T+X_M\frac{\partial M}{\partial T}+X_{\mu^\star}\frac{\partial \mu^\star}{\partial T}~~,
\\
\FB{\frac{\partial n_s}{\partial \mu_B}}_{T}&=&2\int\frac{d^3k}{(2\pi)^3}\FB{\frac{1}{E}-\frac{M^2}{E^3}}\frac{\partial M}{\partial\mu_B}(f^++f^-)+\frac{eB}{2\pi}\sum_{n=0}^{\infty}\alpha_n\int\frac{dk_z}{2\pi}\FB{\frac{1}{E_n}-\frac{M^2}{E_n^3}}\frac{\partial M}{\partial\mu_B}(f_n^++f_n^-)\nn\\
&&\hspace{2cm}+2\int\frac{d^3k}{(2\pi)^3}\frac{M}{E}\TB{\SB{-\frac{1}{T}\FB{\frac{M}{E}\frac{\partial M}{\partial\mu_B}+\frac{\partial \mu^\star}{\partial\mu_B}}}f^-(1-f^-)+\SB{-\frac{1}{T}\FB{\frac{M}{E}\frac{\partial M}{\partial\mu_B}-\frac{\partial \mu^\star}{\partial\mu_B}}}f^+(1-f^+)}\nn\\
&&\hspace{2cm}+\frac{eB}{2\pi}\sum_{n=0}^{\infty}\alpha_n\int\frac{dk_z}{2\pi}\frac{M}{E_n}\TB{\SB{-\frac{1}{T}\FB{\frac{M}{E_n}\frac{\partial M}{\partial\mu_B}+\frac{\partial \mu^\star}{\partial\mu_B}}}f_n^-(1-f_n^-)+\SB{-\frac{1}{T}\FB{\frac{M}{E_n}\frac{\partial M}{\partial\mu_B}-\frac{\partial \mu^\star}{\partial\mu_B}}}f_n^+(1-f_n^+)}\nn\\
&=&X_M\frac{\partial M}{\partial\mu_B}+X_{\mu^\star}\frac{\partial \mu^\star}{\partial \mu_B}
\end{eqnarray}
where
\begin{eqnarray}
X_T&=&2\int \frac{d^3k}{\FB{2\pi}^3}\frac{M}{E}\frac{1}{T^2}\SB{\FB{E+\mu^\star}f^-\FB{1-f^-}+\FB{E-\mu^\star}f^+\FB{1-f^+}}\nn\\
&&\hspace{2cm}+\frac{eB}{2\pi}\sum_{n=0}^{\infty}\alpha_n\int\frac{dk_z}{2\pi}\frac{M}{E_n}\frac{1}{T^2}\SB{\FB{E_n+\mu^\star}f_n^-\FB{1-f_n^-}+\FB{E_n-\mu^\star}f_n^+\FB{1-f_n^+}}~~,
\\
X_M&=&2\int \frac{d^3k}{\FB{2\pi}^3}\TB{\FB{\frac{1}{E}-\frac{M^2}{E^3}}\FB{f^++f^-}-\frac{1}{T}\FB{\frac{M}{E}}^2\SB{f^+\FB{1-f^+}+f^-\FB{1-f^-}}}\nn\\
&&\hspace{2cm}+\frac{eB}{2\pi}\sum_{n=0}^{\infty}\alpha_n\int\frac{dk_z}{2\pi}\TB{\FB{\frac{1}{E_n}-\frac{M^2}{E_n^3}}\FB{f_n^++f_n^-}-\frac{1}{T}\FB{\frac{M}{E_n}}^2\SB{f_n^+\FB{1-f_n^+}+f_n^-\FB{1-f_n^-}}}~~,
\\
X_{\mu^\star}&=&2\int\frac{d^3k}{(2\pi)^3}\frac{1}{T}\frac{M}{E}\SB{f^+\FB{1-f^+}-f^-\FB{1-f^-}}+\frac{eB}{2\pi}\sum_{n=0}^{\infty}\alpha_n\int\frac{dk_z}{2\pi}\frac{1}{T}\frac{M}{E_n}\SB{f_n^+\FB{1-f_n^+}-f_n^-\FB{1-f_n^-}}~~,
\\
Y_T&=&2\int\frac{d^3k}{(2\pi)^3}\frac{1}{T^2}\SB{(E-\mu^\star)f^+(1-f^+)-(E+\mu^\star)f^-(1-f^-)}\nn\\
&&\hspace{2cm}+\frac{eB}{2\pi}\sum_{n=0}^{\infty}\alpha_n\int\frac{dk_z}{2\pi}\frac{1}{T^2}\SB{(E_n-\mu^\star)f_n^+(1-f_n^+)-(E_n+\mu^\star)f_n^-(1-f_n^-)}~~,
\\
Y_M&=&2\int\frac{d^3k}{(2\pi)^3}\frac{1}{T}\frac{M}{E}\SB{-f^+(1-f^+)+f^-(1-f^-)}+\frac{eB}{2\pi}\sum_{n=0}^{\infty}\alpha_n\int\frac{dk_z}{2\pi}\frac{1}{T}\frac{M}{E_n}\SB{-f_n^+(1-f_n^+)+f_n^-(1-f_n^-)}~~,
\\
Y_{\mu^\star}&=&2\int\frac{d^3k}{(2\pi)^3}\frac{1}{T}\SB{f^+(1-f^+)+f^-(1-f^-)}+\frac{eB}{2\pi}\sum_{n=0}^{\infty}\alpha_n\int\frac{dk_z}{2\pi}\frac{1}{T}\SB{f_n^+(1-f_n^+)+f_n^-(1-f_n^-)}~~.
\end{eqnarray}
The derivatives $\frac{\partial M}{\partial T}$, $\frac{\partial \mu^\star}{\partial T}$, $\frac{\partial M}{\partial \mu_B}$, $\frac{\partial \mu^\star}{\partial \mu_B}$ in Eqs.~\eqref{dsbydT}, \eqref{dsbydMu}, \eqref{dnBbydT}, \eqref{dnBbydMu} can be obtained analytically from Eqs.~\eqref{ns1}, \eqref{nB1}, \eqref{ns2}, \eqref{nB2} and are given by the matrix equations as 
\begin{eqnarray}
\begin{pmatrix}
f'-X_M & -X_{\mu^\star} \\
Y_M & Y_{\mu^\star}+\frac{1}{\FB{g_\omega/m_\omega}^2} \\
\end{pmatrix}
\begin{pmatrix}
\frac{\partial M}{\partial T} \\
\frac{\partial \mu^\star}{\partial T}
\end{pmatrix}	
=\begin{pmatrix} X_T \\ -Y_T\end{pmatrix}~~,
\end{eqnarray}
\begin{eqnarray}
\begin{pmatrix}
f'-X_M & -X_{\mu^\star} \\
Y_M & Y_{\mu^\star}+\frac{1}{\FB{g_\omega/m_\omega}^2}
\end{pmatrix}
\begin{pmatrix}
\frac{\partial M}{\partial\mu_B} \\
\frac{\partial \mu^\star}{\partial\mu_B}	
\end{pmatrix}
=\begin{pmatrix} 0\\ \frac{1}{\FB{g_\omega/m_\omega}^2}\end{pmatrix}
\end{eqnarray}
where
\begin{eqnarray}
f&=&-\frac{M-m_N}{\FB{g_\sigma/m_\sigma}^2}+bm_N\FB{m_N-M}^2+c\FB{m_N-M}^3+\frac{qB}{2\pi^2}M\SB{x(1-\text{ln}~x)+\frac{1}{2}\text{ln}\frac{x}{2\pi}+\text{ln}~\Gamma(x)}~~,
\\
f'&=&\frac{\partial f(M)}{\partial M}
=-\frac{1}{\FB{g_\sigma/m_\sigma}^2}-2bm_N(m_N-M)-3c(m_N-M)^2\nn\\
&&\hspace{3.5cm}+\frac{eB}{2\pi^2}\SB{x(1-\text{ln}~x)+\frac{1}{2}\text{ln}~\frac{x}{2\pi}+\text{ln}~\Gamma(x)}+\frac{M^2}{2\pi^2}\SB{\Psi(x)+\frac{1}{2x}-\text{ln}~x}~.
\end{eqnarray}
\bibliographystyle{apsrev4-1}
\bibliography{Ref.bib}

\end{document}